\documentclass[fleqn,10pt,twocolumn]{wlscirep}

\usepackage{blindtext}
\usepackage{dcolumn}% Align table columns on decimal point
\usepackage{bm}% bold math
\usepackage{color, soul, colortbl} %to highlight text ==> \hl{text}
\usepackage{url} %It provides better support for handling and breaking URLs.

\usepackage{booktabs}
\usepackage{multirow} % to do multirow or multicolumn tabs in tables
\usepackage[overload]{empheq}
\usepackage{mathtools}
\usepackage{cases} 
\usepackage{cancel} % to make a strikethrough in math mode (diagonal bar)
\usepackage{mdframed} %box around text         
\usepackage{afterpage} % \afterpage{\clearpage} to flush all floats without pagebreak

\usepackage{enumitem}
\usepackage{graphicx,epsfig}% Include figure files
\usepackage{subfigure}
\usepackage{amsmath}%,amsfonts,multirow,rotate,color}amssymb
\usepackage{xfrac}%simple fraction layout \sfrac{}{}
\usepackage[normalem]{ulem} % crossing out text % use \sout{#1} to strikethrough
\usepackage{mathtools}

\usepackage{dblfloatfix} % The package solves two problems: floats in a twocolumn document come out in the right order and allowed float positions are now [tbp].

\definecolor{airforceblue}{rgb}{0.36, 0.54, 0.66}
\definecolor{brickred}{rgb}{0.8, 0.25, 0.33}
\definecolor{amber}{rgb}{1.0, 0.75, 0.0}
\definecolor{applegreen}{rgb}{0.55, 0.71, 0.0}
\definecolor{magenta}{rgb}{0.965, 0, 0.859}

\title{The dynamics of higher-order novelties}
\newcommand{\deletetext}[1]{\iffalse{{\color{red}{#1}}}\fi}
\newcommand{\newtext}[1]{{{#1}}}
\author[1, 2, 3, 4]{Gabriele Di Bona}
\author[3, 4, 5]{Alessandro Bellina}
\author[4, 5, 6, 7]{Giordano De Marzo}
\author[8]{Angelo Petralia}
\author[9, 10]{Iacopo Iacopini}
\author[1, 7, 11, *]{Vito Latora}
\affil[1]{School of Mathematical Sciences, Queen Mary University of London, London E1 4NS, United Kingdom}
\affil[2]{CNRS, GEMASS, 59 rue Pouchet, F-75017, Paris, France}
\affil[3]{Sony Computer Science Laboratories Rome, I-00184, Rome, Italy}
\affil[4]{Centro Ricerche Enrico Fermi, I-00184 Rome, Italy}
\affil[5]{Dipartimento di Fisica Universit\`a ``Sapienza”, I-00185 Rome, Italy.}
\affil[6]{Sapienza School for Advanced Studies, ``Sapienza'', I-00185 Rome, Italy.}
\affil[7]{Complexity Science Hub Vienna, A-1080 Vienna, Austria}
\affil[8]{Department of Economics and Business, University of Catania, I-95128 Catania, Italy}
\affil[9]{Network Science Institute, Northeastern University London, London, E1W 1LP, United Kingdom}
\affil[10]{Department of Physics, Northeastern University, Boston, MA 02115, USA}
\affil[11]{Dipartimento di Fisica ed Astronomia, Universit\`a di Catania and INFN, I-95123 Catania, Italy}

\affil[*]{Email: v.latora@qmul.ac.uk}

\begin{abstract}
\deletetext{Understanding how humans explore the world in search of novelties is key to foster innovation.}
\newtext{Studying how we explore the world in search of novelties is key to understand the mechanisms that can lead to new discoveries.}
Previous studies analyzed novelties in various exploration processes, defining them as the first appearance of an element.
However, \newtext{novelties} can also be generated by \deletetext{novel association of}\newtext{combining} what is already known.  
We hence define higher-order novelties 
\newtext{as the first time two or more elements appear together}, 
and we introduce higher-order Heaps' exponents as a way to characterize their pace of discovery.
Through extensive analysis of real-world data, we find that processes with the same pace of discovery, as measured by the standard Heaps' exponent, can instead differ at higher orders. 
We then propose to model an exploration process as a random walk
\newtext{on a network in which the possible connections between elements evolve in time}. 
The model reproduces the empirical properties of higher-order novelties, revealing how \newtext{the network we explore changes over time along with the exploration process.}
\end{abstract}

\begin{document}

\flushbottom
\maketitle
\thispagestyle{empty}

%%%%%%%%%%%%%%%%%%%%%%%%%%%%%%%%%%%%%%%%%%%%%%%%%%%%%%%%%%%%%%%%
%INTRODUCTION
%%%%%%%%%%%%%%%%%%%%%%%%%%%%%%%%%%%%%%%%%%%%%%%%%%%%%%%%%%%%%%%%
\section*{Introduction}

As humans, we experience novelties as part of our daily life. By the term {\it novelty} we generally indicate two apparently different things~\cite{tria2014dynamics}. 
On the one hand, we can think of a novelty as the first time we visit a neighborhood, enter a newly launched pub, or listen to a song from an artist we previously did not know. In this case, the novelty represents a discovery for a single individual of a place, an artist or, more in general, an item. On the other hand, there are discoveries that are new to the entire population, as could be a technological advancement or the development of a new drug. However, these two cases are not entirely distinct, as the second set of novelties, those new to everyone, represent\deletetext{ just as} a subset of the first one. 
Analysing how novelties emerge\newtext{,} both at the individual level\deletetext{,} and at the level of the entire population, is key to understand human creativity and the neural and social mechanisms that can lead to new discoveries\deletetext{ and innovation}. 

The increasing availability of data on human behavior and consumption habits has allowed to study how humans explore the world, how novelties emerge in different contexts, and how they are distributed in time~\cite{tria2014dynamics, monechi2017waves, iacopini2018network}. Empirical investigations cover a broad range of different areas~\cite{north2013novelty}\newtext{, }ranging from science~\cite{rzhetsky2015choosing} and language~\cite{loreto2016dynamics,puglisi2008cultural}, to gastronomy~\cite{fink2017serendipity}, goods \deletetext{and }\newtext{or }products~\cite{saracco2015innovation}, network science~\cite{abbas2018popularity}, information~\cite{rodi2017search}, and cinema~\cite{sreenivasan2013quantitative}\newtext{, to name a few relevant examples}.
No matter the topic, one can always represent data coming from real-world exploration processes as sequences of \deletetext{items}\newtext{elements or ``items"} that are sequentially adopted or consumed~\cite{iacopini2021dual}. 
\deletetext{In this way}\newtext{For instance}, the activity of a user \deletetext{of, for example,}\newtext{on} an online digital music platform is turned into a sequence of listened songs, and a novelty is defined as the first time a song, or an artist, appears in the sequence~\cite{dibona2022social}. 
Analogously, articles published in a scientific journal can be turned into a time-ordered sequence of concepts or keywords discovered by the community, and a novelty can be defined, again, as  the first-time appearance of a keyword~\cite{iacopini2018network}.    
Under this framework, evidence shows that---independently of the system they belong to---novelties 
seem to obey the same statistical patterns \deletetext{on}\newtext{in} the way they are distributed and correlated in time~\cite{tria2014dynamics}. 
\newtext{Indeed, a long tradition of works, started by the Yule-Simon processes for text generation~\cite{yule1924evolution,Simon1955skewdistributions}, shows that}  \deletetext{In particular,} most empirical sequences follow Heaps'~\cite{herdan1960type,heaps78, lu2010zipf}, Zipf's~\cite{estoup1916gammes, zipf1929relative, zipf1935language, zipf1949leasteffort, de2021dynamical}, and Taylor's laws~\cite{taylor1961aggregation}.

Along with data-driven investigations, a relevant scientific problem is that of finding plausible mechanisms to reproduce and explain the empirical observations. What are the \deletetext{rules}\newtext{drivers} controlling the appearance of new items in a sequence? How do humans explore the seemingly infinite space of possibilities in search of novelties? Interestingly, an insightful answer comes from biology, \deletetext{when}\newtext{where}, in 1996, Stuart Kauffman introduced the concept of the {\it adjacent possible}~\cite{kauffman1996investigations} (AP)\deletetext{---}\newtext{ referring to }``\textit{all those molecular species that are not members of the actual, but are one reaction step away from the actual}". Inspired by previous works by Packard and Langton~\cite{packard1988adaptation, langton1990computation,langton2003artificial}, the AP provides a fresh view on the problem, for which discoveries (the possible) can only be found among those items which are close (the adjacent) to what is already known (the actual). New discoveries would then generate an expanding space of opportunities that are only available to us in the moment we ``unlock'' what is adjacent to them. 
Kauffman's AP has seen many interesting applications ranging from biology~\cite{bak1993punctuated, kauffman1996investigations} and economics~\cite{saracco2015innovation,armano2017beneficial} to models of discovery and innovation\deletetext{ processes}\newtext{\cite{tria2014dynamics,loreto2016dynamics,iacopini2018network}}. Among these, of particular interest is the recently proposed Urn Model with Triggering (UMT)~\cite{tria2014dynamics, loreto2016dynamics, gravino2016crossing}. Building upon the work of P{\'o}lya~\cite{eggenberger1923polya, hoppe1984polya}, the UMT adds to the traditional {\em reinforcement} mechanism of the P{\'o}lya urn's scheme a {\em triggering} mechanism that expands the space of possible discoveries upon the extraction of each novelty. 
Being able to reproduce the empirical laws\newtext{ and thanks to its simplicity}, 
the UMT has been used to study\newtext{ various systems  with an expanding set of ``items", like} the rise and fall of popularity in technological and artistic productions~\cite{monechi2017waves}, the emergence and evolution of social networks~\cite{ubaldi2021emergence}, and the evolution of the cryptocurrency ecosystem \cite{marzo2022modeling}.
\deletetext{The AP accounts for the emergence of the new starting from the ``edge of what is known''. In this view, o}%
\newtext{O}ne could also picture ideas, concepts, or items as the linked elements of an abstract network. \deletetext{Within this framework}\newtext{In this view}, 
the \deletetext{way we explore the world based on the association of different concepts }\newtext{exploration process }can be modelled as a random walk over this network\newtext{, where the AP accounts for the emergence of the new starting from the ``edge of what is known'' within the network}.
Approaches based on random walks have been used to investigate the cognitive growth of knowledge in scientific disciplines~\cite{iacopini2018network}, and further extended to account for multi-agent systems, where the individual exploration of the agent is enriched by social interactions~\cite{iacopini2020interacting, dibona2022social}.

The idea of the \deletetext{adjacent possible}\newtext{AP}, \deletetext{which can be }modelled either in terms of extractions  from urns or random walks over a network, is of great importance to understand the processes leading to \newtext{novelties}\deletetext{ innovation}. 
There is, however, another important mechanism of creation of the new which is neglected by the frameworks discussed above:  novelties can arise from the  combination of already-known elements. 
For instance, a meaningless sequence of words, if ordered in a different way, may generate elegant poetry\newtext{~\cite{lakoff2008metaphors, pinker2003language}}. Novel \deletetext{sequences}\newtext{combinations} of existing hashtags may lead to new social-media trends\newtext{~\cite{boyd2007social, tufekci2014big}}. Different orderings of the same musical notes may in principle generate an endless number of songs\newtext{~\cite{cope2000algorithmic}}. 
The mechanics of combination of ``pre-existing'' items has been studied in various fields, e.g.\newtext{,} in biology\deletetext{,} \newtext{where new associations of various entities} \deletetext{combinations}\deletetext{ are the keys to} produce new organisms. 
\deletetext{For instance, i}\newtext{I}t has been shown that the immune system recombines existing segments of genes to produce new receptors~\cite{market2003immunesystem,jones2004immunesystem}.
\deletetext{P}\newtext{Also, p}ublications and collaborations in 
science~\cite{fortunato2018science} 
are typically combinations of research ideas~\cite{uzzi2013atypical, ke2015defining, wang2017bias} and expertises~\cite{fontana2020new, wu2019large, alvarez2021evolutionary}.
Similarly, in innovation economics, as originally discussed by Schumpeter~\cite{schumpeter1939business, schumpeter2013capitalism} and confirmed by recent works on the generation of technologies~\cite{mcnerney2011role, abbasiharofteh2020atypical, lambert2020pace}, new combinations of existing factors, \newtext{that interact in a technological production process}, may give rise to innovations, which rule out of the market obsolete products and services~\cite{jin2019emergence, leroi2020revolutions}, thus increasing the probability of reaching further \deletetext{novelties and} innovations (the so-called ``creative destruction''). 

\newtext{In this context, t}\deletetext{T}he aim of this paper is to explore a more general notion of novelty\deletetext{ defined as the}\newtext{, including} novel combinations of existing elements.
We thus investigate the dynamics of ``higher-order'' novelties, i.e., \deletetext{novel combinations of pairs, triplets, etc., of consecutive items}\newtext{novel pairs, triplets, etc., of items} in a sequence. 
In particular, we focus on the Heaps' law, which \deletetext{describes}\newtext{characterizes} the growth \deletetext{in}\newtext{of} the number of novelties \newtext{in the sequence} as a power-law, whose exponent is a proxy for the rate of discovery~\cite{heaps78} in \deletetext{a system}\newtext{the related process}. 
Namely, we introduce higher-order Heaps' laws to characterize the rate at which novel combinations of two and more elements appear in a sequence. 
We then analyse various types of empirical sequences\newtext{,} ranging from music listening records, to words in texts, and concepts in scientific articles, finding that Heaps' laws also hold at higher orders. 
We discover that \deletetext{individual }processes with \deletetext{the same}\newtext{a similar} rate of discovery of single items\deletetext{,} can instead display different rates of  discovery at higher orders, and can hence be differentiated \deletetext{in this way}\newtext{by looking at higher-order novelties}. 
We therefore propose a new model which is capable of reproducing \deletetext{all these}\newtext{various} empirically observed features of higher-order Heaps' laws. 
In our model the process of exploration is described as an edge-reinforced random walk with triggering (ERRWT)\newtext{ on a network}. In our framework, 
the novelties at different orders 
(nodes and links visited for the first time by the walker) 
shape the \deletetext{explored}\newtext{growth of the} network by reinforcing traversed links\newtext{,} while\deletetext{, at the same time,} triggering \newtext{the addition of new elements through }the expansion\newtext{ and exploration} of the adjacent possible. This expansion can happen whenever a node is visited for the first time, making other nodes accessible to the explorer, but also whenever a link is firstly used. In this case, the newly established connection will trigger novel combinations between previously explored nodes. By fitting the contributions of the two mechanisms of reinforcement and triggering, the ERRWT model is able to reproduce well the variety of scaling exponents found in real systems for the Heaps' laws at different orders.

%
%%%%%%%%%%%%%%%%%%%%%%%%%%%%%%%%%%%%%%%%%%%%%%%%%%%%%%%%%%%%%%%%
%NUMERICAL RESULTS 
%%%%%%%%%%%%%%%%%%%%%%%%%%%%%%%%%%%%%%%%%%%%%%%%%%%%%%%%%%%%%%%%
%
\section*{Results}
\label{sec:results}

\subsection*{Higher-order Heaps' laws}
\label{sec:heaps}
%
% HEAPS' LAW
%
An exploration process can be represented as an ordered set of $T$ symbols $ \mathcal{S} = \left\{a_1, a_2, \dots, a_T \right\}$\newtext{ sequentially
explored}. 
Such a set describes the sequence of ``events'' or ``items'' produced along the journey, e.g.\newtext{,} the songs listened by a given individual over time, the list of hashtags posted on an online social network, the list of words in a text, or any other ordered list of items or ideas generated by single individuals or social groups~\cite{tria2014dynamics, tria2018zipf, iacopini2020interacting, iacopini2021dual}. 
Similarly, in the context of some recent modelling schemes of discovery, \newtext{the sequence }$\mathcal{S}$ can \deletetext{represent}\newtext{be made of} the \newtext{colors of }balls extracted from an urn~\cite{tria2014dynamics, tria2018zipf, iacopini2020interacting}, or the  nodes visited over time by a random walker moving \deletetext{over}\newtext{on} a network~\cite{iacopini2018network}. 
Although real-world events have an associated time, here, for simplicity, we focus only on their sequence, i.e.\newtext{,} the relative temporal order of the events, neglecting the precise time at which they happen.
For instance, if a person listens to song $a_1$ at time $t_1$, song $a_2$ at time $t_2$, song $a_i$ at time $t_i$, and so on, with $t_1< t_2< \dots < t_i < \dots$, we neglect these times and only retain the order of the songs in the sequence 
$ \left\{a_1, a_2, \dots, a_T \right\}$.
In other words, we assume that $a_1$ is associated to the discrete time $t=1$, $a_2$ is associated to time $t=2$, and so forth. 

Among the different ways to characterize the discovery rate of a given process, the Heaps' law,  $D(t) \sim t^{\beta}$, describes the power-law growth of the number of novelties as a function of\deletetext{ time}\newtext{ the number of items in the sequence}, i.e., how the number $D(t)$ of novel elements in the sequence $ \mathcal{S}$ scale with the sequence length $t$~\cite{heaps78}. The so-called (standard) Heaps' exponent ${\beta}$, that from now on we indicate as \textit{$1$\textsuperscript{st}-order Heaps' exponent} ${\beta_1}$, is thus a measure of the pace of discovery of the process that generated the considered sequence. 
Given that the number of different elements \deletetext{$D(t)$}\newtext{$D_1(t) \equiv D(t)$} is smaller (or equal) than the total length $t$ of the sequence, the value of ${\beta_1}$ is always bounded in the interval $[0,1]$, with the extreme case ${\beta_1}=1$ reached by a process that generates new elements at a linear rate.

% PAIRS
%
Here, we propose to go one step beyond and look at novelties as novel pairs, \deletetext{triples}\newtext{triplets}, and higher-order combinations of consecutive symbols in a sequence~\cite{sinatra2010networks}. 
For instance, when exploring a network, a novel pair is represented by the first visit of a link.
In order to measure the pace of discovery of these higher-order compounds starting from a sequence of events \deletetext{$\mathcal{S}$}\newtext{$\mathcal{S}_1 \equiv \mathcal{S}$}, we first create
the surrogate sequence of overlapping pairs $\mathcal{S}_2 = \left\{(a_1, a_2), (a_2, a_3), \dots, (a_{T-1},a_T) \right\}$.
Considering for example the sentence \textit{``One ring to rule them all''}, from the sequence of events $\deletetext{\mathcal{S}}\newtext{\mathcal{S}_1} = $ \{\textit{one}, \textit{ring}, \textit{to}, \textit{rule}, \textit{them}, \textit{all} \} we obtain the sequence of overlapping pairs $\mathcal{S}_2 = $ \{(\textit{one}, \textit{ring}), (\textit{ring}, \textit{to}), (\textit{to}, \textit{rule}), (\textit{rule}, \textit{them}), (\textit{them}, \textit{all})\}. From $\mathcal{S}_2$ we can then compute the number $D_2(t)$ of different pairs among the first $t$ ones, with $t \leq T-1$. 
Notice that, in this manuscript, we consider the pairs $(\textit{one}, \textit{ring})$ and $(\textit{ring}, \textit{one})$ as two different pairs, i.e., order matters. 
By construction, we always have $D_1(t) \leq D_2(t) \leq t$, since, on the one hand, for each new element added to $\deletetext{\mathcal{S}}\newtext{\mathcal{S}_1}$ there is a new pair in $\mathcal{S}_2$, and, on the other hand, there cannot be more than $t$ different pairs among $t$ items.
From the power-law scaling $D_2(t) \sim t^{\beta_2}$, we can then extract the value of $\beta_2$, which we refer to as the \textit{$2$\textsuperscript{nd}-order Heaps' exponent}.
%
%
% GENERALIZATION
%
This definition can be naturally extended to any order $n$, considering the sequence $\mathcal{S}_n$ of consecutive overlapping $n$-tuples present in $\deletetext{\mathcal{S}}\newtext{\mathcal{S}_1}$. 
Notice that, if $|\deletetext{\mathcal{S}}\newtext{\mathcal{S}_1}| = T$, then $|\mathcal{S}_n| = T - n + 1$. We can hence compute the number $D_n(t)$ of different tuples among the first $t$ tuples in $\mathcal{S}_n$, and extract the \emph{$n$\textsuperscript{th}-order Heaps' exponent} $\beta_n \in [0,1]$ from $D_n(t) \sim t^{\beta_n}$. 
Notice also that the $n$\textsuperscript{th}-order Heaps' exponent can also be interpreted as the first order Heaps' exponent of a sequence whose events are the overlapping $n$-tuples of the original sequence. 
Finally, it is worth remarking that such an approach is close to the analysis of Zipf's law in linguistic data for $n$-grams or sentences \cite{ha2009extending, ryland2015zipf}. In this context, studies showed that as one moves from graphemes, to words, sentences, and $n$-grams, the Zipf's exponent (reciprocal of the \deletetext{Zipf's}\newtext{Heaps' exponent} for infinitely long sequences~\cite{lu2010zipf}) gradually diminishes. This implies that $n$-grams or sentences are characterized by a larger novelty rate than words, a behavior analogous to what we have discussed above.

\subsection*{Analysis of real-world data sequences}
\label{sec:data_analysis}

\begin{figure*}
\begin{center}
\includegraphics[width=\linewidth]{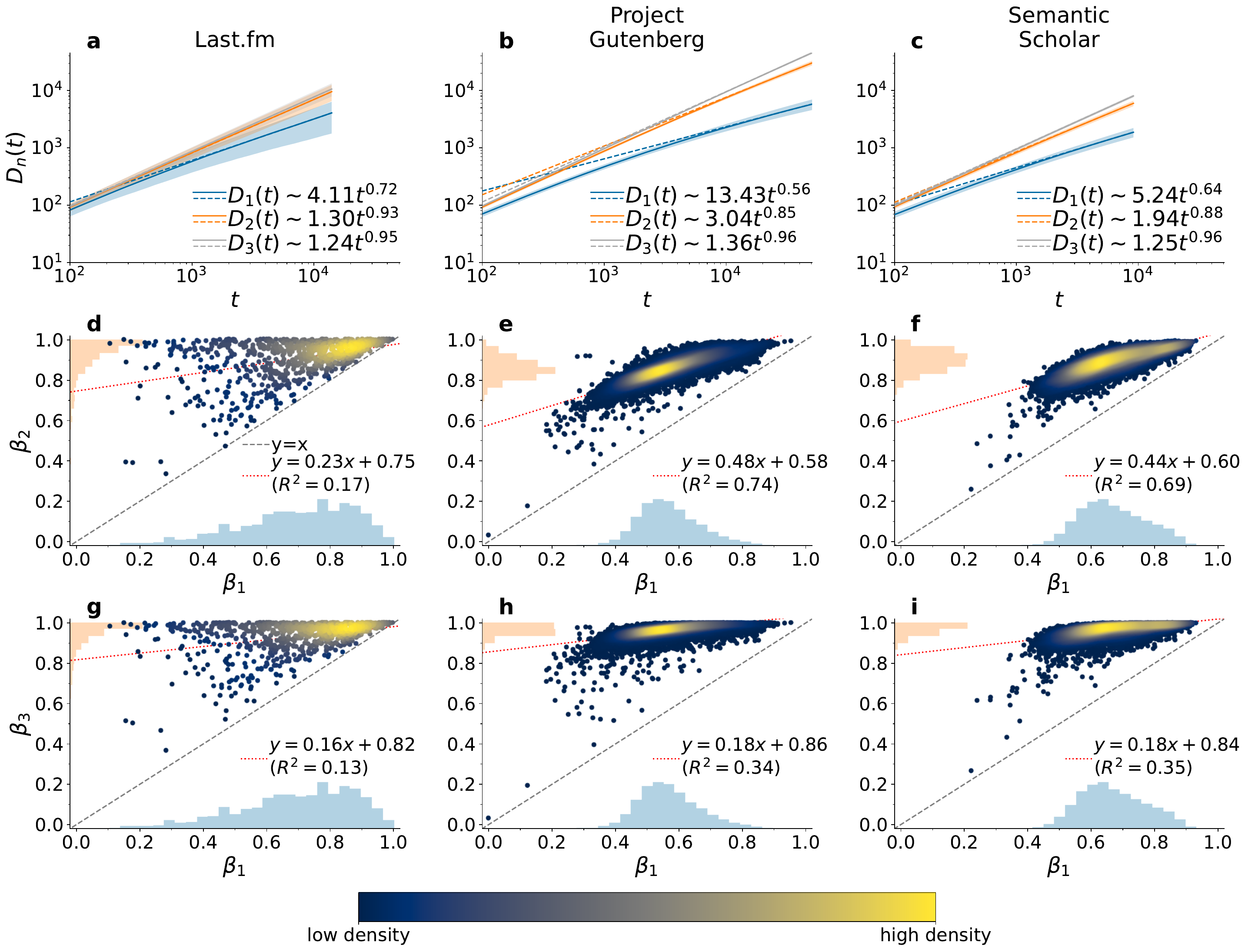}
\caption[Higher-order Heaps' exponents\deletetext{ and their correlations} in real-world data sets.]
{
\textbf{Higher-order Heaps' exponents\deletetext{ and their correlations} in real-world data sets.}
(\textbf{a-c}) Average number $D_n(t)$ of novelties of order $n$, with $n=1,\,2,\,3$, as a function of the sequence length $t$, and fit of the associated Heaps' laws (dashed lines), with estimated exponents shown in the legend. Shaded area represents one standard deviation above and below the average.
(\textbf{d-i}) Scatter plots between the ($1$\textsuperscript{st}-order) Heaps' exponents $\beta_1$ and the $n$\textsuperscript{th}-order exponents $\beta_n$, with $n = 2$ (\textbf{d-f}) and $3$ (\textbf{g-i}).
Each point refers to a different  sequence, with colors representing the density of points (see color bar). 
Each panel also reports histograms of exponents distributions, the bisector $y=x$ (dashed gray line), as well as the fitted linear model (dotted red line) with the value of its coefficient of determination $R^2$.
Each column refers to a different data set: (\textbf{a},\textbf{d},\textbf{g}) Last.fm, (\textbf{b},\textbf{e},\textbf{h}) Project Gutenberg and (\textbf{c},\textbf{f},\textbf{i}) Semantic Scholar, respectively.
}
\label{fig:Heaps_data}
\end{center}
\end{figure*}

We start investigating the emergence of novelties of different orders in empirical exploration processes associated to three different data sets. 
These data sets are substantially different in nature, since they refer, respectively, to songs listened by users of {\it Last.fm}, words in books collected in the {\it Project Gutenberg}, and words of titles of scientific journals from {\it Semantic Scholar} (more details on the data can be found in \textit{Materials and Methods}). 
In Fig.~\ref{fig:Heaps_data}({\bf a-c}) we plot the average temporal evolution of the number $D_n(t)$ of novelties of order $n$, with $n=1,\,2,\,3$, in the three data sets (from left to right, respectively, Last.fm, Project Gutenberg, Semantic Scholar). 
In order to avoid spurious effects due to different lengths of the sequences, we restrict \deletetext{the}\newtext{these} averages to the sequences of length $T$ greater than the median length $\tilde{T}$ in the corresponding data set (see Fig.~S1 in the Supplementary Information (SI) for their distribution). Each continuous curve, plotted up to \deletetext{time}\newtext{length} $\tilde{T}$, is obtained by averaging $D_n(t)$ over all such sequences, while the shaded area represents one standard deviation above and below the mean.
We also perform power-law fits (see \textit{Materials and Methods} for details on the procedure), and plot the resulting curves as dashed lines\newtext{, with the fitted function shown in the legend}.
Focusing first on the broadly-studied ($1$\textsuperscript{st}-order) Heaps' law, notice how the power-law fit is only accurate in the last part of the sequence. 
This highlights that the Heaps' law starts after a transient phase, where most of the events are new for the individual, as also reported in Ref.~\cite{tria2014dynamics} and similarly reported in other contexts~\cite{csanyi2004structure,glanzel2007characteristic,milojevic2010modes,milojevic2010power,gerlach2013stochastic}.
Secondly, notice how the $n$\textsuperscript{th}-order Heaps' law, with $n=2,3$, is valid across the data sets, but with different values of the fitted exponents, especially for $n=2$.
Finally, as expected from their definition, the fitted Heaps' exponents of order $n+1$, i.e., $\beta_{n+1}$, are higher than the lower-order ones, that is, $\beta_{n+1} \geq \beta_n$.

To explore the gain in information brought by the higher-order Heaps' exponents with respect to the 1\textsuperscript{st}-order Heaps', we now look directly at individual sequences. \newtext{In }Figure~\ref{fig:Heaps_data}({\bf d}-{\bf i}) \deletetext{shows}\newtext{we show} the scatter plots of $\beta_2$ ({\bf d}-{\bf f}) and $\beta_3$ ({\bf g}-{\bf i}) against $\beta_1$, where each point refers to a single sequence from Last.fm ({\bf d},{\bf g}), Project Gutenberg ({\bf e},{\bf h})\newtext{,} or Semantic Scholar ({\bf f},{\bf i}), with colors representing \deletetext{the density of points}\newtext{how dense points are} (see color bar at the bottom of the figure).
Here, \deletetext{we have only considered}\newtext{we filter out} sequences whose fitted exponent has a standard error \deletetext{below}\newtext{above} the 0.05 threshold (see Table~S1 in SI for more details)\newtext{, for which the Heaps' law cannot be considered valid}. 
This filtering removes \newtext{only }30 (3.37\%), 8 (0.04\%), and 5 (0.03\%) sequences in the three data sets, respectively. \newtext{Furthemore, we have removed sequences for which the extracted value of $\beta_2$ is higher than the associated value of $\beta_1$, or for which $\beta_3 > \beta_2$, since $D_{n}(t) \leq D_{n+1}(t)$ as previously discussed. This filtering removes 53 (6.16\%), 7 (0.04\%), 6 (0.03\%) in the three data sets, retaining a total of 807, 19\ 622, and 18\ 909 sequences, respectively.
} 
\deletetext{This shows that, in almost all cases, we can consider the Heaps' law assumption to be valid. }%
Looking at \deletetext{the plots, we notice that some cases have a higher density of points compared to others.
For example, in }\newtext{Figure~\ref{fig:Heaps_data}}(\textbf{d}), we see how users of Last.fm sharing the same value of $\beta_1$ can have very different values of $\beta_2$.
Conversely, the other two data sets present stronger correlation between $\beta_2$ and $\beta_1$.
\begin{figure*}
\begin{center}
\includegraphics[width=\linewidth]{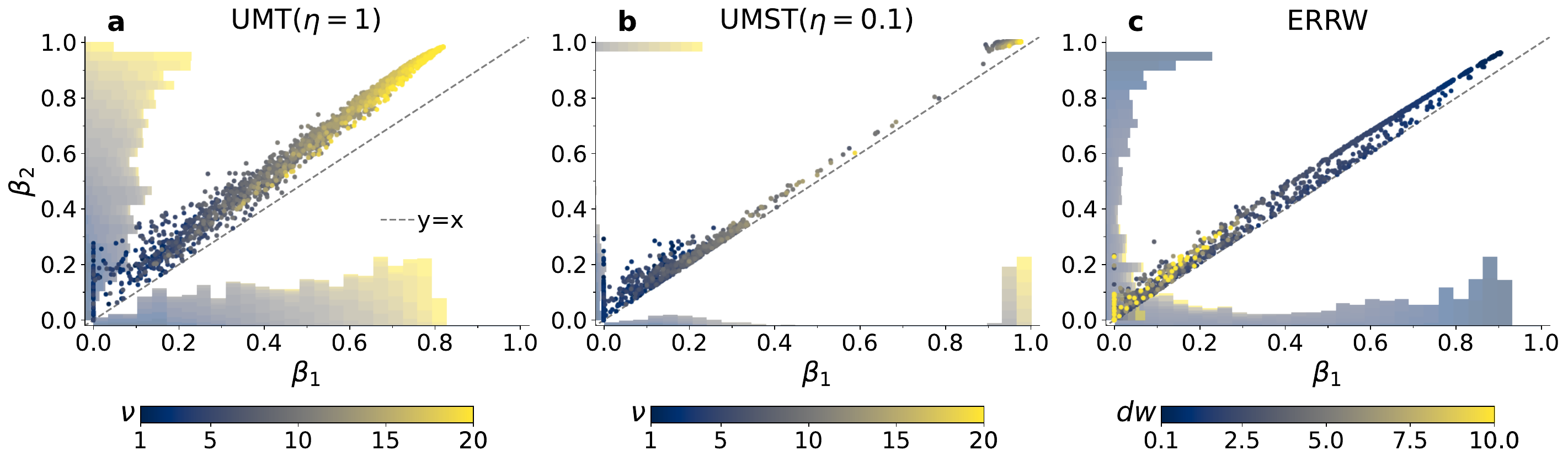}
\caption[Higher-order Heaps' exponents in existing models.]
{
\textbf{Higher-order Heaps' exponents in existing models.} 
Scatter plots of the (1\textsuperscript{st}-order) Heaps' exponent $\beta_1$ against the $2$\textsuperscript{nd}-order exponent $\beta_2$ in:  
(\textbf{a}) the urn model with triggering (UMT), no semantic correlations ($\eta=1$), and $\rho=20$, $\nu = 1,\,2, \dots,\, 20$; (\textbf{b}) the urn model with semantic triggering (UMST) with $\eta=0.1$ and $\rho=4$, $\nu = 1,\,2, \dots,\, 20$; (\textbf{c}) the edge-reinforced random walk (ERRW) on a small-world network (average degree $\langle k \rangle = 4$ and rewiring probability $p = 0.1$~\cite{newman1999scaling}) with edge reinforcement $\rho$ ranging geometrically from $0.1$ to $10$.
Each point refers to a different simulation of the related model, with colors representing the value of the free parameter (see color bar). 
Each panel also reports histograms of exponent distributions on the respective axes, and the bisector $y=x$ (dashed gray line).
All simulations have run for $10^5$ time steps. }
\label{fig:Heaps_simulations_existing_models}
\end{center} 
\end{figure*}
To quantitatively characterize this, we fit a linear model with an ordinary least squares method, displayed in each plot as a red dotted line. In the legend we also report the value of the related coefficient of determination $R^2$, which represents the percentage of variance of the dependent variable explained by the linear fit with the independent variable.
\deletetext{For users of Last.fm, at both orders $n=2$ and $3$, we quantitatively confirm that points are much more spread around the linear fit, since the values of $R^2$ are very low, between $0.11$ and $0.16$. 
In the other two data sets there is instead a higher correlation between $\beta_1$ and both $\beta_2$ ($R^2$ around 0.70) and $\beta_3$ ($R^2$ around 0.35). }%
\newtext{For the Last.fm data set, points are much more spread around the linear fit compared to the other two data sets, as also confirmed by the values of $R^2$, indicating a greater variability in listening habits compared to writing habits.}
Moreover, the values of the parameters of the linear fit greatly change across data sets and orders.
In particular, \deletetext{in (\textbf{d}) there is}\newtext{Last.fm is characterized by} a much lower slope and intercept compared to the other data sets for the same order\deletetext{ in (\textbf{e-f})}. 
Furthermore, we notice how, for each data set, \deletetext{the higher the order, the lower the fitted slope ---and the higher the intercept of the linear model. }\newtext{the values of $\beta_3$ are much much higher and more spread than the respective values of $\beta_1$ and $\beta_2$, resulting in lower values of $R^2$ in the linear fit between $\beta_1$ and $\beta_3$.}

\deletetext{Finally, on}\newtext{At} an aggregate level, we observe that at all orders the distribution of the Heaps' exponents are very different across data sets (see Fig.~S2 in SI for a comparative figure, while further statistical information on the Heaps' exponents distribution can be found in Table~S2 in SI\deletetext{.}).
The exponents are more spread in Last.fm, which also shows a higher average of $\beta_1$ and $\beta_2$, but a lower one for $\beta_3$ compared to the other data sets. 
Distributions for Project Gutenberg and Semantic Scholar, which are both related to linguistic data, are more peaked\deletetext{ ---}\newtext{. Such peaks appear }at higher values for the latter data set.
This could be the result of how titles of scientific papers are written with respect to books or poems, that is, concentrating the whole message of a scientific work in a few significant\newtext{ and specialised} words, avoiding stop-words and repetition. In addition, scientific advancements tend to favor the combinations of previously existing scientific concepts to form new ones, while the same does not apply to non-scientific literature in general, where instead similar constructions tend to be repeated across the piece.

Finally, similar results are obtained also for more coarse-grained sequences generated by using artists and stemmed words instead of songs and words (see Fig~S3 in SI).
\newtext{Furthermore, in Fig~S4 in SI we analyse higher-order novelties in the collective sequences obtained by randomly concatenating all the  individual  sequences of each data set~\cite{tria2014dynamics}. 
}

%%%%%%%%%%%%%%%%%%%%%%%%%%%%%%%%%%%%%%%%%%%%%%%%%%%%%%%%%%%%%%%%%%%%%%%%%%%
% EXISTING MODELS
%%%%%%%%%%%%%%%%%%%%%%%%%%%%%%%%%%%%%%%%%%%%%%%%%%%%%%%%%%%%%%%%%%%%%%%%%%%

\subsection*{Analysis of existing models}
\label{sec:model_analysis}

After studying higher-order Heaps' laws in real data, we check whether the observed patterns can be\deletetext{ also} reproduced by the available models of discovery processes.
We start \newtext{this analysis }from the Urn Model with Triggering (UMT). In such a model, sequences of events are generated by the extraction of \deletetext{coloured}\newtext{colored} balls from an urn~\cite{tria2014dynamics}, \newtext{where }different \deletetext{colours}\newtext{colors} correspond\deletetext{ing} to different events or items being discovered or adopted. \newtext{Here, an event in the sequence is simply represented by the color extracted. }In the UMT, for each extracted ball, the corresponding color is reinforced by adding $\rho$ additional balls, of the same color, to the urn. At the same time, whenever a novel color is drawn, the discovery triggers the addition of $\nu+1$ balls of new different colors to the urn (see detailed model definition in \textit{Materials and Methods}). \newtext{The reinforcement process ensures the wide-spread adoption of items or concept that were frequently adopted in the past. Conversely, the triggering mechanism mimics the adjacent possible expansion, since each novelty makes the space of possible colors expand. Intuitively, these two parameters modulate the exploit-explore tendency of the system, with a more pronounced exploratory behavior for larger $\nu/\rho$ ratios.}

Previous studies have shown that the $1$\textsuperscript{st}-order Heaps' law is verified in sequences \deletetext{obtained with the UMT}\newtext{generated by UMT simulations}~\cite{tria2014dynamics, loreto2016dynamics}. In particular, the number of novelties in the model grows asymptotically as $D_1(t)\sim t^{\frac{\nu}{\rho}}$ when $\nu<\rho$, while a linear \deletetext{behaviour}\newtext{behavior} is found \deletetext{in the other cases}\newtext{for $\nu>\rho$}. We hence focus on the most interesting case\deletetext{, that is for} $\nu\leq\rho$, studying how variations of the two parameters $\rho $ and $\nu$, respectively representing the reinforcement and the increase in size of the adjacent possible, affect the Heaps' law at various orders. 
Since the pace of discovery effectively depends only on the fraction $\nu/\rho$, we fix $\rho = 20$ and numerically simulate the UMT with $\nu = 1,\,2,\,3,\dots,\,20$ for $T=10^5$ time-steps\newtext{, obtaining sequences of length comparable to the data sets (see Fig.~S1 in SI)}. 
For each set of parameters we run $100$ simulations, generating a total of $2 \times 10^{3}$ synthetic sequences. 
Then, for each generated sequence, we compute the temporal evolution of the number of novelties $D_n(t)$, and estimate a power-law fit, extracting the related $n$\textsuperscript{th}-order Heaps' exponent $\beta_n$.
In Fig.~\ref{fig:Heaps_simulations_existing_models}(\textbf{a}), we show how the extracted values of $\beta_2$ change with respect to $\beta_1$ across simulations.
The color represents the value of the parameter $\nu$, as \deletetext{shown in}\newtext{indicated by} the color bar.
We observe that, although the exponents \deletetext{are distributed all}\newtext{span} \deletetext{across }the interval $(0,1)$, the points $(\beta_1,\beta_2)$ are \newtext{aligned} just above the bisector (gray dashed line).
\deletetext{Moreover}\newtext{In other words}, \deletetext{for a certain value of $\beta_1$, the model produces very similar values of $\beta_2$ that do not vary much }\newtext{the values of $\beta_2$ are highly correlated with the related values of $\beta_1$}. % across different simulations
We can derive an analytical approximation of the higher-order Heaps' exponents for this model\deletetext{. As }\newtext{, as} 
we show in Sec.~S3.2 of the SI\deletetext{, for the UMT }\newtext{. We obtain that }the number of unique pairs\newtext{ for the UMT approximately} grows as
\begin{equation}\label{eq:analytical_integration_D2}
    D_2(t) \approx a \,t^{\beta_2}, \quad \text{ with } \quad
     \beta_2 = \beta_1 + \frac{c}{d+\log(t)} ,
\end{equation}
where $a$, $c$, $d>0$ depend on the parameters $\rho$ and $\nu$, and $\beta_1 = \nu/\rho$.
Although the predicted $2$\textsuperscript{nd}-order exponent is slightly higher than the 1\textsuperscript{st}-order one, their difference just depends on the sequence length, and vanishes at larger times.
\deletetext{In other words, the increased value of the higher-order Heaps' exponent is only due to a finite time effect, and the UMT struggles in reproducing the empirical patterns discussed in Fig.~\ref{fig:Heaps_data}. }%
\newtext{Therefore, the difference between $\beta_1$ and $\beta_2$
observed in the simulations is only due to finite time effects, revealing how the UMT cannot reproduce the empirical patterns of  Fig.~\ref{fig:Heaps_data}.
Due to the same reason, notice how the fitted values of $\beta_1$ in the simulations of the UMT are lower than the asymptotically expected value of $\beta_1 = \nu/\rho$ for high values of $\nu$, as also shown in Fig~S6(a) in SI.} 

\medskip

We repeat the analysis for \newtext{two other generative models for discovery and exploration processes, i.e., }the Urn Model with Semantic Triggering (UMST)~\cite{tria2014dynamics} and the Edge-Reinforced Random Walk (ERRW)~\cite{iacopini2018network}, which have\deletetext{ also} been proved to generate \deletetext{discovery }sequences obeying to the Heaps' law. 
These models share the same foundations of the UMT, but with some crucial differences. The UMST builds on top of the UMT introducing also semantic groups for colors\deletetext{ (topic common to different items)}. This addition effectively diminishes the probability to draw colors outside\deletetext{ of} the semantic group of the last extracted color by a factor \deletetext{$\eta$}\newtext{$\eta\leq 1$}.
The ERRW\newtext{, instead,} is formulated as a network exploration rather than a process of extractions from an urn. Instead of a sequence of extracted balls, the ERRW \deletetext{features a set of}\newtext{generates a sequence made of the} nodes sequentially visited by \deletetext{a}\newtext{the edge-reinforcing} random walker over a weighted network, where the weight of the visited edges are reinforced \deletetext{at each time }by $\rho$\newtext{ when crossed}. A full description of the models can be found in \textit{Materials and Methods}.

We simulate the UMST with parameters $\eta=0.1$\newtext{ (semantic parameter)}, $\rho=4$\newtext{ (reinforcement parameter)}, $\nu = 1,\,2, \dots,\, 20$\newtext{ (triggering parameter)}, while \newtext{the simulations of }the ERRW run over \deletetext{a }small-world \deletetext{network}\newtext{networks~\cite{watts2004small,gravino2012complex}} (with average degree $\langle k \rangle = 4$ and rewiring probability $p = 0.1$\newtext{ following the procedure of Ref.~\cite{newman1999scaling}}), with edge reinforcement $\rho$ ranging from $0.1$ to $10$.
Similarly to the \deletetext{exploration of the }UMT, we perform 100 simulations for each set of parameters\newtext{,} and report the results in Fig.~\ref{fig:Heaps_simulations_existing_models}(\textbf{b-c}).
For both UMST and ERRW, we find that the values of $\beta_2$ do not differ much from their corresponding value of $\beta_1$---as shown by the great proximity of the points $(\beta_1,\, \beta_2)$ to the bisector.
This means that also these models fail to reproduce the empirical variability of higher-order Heaps' exponents with respect to the $1$\textsuperscript{st}-order one.
Moreover, we notice in (\textbf{b}) that for the UMST we only obtain exponents with either very low (up to 0.4) or very high (close to 1) values.
\deletetext{It seems thus that there is}\newtext{We indeed see} an abrupt transition between \deletetext{the two cases}\newtext{these two extremes}, with the model not able to cover the values in between\deletetext{. This is instead a crucial point when we are confronted with the empirical values }\newtext{, which are instead present in the empirical data }reported in Fig.~\ref{fig:Heaps_data} (see also the relation with analytical results in Fig~S6 in SI). 
\newtext{Further simulations of both the UMST and ERRW with other sets of parameters are reported in Fig.~S7 of the SI. Also in these other cases, these models are unable to generate sequences with $\beta_2$ different from $\beta_1$.}

\medskip

Overall, \newtext{the analyses above indicate} that, while the existing models of discovery and innovation dynamics are able to reproduce the empirically observed pace of discovery of new items (singletons)\newtext{ as captured by the $1$\textsuperscript{st}-order Heaps' law}, they fail \deletetext{when it comes to capturing}\newtext{ to capture} the distributions of \deletetext{the }Heaps' exponents of higher order\deletetext{ and their correlations}.

\subsection*{
\newtext{The ERRWT}: a model for higher-order Heaps' laws}
\label{sec:new_model}

\deletetext{We now introduce a model that generates synthetic sequences displaying different Heaps' exponents at various orders. }%
\newtext{In order to fill the gap between empirical observations and models, we introduce here a new model that can generate synthetic sequences with tunable discovery paces both at the first order and at higher orders.}
As for the previously discussed ERRW, our model is formulated 
\newtext{in terms of network exploration. Namely, in the model}: ({\it i}) the items to be explored correspond to the nodes of the network\deletetext{;}\newtext{,} ({\it ii}) \newtext{the }links between nodes represent semantic associations between items that one can use to move from one to another\deletetext{;}\newtext{, and} ({\it iii}) the exploration process is modelled as a random walk over the network and the sequence is obtained from the ordered list of visited nodes.
Under these assumptions, the first visit of a node corresponds to a $1$\textsuperscript{st}-order novelty, while the first \newtext{visit of a link corresponds to a 
$2$\textsuperscript{nd}-order novelty}. Such definition can be 
\newtext{straightforwardly} 
extended to higher orders. In this manuscript, for simplicity, we limit our attention to the first two orders.
The ERRW proposed in Ref.~\cite{iacopini2018network} consists of a \newtext{random }walk on a network \newtext{whose topology is fixed, i.e., the links cannot change in time, but the link weights can be modified by the passage of the random walker.} By contrast, in our model \newtext{not only the weights, but the entire network structure co-evolves} with the exploration process\newtext{,} and new \newtext{nodes and new} links can be triggered. 
Thus, in analogy with the UMT~\cite{tria2014dynamics}, we name the model {\em Edge-Reinforced Random Walk with
Triggering} (ERRWT). More specifically, \newtext{on top of the edge reinforcement mechanism of the ERRW, }the model is based on two different triggering  mechanisms that add new edges and new nodes every time a novelty appears. 
\deletetext{As per the UMT and the ERRW, }\newtext{Similarly to the previous models analysed before, the first visit of a node}  
triggers the expansion of the adjacent possible, as new nodes, \deletetext{the nodes of the newly 
visited node}\newtext{neighbors of the discovered node}, become now accessible. As an example of this mechanism, think for instance to the invention of the transistor, which made it possible to create mobile phones, among other things.
\newtext{Moreover, differently from the previous models, here the first visit of an edge is also considered a novelty, and as such, it triggers new edges. }\deletetext{Concerning the triggering of new edges, the }\newtext{The }idea is that whenever two elements are associated for the first time, new possible combinations involving one of these elements are then triggered. For instance, once photo-cameras and mobile phones were firstly combined, this association made clear that many more functions could be added to the latter, e.g., a music player, a game console, a GPS, etc. 
\newtext{More formally, the model considers that the adjacent possible can be expanded at various orders, i.e., not just by introducing new nodes, but also by triggering new links.}

\begin{figure}
\begin{center}
\includegraphics[width=\linewidth]{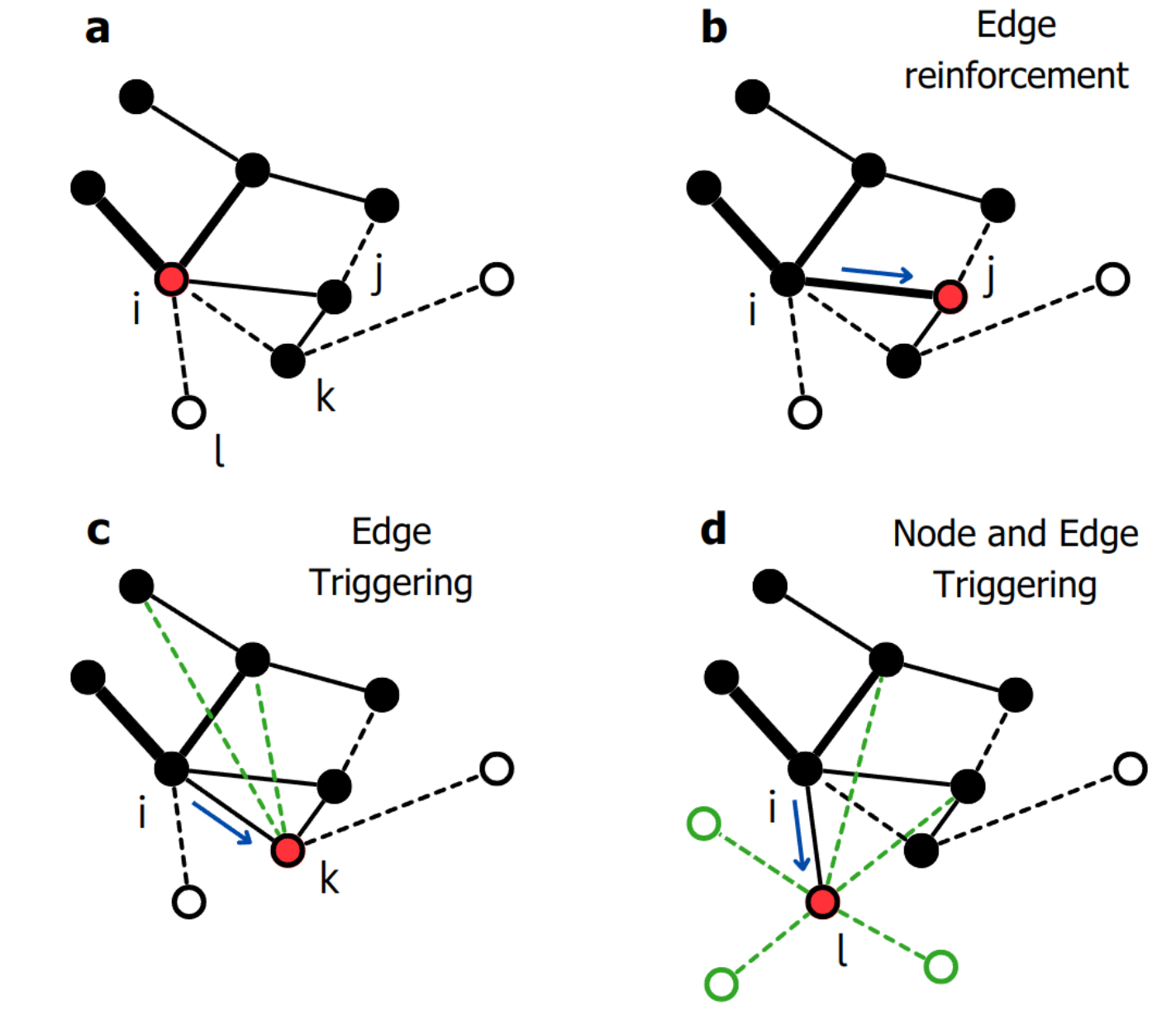}
\caption[The Edge-Reinforced Random Walk with Triggering (ERRWT) model.]
{
\textbf{The Edge-Reinforced Random Walk with Triggering (ERRWT) model.}
An exploration process is modelled as a random walk on a growing weighted network. (\textbf{a})
At time $t$, the walker is at the red node $i$. Nodes that have been already visited by the walker are colored in black, in white those left to be visited. Similarly, traversed (old) and not-traversed (new) links are respectively depicted with continuous and dashed lines, whose widths represent their weights.
At time $t+1$, the walker can move to each of the neighbours of $i$, e.g.\ nodes $j$, $k$, or $l$, with a probability proportional to the weight of the respective link.  
(\textbf{b}) \deletetext{If the walker moves to $j$, since the link $(i,\,j)$ is old, its weight is reinforced by $\rho$ (edge reinforcement);}%
\newtext{If the walker moves to $j$, the weight of the link 
$(i,\,j)$ is reinforced by $\rho$ (Edge Reinforcement mechanism), but no new nodes or links are added to the network, since the link $(i,\,j)$ is old;}
(\textbf{c}) if the walker moves to node $k$, since link $(i,\,k)$ is new but node $k$ is old, in addition to the edge reinforcement, $\nu_2 + 1 = 2$ new edges (in green) between $k$ and old nodes are added to the network (Edge Triggering mechanism);
(\textbf{d}) finally, if the walker moves to $l$, since both the link $(i,\,l)$ and the node $l$ are new, in addition to the edge reinforcement and the edge triggering, $\nu_1+1 = 3$ new nodes (in green) are added to the network and connected to $l$ (Node and Edge Triggering mechanism). 
}
\label{fig:new_model_illustration}
\end{center}
\end{figure}

The basic mechanisms of the ERRWT model are illustrated in Fig.~\ref{fig:new_model_illustration}. 
Suppose that at a given time $t$, the walker is located at node $i$ of a network. At this point, some nodes and links,
represented by full circles and solid lines in 
Fig.~\ref{fig:new_model_illustration}(\textbf{a}),
have already been visited, while others, shown as empty circles and dashed lines, are part of the adjacent possible. 
In Fig.~\ref{fig:new_model_illustration}(\textbf{b}), the walker 
\newtext{moves from node $i$ to node $j$, crossing in this way} an already explored link. 
\newtext{Consequently, the weight of such link is increased by a positive quantity $\rho$}, meaning that the association \deletetext{of}\newtext{between} the two nodes $i$ and $j$ becomes \deletetext{more likely}\newtext{stronger and thus more likely to be used again}. This is the same edge reinforcement mechanism adopted in the ERRW model\deletetext{ in Ref.}~\cite{iacopini2018network}.  
\deletetext{If}\newtext{In} addition to this, if instead 
\newtext{the walker moves from node $i$ to node $k$,} traversing an edge for the first time, \deletetext{along with the edge reinforcement the process triggers also the creation of new edges}\newtext{as displayed in Fig.~\ref{fig:new_model_illustration}(\textbf{c}), this event is considered a 2\textsuperscript{nd}-order novelty and triggers 
the creation of new edges}. In particular\deletetext{, as displayed in Fig.~\ref{fig:new_model_illustration}(\textbf{c})}, $\nu_2+1$ new edges connecting node $k$ 
to other already-visited nodes are created\newtext{ (green dashed lines)}. 
Finally, the third mechanism of the ERRWT model is analogous to the triggering mechanism of the UMT model. 
As illustrated in  Fig.~\ref{fig:new_model_illustration}(\textbf{d}), 
\newtext{when the walker moves from $i$ to a node $l$, visiting node $l$ for the first time,} this event triggers the expansion of node $l$'s adjacent possible \newtext{with the addition of new nodes and new links. Namely,}  
$\nu_1+1$ new nodes are added to the network and connected to the node $l$ itself. In addition to this, $\nu_2+1$ new links to already known elements are created, since whenever a node is explored for the first time, also the link leading to it is explored for the first time. 
More details about the ERRWT model can be found in \textit{Materials and Methods}.

\begin{figure*}
\begin{center}
\includegraphics[width=\linewidth]{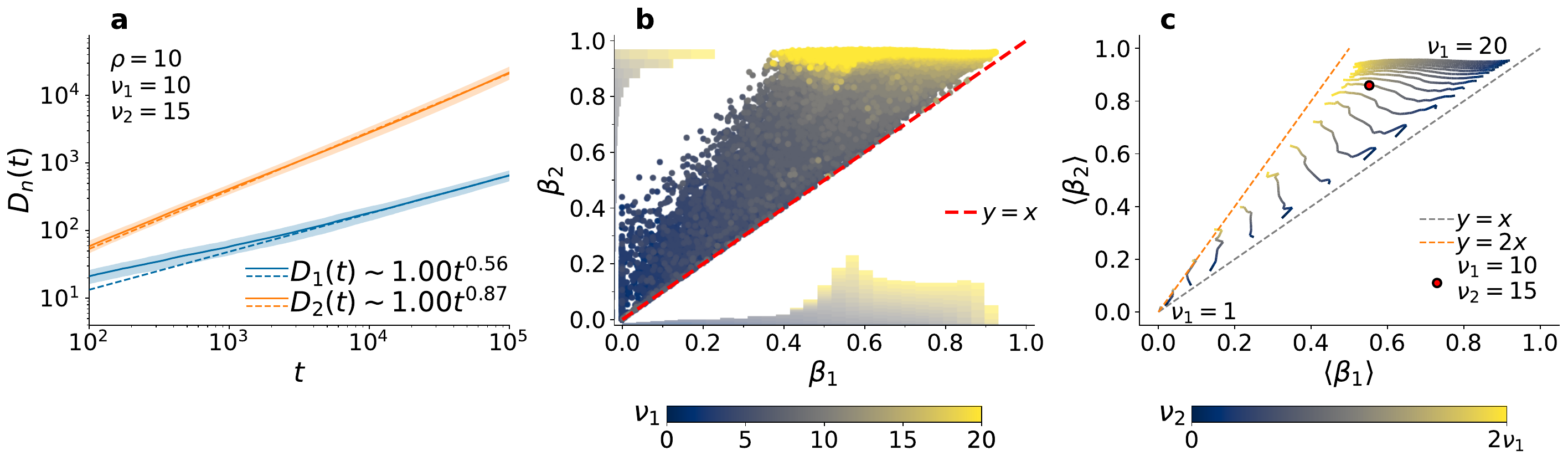}
\caption[Higher-order Heaps' exponents in the ERRWT model.]
{
\textbf{Higher-order Heaps' exponents in the ERRWT model.} 
(\textbf{a}) Average number $D_n(t)$ of novelties of order $n$, with $n=1$ and $2$, as a function of the sequence length $t$ for simulations of the ERRWT model with parameters $\rho = 10$, $\nu_1 = 10$, $\nu_2 = 15$, and fit of the associated Heaps' laws (dashed lines), with estimated exponents shown in the legend. 
% \vito{togliere terzo ordine dal pannello a },
Shaded areas represent one standard deviation above and below the average.
(\textbf{b}) Scatter plot between the (standard) Heaps' exponent $\beta_1$ and the $2$\textsuperscript{nd}-order exponent $\beta_2$.
Each point refers to a different simulation of the model, with colors representing the corresponding value of the parameter $\nu_1$ ranging from 0 to 20 (see color bar), while $\rho = 10$ and $\nu_2 = 0,\,\dots,\,2\nu_1$.
(\textbf{c}) Variation of the average $n$\textsuperscript{th}-order Heaps' exponents $\beta_n$, with $n=1,\,2$. Each curve refers to a different value of $\nu_1$, increasing from 1 to 20 from bottom left to top right, while the color represents the value of $\nu_2$ (see color bar). The set of parameters used in (\textbf{a}) is here highlighted in with a red dot.
}
\label{fig:new_model_higher_order_heaps}
\end{center}
\end{figure*}

\medskip
Balancing edge reinforcement and node and edge triggering mechanisms through the parameters $\rho$, $\nu_1$ and $\nu_2$ 
\newtext{of the ERRWT model}, 
it is possible to control the pace of discovery of new nodes and edges, and consequently tuning the exponents of the $1$\textsuperscript{st}-order and the $2$\textsuperscript{nd}-order Heaps' law associated to the sequences produced by the model.
To systematically explore this, we simulate the ERRWT model with parameters $\rho=10$, $\nu_1 = 0,\,1,\, \dots,\, 20$, and $\nu_2 = 0,\,1,\, \dots,\, 2\nu_1$, running 100 simulations for each set of parameters.
Higher values of $\nu_2$ have not been considered since they produce the same exponents as those for $\nu_2=2\nu_1$.
\newtext{In }Fig.~\ref{fig:new_model_higher_order_heaps}(a) \newtext{we }report\deletetext{s} the increase in the number of 
$1$\textsuperscript{st}-order and $2$\textsuperscript{nd}-order novelties (continuous lines)\newtext{ for a specific set of parameters as an example}. The power-law fits (dashed lines) highlight that the Heaps' law is verified at \deletetext{higher-orders}\newtext{the 2\textsuperscript{nd} order} too, leading to an increase of the exponent\deletetext{s} values (from $\beta_1=0.56$ to $\beta_2=0.87$)\deletetext{ as \deletetext{we increase }
\newtext{with the considered order}}. The relation between the different orders is explored in the scatter plot between the $1$\textsuperscript{st}- and $2$\textsuperscript{nd}-order Heaps' exponents reported in Fig.~\ref{fig:new_model_higher_order_heaps}(b). Each point refers to a different simulation, and we use the color to indicate the value of the \deletetext{used }parameter $\nu_1$\newtext{ used} (see color bar). 
We notice that the ERRWT model can produce a wide range of values for the exponents at both orders, and that the 
$2$\textsuperscript{nd}-order exponents are not trivially correlated to the $1$\textsuperscript{st}-order ones, as it happened in the models\newtext{ considered in the previous section}. 
This is even more clear when we look at Fig.~\ref{fig:new_model_higher_order_heaps}(c), where the Heaps' exponents are averaged across simulations for each set of parameters. Each curve in the figure refers to a different value of $\nu_1$, with $\nu_1$ increasing from 1 to 20 from bottom left to top right of the panel. The color represents instead different values of the parameter $\nu_2$ from 0 to $2\nu_1$. 
For reference, we also flag using a red dot the pair of exponents related to the parameters used in Fig.~\ref{fig:new_model_higher_order_heaps}(a).
We can immediately notice how the 1\textsuperscript{st}- and 2\textsuperscript{nd}-order Heaps' exponents increase as $\nu_1$ gets larger. More interestingly, we observe the combined role of the two parameters. 
For each curve, by increasing $\nu_2$\newtext{, therefore triggering new links in the network,} the difference between $\beta_1$ and $\beta_2$ becomes larger, and the point $(\beta_1,\, \beta_2)$ moves away from the bisector, in a way that depends on the specific value of $\nu_1$. In particular, for low values of $\nu_1$, the curves are almost vertical, with only $\beta_2$ increasing. Instead, for higher values of $\nu_1$, especially when $\nu_1 \geq \rho$, an increase of $\nu_2$ produces a decrease of $\beta_1$, while the value of $\beta_2$, which is close to its upper bound value $1$, does not change. \newtext{Intuitively, this happens because the creation of more and more new links between explored nodes increases the chance to exploit nodes already discovered, while still exploring never traversed links.}

It is also possible to perform an analytical investigation of a simplified version of the ERRWT model, which leads to results \deletetext{which are }in agreement \newtext{with the simulations }(see Sec.~S4 in SI). In particular, for such a simplified model, we can prove that the values of the asymptotic Heaps' exponents $\beta_1$ and $\beta_2$ depend on the two ratios $\nu_1/\rho$ and $\nu_2/\rho$. Moreover, we find that, for $\nu_1/\rho > 1$, the $2$\textsuperscript{nd}-order Heaps' exponent is asymptotically equal to $1$, while the $1$\textsuperscript{st}-order one depends on $\nu_1/\nu_2$, \deletetext{in agreement with our numerical results}\newtext{as seen in Fig.~\ref{fig:new_model_higher_order_heaps}(c)}.
Finally, the exponents are asymptotically bounded by $\beta_1 \leq \beta_2 \leq 2\beta_1$, as also observed in the simulations in Fig.~\ref{fig:new_model_higher_order_heaps}(c). 
This also explains why the exponents do not change when we increase $\nu_2$ above $2\nu_1$.

\begin{figure*}
\begin{center}
\includegraphics[width=\linewidth]{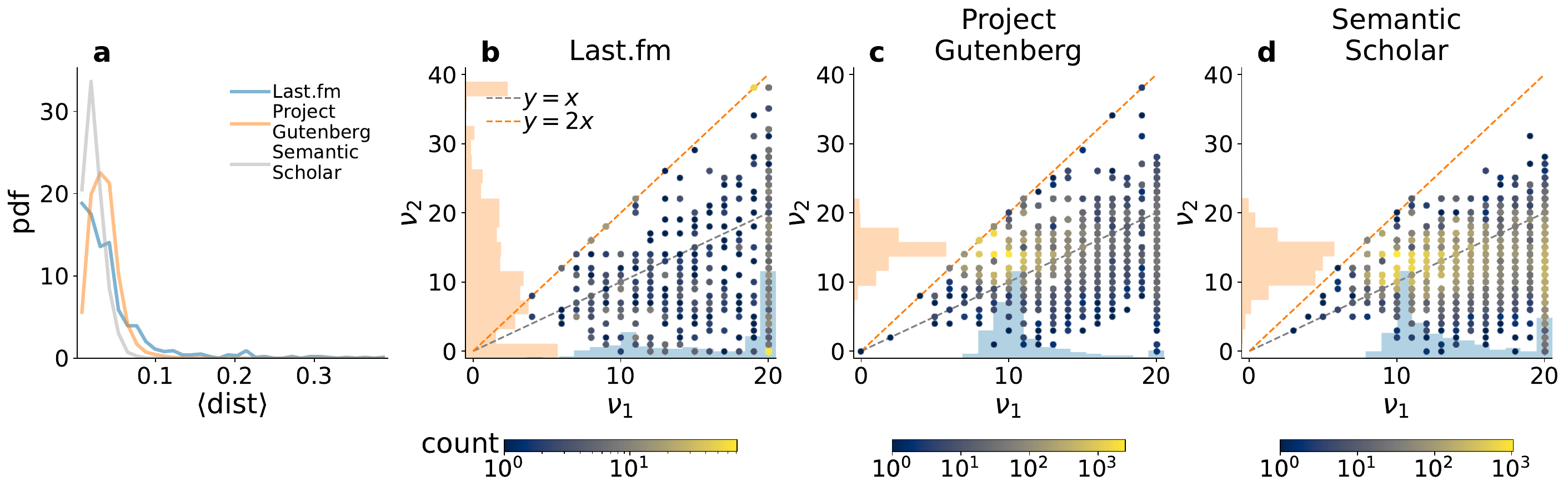}
\caption[Fitting the ERRWT model to real-world data sets.]
{
\textbf{Fitting the ERRWT model to real-world data sets.} 
(\textbf{a}) Distribution of the average distance between the pair of exponents $(\beta_1,\, \beta_2)$ of a real sequence and the pair $(\beta_1',\, \beta_2')$ obtained by the best fitting ERRWT model.
(\textbf{b-c}) Scatter plots of the best-fitted parameters $\nu_1$ and $\nu_2$ of the model across the sequences of the three data sets, respectively  Last.fm (\textbf{b}), Project Gutenberg (\textbf{c}), and Semantic Scholar (\textbf{d}). 
The color of a point refers to the number of sequences with that pair of parameters in the best fitting ERRWT model (see color bar).
}
\label{fig:new_model_distance_data}
\end{center}
\end{figure*}

\subsection*{Comparison between ERRWT and real-world data}

To show that the ERRWT model is able to reproduce the properties observed in real-world processes, we now fit \deletetext{it}\newtext{the parameters of the model} to the three data sets analyzed (Last.fm, Project Gutenberg and Semantic Scholar).
Given an empirical sequence and its pair of 1\textsuperscript{st}- and 2\textsuperscript{nd}-order Heaps exponents $(\beta_1,\, \beta_2)$, 
we compute the Euclidean distance between the pair $(\beta_1,\, \beta_2)$ and each of the pairs of exponents $(\beta_1',\, \beta_2')$ obtained by simulating the ERRWT model using the sets of parameters considered in the previous section. We then select the best model parameters by minimizing the average distance over 100 simulations for each set, and repeat the procedure for all the sequences of the three data sets.
Figure~\ref{fig:new_model_distance_data}(a) shows the probability density distribution of the distances between the empirical sequences and the simulations of the best-performing ERRWT model. 
Notice how these distances are almost all below 0.1, that is \deletetext{to }the uncertainty we expect on the values of the parameters. Indeed, being $\nu_1$, $\nu_2$ integers and $\rho = 10$, the maximum precision we can gain on the estimate of the best parameters is \newtext{about }$1/\rho = 0.1$.
The percentage of sequences with higher distance than this threshold is 7.67\%, 0.73\%, and 0.05\% for Last.fm, Project Gutenberg, and Semantic Scholar, respectively.  
The scatter plots of the best-fitted parameters $\nu_1$ and $\nu_2$ for the three data sets are shown in Fig.~\ref{fig:new_model_distance_data}(b-d). The colors here indicate the number of empirical sequences which are best represented by each pair of parameters $\nu_1$ and $\nu_2$.  
We notice that most of the sequences of Last.fm are characterized by relatively large values of $\nu_1$. Since $\nu_1$ is related to the triggering of new nodes, this result indicates that the discovery of a new song exposes the user to a large variety of related songs, previously not accessible, which can now be discovered. 
Conversely, the parameter $\nu_2$, which controls the triggering of new edges between already existing items in the model, takes values in a larger range, predominantly skewed towards the lower end.
This suggests that, once a new association \deletetext{of}\newtext{between} two songs is established by a user, there is a high probability that the same association will be repeated over and over.
Consequently, the user will preferably listen to songs in a similar order, instead of creating new associations. \newtext{We point out that we cannot distinguish if this is due to individual preferences or is pushed by the presence of recommender systems in music listening platforms.}
In the case of Project Gutenberg, most sequences have $\nu_2 > \nu_1$. 
This implies that writers tend to frequently generate new word associations\newtext{ instead of using words never used before in the text}, highlighting the incredible variety of expressions we can make by combining a limited set of words. 
Finally, Semantic Scholar exhibits values of $\nu_1$ and $\nu_2$ similar to those found in the Project Gutenberg data set. 
However, some sequences of Semantic Scholar have a relatively high value of $\nu_1$ with respect to $\nu_2$.
This is an indication that, when choosing words for titles, authors tend to use more original words, while the pace of creation of new word associations remains similar.

%
%
%%%%%%%%%%%%%%%%%%%%%%%%%%%%%%%%%%%%%%%%%%%%%%%%%%%%%%%%%%%%%%%%
% DISCUSSION
%%%%%%%%%%%%%%%%%%%%%%%%%%%%%%%%%%%%%%%%%%%%%%%%%%%%%%%%%%%%%%%%
%
\section*{Discussion}
\label{sec:discussion}

\newtext{ The extraction of the Heaps' exponent from empirical sequences has recently allowed to characterize the pace at which discoveries occur in different contexts}~\cite{tria2014dynamics,monechi2017waves, cattuto2007semiotic, cattuto2007vocabulary, iacopini2018network}.
However, there is more and more evidence that discoveries are often made from novel combinations of already known elements~\cite{uzzi2013atypical, wang2017bias, fontana2020new, abbasiharofteh2020atypical, lambert2020pace}.
\newtext{In this manuscript we have proposed to explore higher-order Heaps' laws and to extract higher-order Heaps' exponents as a way to characterize novel combinations in an exploration process. 
The key idea is to look at novelties not only as discoveries  of single items, 
but also as the first appearances of new combinations of two or more items. More precisely, the Heaps' exponent of order $n$ measures the rate of discovery of novelties of order $n$, i.e., combinations of $n$ items.} 
Notice that 
\newtext{our approach 
differs and complements other measures of the pace of discovery recently proposed.} For example, 
the authors of Refs.~\cite{fink2017serendipity, fink2019much}~have investigated the number of all possible valid combinations that can be created by using the elements acquired so far as a proxy of the level of innovation of a given system.   
In this way, the potential for new discoveries is accounted for, rather than the actual number of novel combinations observed in a system and their rate of appearance. 

As we have \newtext{shown through the analysis of empirical sequences of music
listening records}, higher-order Heaps' exponents can be used 
\newtext{to further classify 
users of Last.fm with the same rate of discovery} of new songs \deletetext{and}\newtext{or new} artists. The higher-order Heaps' exponent can indeed tell apart different ways to explore the same set of songs in terms of number of \newtext{different associations of consecutive pairs or triples of songs.}   
Analogously, \newtext{we found that higher-order Heaps' exponents can uncover different patterns in the use of words in different texts}.   
Titles of peer-reviewed papers published in scientific journals show more creative combinations of words, than the texts of narrative books. They indeed exhibit many more new $n$-grams, even if the total set of words used is similar in length. 
Overall, our analysis shows that the space of possibilities grows in a complex way, which does not depend solely on the balance between old items to exploit and new ones to explore, but also on the structure of their associations.
\newtext{Notice, however that in our framework we have considered all associations of consecutive items in a sequence as possible discoveries. 
This is certainly a strong assumption, which might  
not always be valid. 
For instance, in the context of a written text, the last word of a paragraph and the first word of the following paragraph, are not necessarily related to each other and could be discarded from the sequence of pairs to be analyzed. The problem can be solved by 
filtering out all such cases, as we have done for books in Gutenberg in Fig.~S5, obtaining similar results in terms of 1\textsuperscript{st}-order and higher-order Heaps' exponents. In other contexts, it might be necessary to better tailor the precise definition of ``novel combination'' according to the nature of the sequence being analysed and the underlying research question.}

\deletetext{We have hence}\newtext{We have then} focused our attention in understanding the underlying mechanisms \deletetext{at the core of such differences in the pace of discovery of}\newtext{that can trigger} higher-order novelties. 
\deletetext{ 
In particular, we have \newtext{first }extracted higher-order Heaps' exponents from synthetic sequences generated through existing models of discovery and exploration, from the urn model with \newtext{(semantic) }triggering~\cite{tria2014dynamics} to the edge-reinforced random walk~\cite{iacopini2018network}.
On the one hand, these models are able to reproduce different behaviors in terms of $1$\textsuperscript{st}-order Heaps' exponents. On the other hand, however, we find that they are not able to reproduce higher-order \deletetext{ones}\newtext{Heaps' exponents different from the 1\textsuperscript{st}-order ones}.\newtext{ In other words, the pace of higher-order discoveries solely depends on the 1\textsuperscript{st}-order one in these models.} 
}
We have proposed a new modelling framework, the ERRWT which takes into account not only the  exploration rate of new items, but also the propensity to explore the same content in a more creative way. 
\newtext{Considering that pairs of items can be seen as the links of a network, the model is based on a process of network exploration and on the co-evolution of the network structure with the dynamics of the exploration process.} 
The model considers  
a reinforcement of the visited links and the triggering of new nodes and links whenever new \newtext{nodes or links} are explored. 
\newtext{Not only the ERRTW model 
is able to reproduce the higher-order Heaps' exponents extracted from real data, but also provides a new intuition of how the space of possibilities grows over time, shedding light on the underlying mechanism in which novel elements and combinations emerge.} 

We acknowledge there are various ways in which our model can be improved and generalized. For example, future work should investigate the interplay between initial knowledge, either of the individual or of \deletetext{the society}\newtext{a group}, and the pace of discovery at various orders during the exploration process\newtext{, or the influence of recommendation algorithms}. 
In our model, we have supposed that \newtext{the links all start 
with the same weight, which can be a too strong} assumption in certain contexts.
Moreover, we have assumed to trigger new links \newtext{with a uniform probability}. It would be interesting to study cases in which the space has some preferential pathways, for example represented by an underlying network structure. This could be implemented in our model by limiting the addition of new links to only those permitted by an underlying network
\newtext{given as an input to the model. Alternatively,}  
more complex ways to trigger edges, such as preferential attachment mechanisms,   
could be considered 
~\cite{barabasi1999emergence,bianconi2001competition}. 
Finally, we have not considered the presence of semantic correlations in the temporal sequence of visited items, which can be a consequence of the interplay between the network topology and a predisposition to move within items semantically close to the recent ones, reinforcing a clustered structure. It would indeed be interesting to use higher-order Heaps' exponents and the ERRWT model to study phenomena related to waves of novelties~\cite{monechi2017waves} and popularity~\cite{monechi2017significance}. Moreover, the ERRWT model could be extended to a multi-agent model to study how different agents would cooperate and diffuse knowledge~\cite{iacopini2020interacting, dibona2022social}, also taking into account the presence of a limited attention capacity and memory that could influence the rise and fall of popular items~\cite{castaldo2022junk}.
We believe that our model can be directly used to answer these questions and, more in general, to better understand the fundamental mechanism behind innovation and creativity.

%%%%%%%%%%%%%%%%%%%%%%%%%%%%%%%%%%%%%%%%%%%%%%%%%%%%%%%%%%%%%%%%%%%%%%%%%%%%%%%%%%%%%%%%%%
%%%%%%%%%%%%%%%%%%%%%%%%%%%%%%%%%%%%%%%%%%%%%%%%%%%%%%%%%%%%%%%%%%%%%%%%%%%%%%%%%%%%%%%%%%
%%%%%%%%%%%%%%%%%%%%%%%%%%%%%%%%%%%%%%%%%%%%%%%%%%%%%%%%%%%%%%%%%%%%%%%%%%%%%%%%%%%%%%%%%%
%%%%%%%%%%%%%%%%%%%%%%%%%%%%%%%%%%%%%%%%%%%%%%%%%%%%%%%%%%%%%%%%%%%%%%%%%%%%%%%%%%%%%%%%%%

\medskip

%
%%%%%%%%%%%%%%%%%%%%%%%%%%%%%%%%%%%%%%%%%%%%%%%%%%%%%%%%%%%%%%%%
% Data and methods 
%%%%%%%%%%%%%%%%%%%%%%%%%%%%%%%%%%%%%%%%%%%%%%%%%%%%%%%%%%%%%%%%
%
\section*{Methods}
\label{sec:methods}
%
% DATA
%
\subsection*{Data}
\label{sec:data}
In this work we consider three different data sets\newtext{:} \deletetext{on }music listening records (\textit{Last.fm}), books (\textit{Project Gutenberg}), and scientific articles (\textit{Semantic Scholar}). 
    
    \medskip
    \emph{Last.fm} is a digital platform for music born in 2002, famous for logging all listening activities of its users, providing both personal recommendations and a space to interact with other users interested in music~\cite{lastfm_about}. In this manuscript, we use a data set presented in Ref.~\cite{lastfm360k1k} and available at Ref.~\cite{lastfm1k_dataset}. 
    This data set has been obtained and used according to terms and conditions of Last.fm~\cite{lastfm_about}, and can only be used for non-commercial use~\cite{lastfm1k_dataset}.
    It contains all listening records of about 1000 users. In order to have sequences long enough for statistically relevant fits, only users with more than 1000 logs have been retained. The final data set contains 890 users having a median number of listened records of 13\ 985. 
    Each record contains the timestamp at which a user listened to a given song. 
    In the database, each song is associated to a title, the artist's name and a unique MusicBrainz Identifier (MBID), which can be used to obtain additional metadata~\cite{musicbrainz}.
    Using this information, we are able to create, for each user, a temporally ordered sequence of songs together with the associated sequence of artists. 
    \newtext{It is worth noting that the behavior of each user might be influenced by additional factors, such as recommendation algorithms. While the specifics of these procedures are not known in details, we expect that they would not drastically alter our main findings~\cite{monechi2017waves}. They may, for example, impact only the numerical values obtained, without affecting the fundamental mechanisms captured by our modelling approach.}

    \medskip
    \emph{Project Gutenberg} is an open access text corpus containing more than 50\ 000 books of different nature~\cite{hartproject}. 
    This corpus is made of public domain books, with expired copyrights, which can therefore be disseminated freely and legally.
    Here, we make use of the Standardized Project Gutenberg Corpus~\cite{gerlach2020standardized}, which allows to download and process an updated version of the open corpus. 
    Using Google's Compact Language Detector 3 (\texttt{cld3} package in Python), we filter out all non-English texts. We then discard all texts with less than 1000 words, retaining a total of 19\ 637 books with a median number of 50\ 726 words. 
    A sequence of events for each book is hence created with the lemmatized words, disregarding punctuation and putting all characters in lower case. 
    We also extract stems from each word using the English Snowball stemmer~\cite{porter2001snowball}---a more accurate extension of the Porter stemmer~\cite{porter1980algorithm}---, which is not as aggressive as the Lancaster stemmer~\cite{paice1990another}. 
    
    \medskip
    \emph{Semantic Scholar} is a recent project with the scope of facilitating scientific analysis of academic publications~\cite{semanticscholar}. It provides monthly snapshots of research papers published in all fields, publicly and freely accessible through the \emph{Semantic Scholar Academic Graph} (S2AG, pronounced ``stag")~\cite{semanticscholar2018}.
    This database ($1$\textsuperscript{st} Jan. 2022 snapshot) contains about 203.6M papers, 76.4M authors, and 2B citations, obtained in accordance with the project terms and conditions~\cite{semanticscholar_license}. 
    It also classifies each paper into one or more fields of study~\cite{s2_fos}, for a total of 19 different fields. 
    For simplicity, we associate each paper to its first (and most relevant) field of study. 
    To create the sequences to analyze, for each field we consider the first 1000 journals in terms of number of English papers. Then, for each journal, we order the published papers based on the respective year of publication, volume, issue, and first page. When some of this information is not available, the Semantic Scholar unique ID of the paper is also used in the ordering process. Thus, for each paper, we extract and lemmatize their title, similarly to what done for the Project Gutenberg.
    Finally, a sequence of events is created for each selected journal, concatenating the lemmatized words in the titles of each paper in their temporal order, for a total of 19\ 000 sequences with median length of 9\ 114.5.
    Associated to \deletetext{this}\newtext{each} sequence, we also consider the sequence of stemmed words for further analysis\newtext{, similarly to the Project Gutenberg corpus.}.

\subsection*{Power-law fit}
\label{sec:fit}
Fundamental for the estimation of the higher-order Heaps' exponent of a sequence is the power-law fitting procedure for the number of novel $n$-tuples $D_n(t)$ as a function of the sequence length $t$, with $n \geq 1$. 
The sequences analyzed in this \deletetext{work}\newtext{manuscript} come from very different contexts, from empirical data sets to model simulations. We thus need to take into consideration all those cases that show a transient regime---whose length might also depend on the system structure~\cite{iacopini2020interacting}---in which the pace of discovery can fluctuate before reaching its stationary value.
\deletetext{W}\newtext{Therefore, w}e fit each sequence according to the following procedure. To reduce computational times, we first logarithmically sample 1000 \newtext{real }points\deletetext{ from each sequence} in the range $[1, T]$\newtext{, where $T$ is the length of the sequence}. 
\deletetext{Considering their integer part and discarding all duplicates, we obtain a set of $k$ integer times $\{t_i\}_{i=1,\dots,k}$ between $1$ and $T$. }% 
\newtext{Considering their integer part, we discard all the duplicates that may be produced when some sampled points differ only for the decimal part. We thus obtain a set of $k$ integer times $\{t_i\}_{i=1,\dots,k}$ between $1$ and $T$. Due to the removal of duplicates, $k$ can be equal or smaller than $1000$.}
If $T \geq 1000$, that is the case of all sequences \deletetext{used}\newtext{analyzed} in this manuscript, then this process results in $k \geq 424$ points.
Taking into account that the associated sequence of $n$-tuples has length $T-n+1$, we thus consider the points $\left\{(t_i-n+1,\,D_n(t))\right\}_{i=1,\dots,k}$ in logarithmic scale, i.e., 
\begin{equation}\label{eq:power_law_fit_points}
    (x_i, y_i) = (\log_{10}(t_i-n+1),\,\log_{10}(D_n(t)))\,,
\end{equation}
with $i=1,\dots,k$.
In order to neglect the initial transient regime, but still have enough points for a sufficiently significant fit, we select only the last 100 of such points. 
We hence look for the best fit of $\{(x_i, y_i)\}_{i=k-100+1,\dots,k}$ by optimizing the linear function $y = a + b\,x$, with $a \geq 0$, using the tool \texttt{curve\_fit} of the Python package \texttt{Scipy}~\cite{virtanen2020scipy}. \newtext{The constraint $a \geq 0$ is necessary to avoid that the fitted Heaps' exponent is greater than 1, which could happen when the initial transient regime differs significantly from the asymptotic one and could hence produce wrong fits.}
\newtext{Finally, }if $\overline{a}$ and $\overline{b}$ are the best parameters, then the power-law fit of the Heaps' law is $D_n(t) \approx 10^{\overline{a}}\,t^{\overline{b}}$, that is, the $n$\textsuperscript{th}-order Heaps' exponent is approximated by the slope $\overline{b}$ of the fit.

\subsection*{Urn Model with (Semantic) Triggering}
\label{sec:UMT_explanation}
The Urn Model with Triggering (UMT) is a \newtext{random }generative model \deletetext{of a}\newtext{for} discovery process\newtext{es}, producing a sequence of extractions of balls of various colors\deletetext{, representing different events,} from an urn. 
First introduced in Ref.~\cite{tria2014dynamics}, it successfully reproduces the main features of empirical discovery\deletetext{ and innovation} processes~\cite{tria2014dynamics, loreto2016dynamics, tria2018zipf, tria2020taylor}.
The UMT can be thought as an extension of P\'{o}lya Urn processes~\cite{eggenberger1923polya,polya1930quelques,johnson1977urn,hoppe1984polya,mahmoud2009polya} that includes the concept of {\it adjacent possible}~\cite{kauffman1996investigations} in the way a novelty can trigger further ones~\cite{gravino2016crossing, monechi2017waves}. 
Differently from the classic urn of P\'{o}lya in which only balls of existing colors can be added to the urn, the UMT features a growing number of colors, that is, the set of possible events expands together with the exploration process. It is hence the process itself that shapes the content of the urn by reinforcing elements already discovered and adding new possibilities.

Supposing that the urn initially contains $N_0$ balls of different colors, the UMT works as follows. At each discrete time-step $t$, a ball is randomly drawn from the urn with uniform probability, and its color is marked in a temporally-ordered sequence of events $\mathcal{S}$ at position $t$. The extracted ball is then put back in the urn together with other $\rho$ copies of the same color, in a {\it rich-get-richer} manner~\cite{barabasi1999emergence}. This mechanism ensures that frequently adopted items, visited places, or exploited concepts will be more and more likely to be adopted, visited, or exploited in the future. 
Furthermore, if the color of the extracted ball has never appeared before in $\mathcal{S}$, this event is considered to be a novelty. As a consequence it triggers new possibilities, represented by the addition of $\nu+1$ balls---each of a new different color---into the urn. This triggering mechanism thus ensures the expansion of the space of possibilities.

\medskip

In a different version of the model, the Urn Model with Semantic Triggering (USMT), the sequences produced contain semantic correlations between consecutive extractions, as seen in the data~\cite{tria2014dynamics}. 
The UMST works similarly to the UMT, but with the introduction of semantic groups\newtext{ for colors}. In particular, at each triggering event, supposing that the triggering color belongs to the group $A$, the new $\nu+1$ colors are assigned to a common new group $B$, semantically related to the triggering color.
Therefore, a color $i$ of label $A$ is semantically related to all other colors of label $A$ (siblings), the color that triggered the addition of $A$ in the urn (parent), as well as all colors of label $B$ that have been triggered by $i$ (children).
Taking this into consideration, at each extraction, the probability to extract each color changes depending on a fixed parameter $\eta \in [0,\,1]$. A ball has weight 1 if its color is semantically related to the one extracted on the previous time-step, otherwise it has weight $\eta$.
Notice that we can recover the original UMT by simply considering $\eta = 1$.

\newtext{Finally, a}\deletetext{A}s shown in Ref.~\cite{tria2014dynamics}, the effect of $N_0$ is negligible at large times. 
For simplicity, we thus consider $N_0 = 1$ in our simulations of both UMT and UMST.

%%%%%%%%%%%%%%%%%%%%%%%%%%%%%%%%%%%%%%%%%%%%%%%%%%%%%%%%%%%%%%%%%%%%%%%%%%%%%%%%%%%%%%%%%%
% ERRW
%%%%%%%%%%%%%%%%%%%%%%%%%%%%%%%%%%%%%%%%%%%%%%%%%%%%%%%%%%%%%%%%%%%%%%%%%%%%%%%%%%%%%%%%%%
\subsection*{Edge-Reinforced Random Walk}
Given a weighted connected graph $G=(\mathcal{V},\,\mathcal{E})$ with $N=|\mathcal{V}|$ \deletetext{nodes}\newtext{vertices (nodes)} and $M=|\mathcal{E}|$ \deletetext{links}\newtext{edges (links)}, the Edge-Reinforced Random Walk (ERRW) is a dynamical process that reinforces the weights of the visited edges in $\mathcal{E}$, leading to Heaps' laws~\cite{iacopini2018network}.
The weights of the links in the network\deletetext{s} quantify the strength of the relationship among nodes, and are encoded in a time-varying adjacency matrix $W^t \equiv \{w_{ij}^t\}$. This matrix features non-zero entries $w^t_{ij}$ when at time $t$ the link connecting node $i$ and node $j$ is different from zero. Let us assume that at time $t=0$ each link $(i,j) \in \mathcal{E}$ has weight $w_{ij}^0 = 1$, while all other weights are set to zero.
At each time step, a walker at node $i$ walks to a neighboring node $j$ with a probability that is proportional to the weight of the \deletetext{corresponding link}\newtext{outgoing links}, i.e., $\mathbb{P}(i \to j) = w_{ij}^t/\sum_l w_{il}^t$.
After moving to the \newtext{randomly }chosen node $j$, a reinforcement $\rho$ is added to the weight of the traversed edge $(i,j)$, i.e., $w_{ij}^{t+1} = w_{ij}^t + \rho$. 
\deletetext{Given}\newtext{Starting from} an underlying structure\newtext{ given by the graph $G$}, the ERRW can generate sequences of visited nodes \deletetext{associated to a different }\newtext{with a tunable }pace of discovery \deletetext{by tuning}\newtext{obtained by properly calibrating} the reinforcement parameter $\rho$~\cite{iacopini2018network}. 
\deletetext{The}\newtext{Because of the} interplay between structure and dynamics\deletetext{ means that}\newtext{,} different structures might require different values of the reinforcement parameter to reach the same pace of discovery\deletetext{ (Heaps' law)}. 
For example, higher values of $\rho$ must be chosen for a \newtext{denser }graph\deletetext{ with a higher average degree}. This is similar to what happens in the UMT, in which we need higher values of the reinforcement parameter $\rho$ to obtain the same pace of discovery as we increase the triggering parameter $\nu$.

%%%%%%%%%%%%%%%%%%%%%%%%%%%%%%%%%%%%%%%%%%%%%%%%%%%%%%%%%%%%%%%%%%%%%%%%%%%%%%%%%%%%%%%%%%
% NEW MODEL
%%%%%%%%%%%%%%%%%%%%%%%%%%%%%%%%%%%%%%%%%%%%%%%%%%%%%%%%%%%%%%%%%%%%%%%%%%%%%%%%%%%%%%%%%%
\subsection*{Edge-Reinforced Random Walk with Triggering}
\label{sec:ERRWT}
In this manuscript we propose a generative model of a discovery process based on the exploration of a growing network, i.e., the Edge-Reinforced Random Walk with Triggering (ERRWT), which can be considered as a UMT-inspired extension of the ERRW model. 
For this model, any initial connected network $G^0 = (\mathcal{V}^0,\,\mathcal{E}^0)$ with $N^0 = |\mathcal{V}^0| \geq 1$ nodes and $M^0=|\mathcal{E}^0|$ links can be used. Let us suppose that the nodes of the graph are indexed, that is, $\mathcal{V}^0 = \{1,\,2,\,\dots,\,N_0\}$.
Similarly to the ERRW model, we assume that all initial links $(i,j) \in \mathcal{E}^0$ have weight $w_{ij}^0 = 1$.
The initial node to start the exploration process is randomly selected from $\mathcal{V}^0$.
We let the graph evolve during the process, adding new nodes and links. Let $G^t = (\mathcal{V}^t,\,\mathcal{E}^t)$ be the graph at time $t$.
The structure of the growing network is encrypted in the time-varying weighted adjacency matrix $W^t \equiv \{w_{ij}^t\}$, where $w_{ij}^t$ represents the weight of the link $(i,\, j)$ at time $t$. We assume here that $G^t$ is an undirected graph, so the matrix $W^t$ is symmetric, and any variation of $w_{ij}^t$ affects $w_{ji}^t$ too.
Supposing that at time $t$ the ERRWT is positioned on node $i$ of $G^t$, the model obeys to the following rules.
\begin{itemize}
    \item \textit{Choice of next node.} The ERRWT randomly moves to a neighbouring node $j$ of the current node $i$. The probability to move to node $j$ depends on the weight of the outgoing links of $i$, i.e., 
    \begin{equation}
        \mathbb{P}(i \to j) = \frac{w_{ij}^t}{\sum_l w_{il}^t}.
    \end{equation}
    \item \textit{Edge reinforcement.} The weight of the chosen edge $(i,j$) is reinforced by $\rho$, that is, 
    \begin{equation}
        w_{ij}^{t+1} = w_{ij}^t + \rho.
    \end{equation}
    \item \textit{Edge triggering.} If the walker never traversed the chosen edge $(i,j)$ before\deletetext{ this time}, i.e., it is a new link, then $\nu_2+1$ new possible links are added to the network. These links are connections of unitary weight between $j$ and previously visited nodes $l=l_1,\,\dots,\,l_{\nu_2}$ in $\mathcal{V}^t$, for which the link $(j, \, l)$ has never been traversed by the walker. If one of these edges already exists in the space of possibilities, its weight is reinforced by one more unit, otherwise, it is added to $\mathcal{E}^{t+1}$.
    In other words, we have
    \begin{equation}
        \begin{split}
            &w_{jl}^{t+1} = w_{jl}^t + 1, \quad l = l_1,\,\dots,\,l_{\nu_2} \,\,|\,\, l \text{ old}, (j,l) \text{ new}.\\
        \end{split}
    \end{equation}
    \item \textit{Node triggering.} If the walker never visited the chosen node $j$ before\deletetext{ this time}, i.e., it is a new node, then $\nu_1 + 1$ new\newtext{ possible} nodes are added to the network; these are connected to node $j$ with unitary weights.
    Mathematically, we have
    \begin{equation}
        \begin{split}
            &\mathcal{V}^{t+1} = \mathcal{V}^t + \{l\}_{l=|\mathcal{V}^t| + 1,\,\dots,\, |\mathcal{V}^t| + \nu_1+2}\\
            &w_{jl}^{t+1} = 1, \quad l=|\mathcal{V}^t| + 1,\,\dots,\, |\mathcal{V}^t| + \nu_1+2.
        \end{split}
    \end{equation}
    Notice that if the chosen node $j$ is new, then also the traversed edge $(i,j)$ is necessarily new as well. Therefore, in this case there is also a triggering of $\nu_2+1$ edges from $j$ to other previously visited nodes, as described before.
\end{itemize}

Finally, in this manuscript, we let \newtext{the initial graph }$G_0$ be a small graph that emulates the triggering mechanism introduced, shown in Fig.~S11 in SI. 
This is a regular tree with branching parameter $\nu_1 + 1$ and two levels, where \newtext{only }the leaves are considered new, \deletetext{while}\newtext{since} all other nodes have already triggered.
In other words, a root node has triggered $\nu_1 + 1$ nodes connected to it, and again these nodes have also triggered each $\nu_1 + 1$ other nodes.
Therefore, we initially suppose that the triggered nodes, which are $\nu_1 + 2$ in number, are all known to the walker at the start of the simulation, and do not trigger again when later explored.
Moreover, we assume that all links are new to the walker and have unitary weight.
This initialization makes sure that in the initial stages of the simulation there are enough possible links between already known nodes.
As we show in Sec.~S4 in SI where we test different initial graphs, the initialization procedure only affects thermalization times, and becomes irrelevant asymptotically. 

%%%%%%%%%%%%%%%%%%%%%%%%%%%%%%%%%%%%%%%%%%%%%%%%%%%%%%%%%%%%%%%%%%%%%%%%%%%%%%%%%%%%%%%%%%
%%%%%%%%%%%%%%%%%%%%%%%%%%%%%%%%%%%%%%%%%%%%%%%%%%%%%%%%%%%%%%%%%%%%%%%%%%%%%%%%%%%%%%%%%%
%%%%%%%%%%%%%%%%%%%%%%%%%%%%%%%%%%%%%%%%%%%%%%%%%%%%%%%%%%%%%%%%%%%%%%%%%%%%%%%%%%%%%%%%%%
%%%%%%%%%%%%%%%%%%%%%%%%%%%%%%%%%%%%%%%%%%%%%%%%%%%%%%%%%%%%%%%%%%%%%%%%%%%%%%%%%%%%%%%%%%

% \medskip

\subsection*{Data availability}
The data used in this manuscript is publicly available at Refs.~\cite{lastfm1k_dataset, gerlach2020standardized, semanticscholar2018}, and has been obtained and used according to their terms and conditions.

\subsection*{Code availability}
All the code used to download, process and analyse the data and the models can be found at Ref.~\cite{github_higher_order_heaps_laws}.

%%%%%%%%%%%%%%%%%%%%%%%%%%%%%%%%%%%%%%
%%%%%%%%%%%% BIBLIOGRAPHY %%%%%%%%%%%% 
%%%%%%%%%%%%%%%%%%%%%%%%%%%%%%%%%%%%%%

% \section*{References}
{
\footnotesize
\bibliography{biblio.bib}
}

\subsection*{Acknowledgements}
G.D.B. acknowledges support from the French Agence Nationale de la Recherche (ANR), under grant ANR-21-CE38-0020 (project ScientIA). 
A.P. and V.L. acknowledge support from the PNRR GRInS Project. 
I.I. acknowledges partial support from the James S. McDonnell Foundation $21^{\text{st}}$ Century Science Initiative ``Understanding Dynamic and Multi-scale Systems''.
All computations have been performed via the High Performance Computing (HPC) cluster provided by Queen Mary University of London~\cite{apocrita2017qmul}. 
\newtext{A.P. wishes to thank Roberto di Mari for several comments and suggestions.}

\subsection*{Author contributions}
G.D.B, I.I., A.P., and V.L. designed the study. A.P. performed a preliminary investigation, collected in an early draft. G.D.B. carried out the data collection and performed the numerical simulations. G.D.B., A.B., and G.D.M. carried out the analytical calculations. G.D.B, A.B., I.I., and V.L. wrote 
the manuscript. 
All authors contributed to analyze the data, discuss the results, define the proposed model, and revise the manuscript.

\subsection*{Competing interests}
The authors declare no competing interests.

\subsection*{Additional information}
{\bf Supplementary Information} is attached to this manuscript.

\clearpage
\newpage

%%%%%%%%%%%%%%%%%%%%%%%%%%%%%%%%%%%%%%
%%%%%%% SUPPLEMENTARY MATERIAL %%%%%%%
%%%%%%%%%%%%%%%%%%%%%%%%%%%%%%%%%%%%%%

%%%%%%%%%% Prefix a "S" to all equations, figures, equations, tables and reset the counter %%%%%%%%%%
\setcounter{figure}{0}
\setcounter{table}{0}
\setcounter{equation}{0}
\setcounter{section}{0}
\makeatletter
\renewcommand{\thefigure}{S\arabic{figure}}
\renewcommand{\theequation}{S\arabic{equation}}
\renewcommand{\thetable}{S\arabic{table}}
\renewcommand{\thesection}{S\arabic{section}}
%%%%%%%%%% Prefix a "S" to all equations, figures, equations, tables and reset the counter %%%%%%%%%%

\setcounter{secnumdepth}{2} 
\onecolumn

% \widetext
\begin{center}
	\textbf{\Large Supplementary Information for\\``The dynamics of higher-order novelties"}
\end{center}

\section{Higher-order Heaps' exponents in the data sets}
\label{sec:SI_Heaps_data}

\begin{figure}[!hb]
\begin{center}
\includegraphics[width=0.5\linewidth]{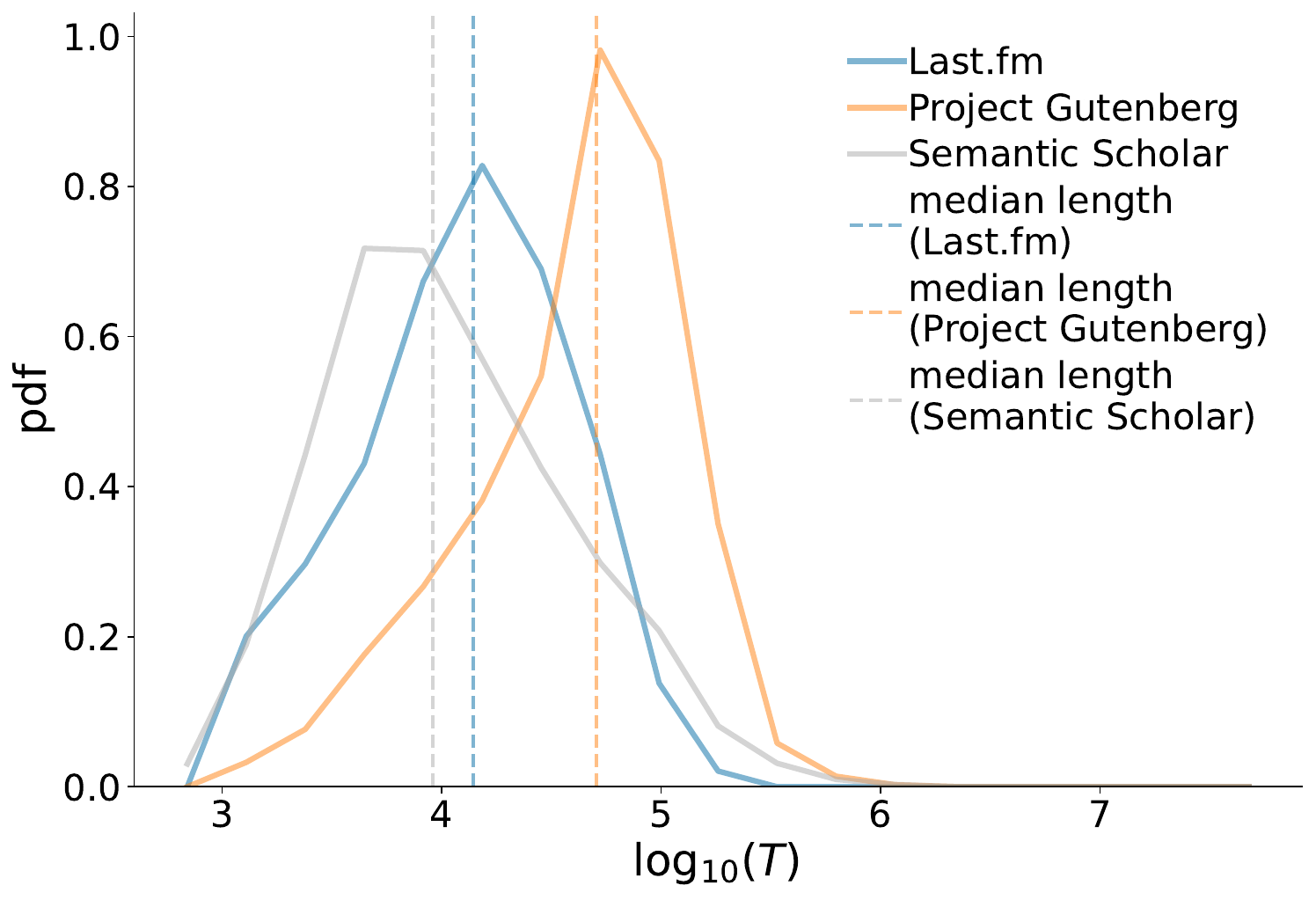}
\caption[Length distribution of the sequences in the data sets.]
{
\textbf{Length $T$ distribution of the sequences in the data sets.} 
Probability density function of the length $T$ of all sequences in the three data sets (Last.fm in blue, Project Gutenberg in orange, Semantic Scholar in green). 
Moreover, the median lengths, respectively equal to 13\ 985, 50\ 726, and 9\ 114.5, are shown in the plot as vertical dashed lines, with the color corresponding to each data set\newtext{, as shown in the legend}.
}
\label{fig:SI_Heaps_data_lengths}
\end{center}
\end{figure}

%%%%% Table SI: Heaps exponents fits standard error
\begingroup
\definecolor{Gray}{gray}{0.9}
% \newcolumntype{A}{>{\centering\arraybackslash}m{0.14\linewidth}}
% \newcolumntype{B}{>{\centering\arraybackslash}m{0.1\linewidth}}
% \setlength{\tabcolsep}{10pt}
\begin{table}[!ht]
\begin{center}
\rowcolors{0}{white}{Gray}
\begin{tabular}{cccccccc}
\toprule
         data set ($\beta_n$) &    min &  $1$\textsuperscript{st} perc. &  $25$\textsuperscript{th} perc. &  median &  $75$\textsuperscript{th} perc. &  $99$\textsuperscript{th} perc. &    max \\
\midrule
          Last.fm ($\beta_1$) & 0.0007 &                         0.0016 &                          0.0052 &  0.0079 &                          0.0129 &                          0.0693 & 0.1988 \\
          Last.fm ($\beta_2$) & 0.0000 &                         0.0001 &                          0.0026 &  0.0047 &                          0.0091 &                          0.0510 & 0.1497 \\
          Last.fm ($\beta_3$) & 0.0000 &                         0.0000 &                          0.0019 &  0.0038 &                          0.0073 &                          0.0388 & 0.1366 \\
          Last.fm ($\beta_4$) & 0.0000 &                         0.0000 &                          0.0015 &  0.0031 &                          0.0064 &                          0.0343 & 0.1294 \\
Project Gutenberg ($\beta_1$) & 0.0000 &                         0.0010 &                          0.0021 &  0.0029 &                          0.0043 &                          0.0169 & 0.0727 \\
Project Gutenberg ($\beta_2$) & 0.0003 &                         0.0005 &                          0.0010 &  0.0014 &                          0.0020 &                          0.0087 & 0.0522 \\
Project Gutenberg ($\beta_3$) & 0.0001 &                         0.0002 &                          0.0004 &  0.0006 &                          0.0009 &                          0.0064 & 0.0484 \\
Project Gutenberg ($\beta_4$) & 0.0000 &                         0.0001 &                          0.0001 &  0.0002 &                          0.0005 &                          0.0051 & 0.0444 \\
 Semantic Scholar ($\beta_1$) & 0.0003 &                         0.0008 &                          0.0018 &  0.0025 &                          0.0035 &                          0.0115 & 0.1279 \\
 Semantic Scholar ($\beta_2$) & 0.0002 &                         0.0004 &                          0.0010 &  0.0014 &                          0.0021 &                          0.0093 & 0.1677 \\
 Semantic Scholar ($\beta_3$) & 0.0000 &                         0.0002 &                          0.0005 &  0.0008 &                          0.0013 &                          0.0078 & 0.1698 \\
 Semantic Scholar ($\beta_4$) & 0.0000 &                         0.0001 &                          0.0003 &  0.0005 &                          0.0009 &                          0.0068 & 0.1620 \\
\bottomrule
\end{tabular}
\caption[Statistics on the standard error of the fitted higher-order Heaps' exponents in the empirical data.]
{\textbf{Statistics on the standard error of the fitted higher-order Heaps' exponents in the empirical data.}
Various statistics on the standard error, or standard deviation of the estimator, of the fitted $n$\textsuperscript{th}-order Heaps' exponents $\beta_n$ of the sequences in the three data sets, with $n = 1$, $2$, $3$, and $4$. 
Notice how the standard deviation of the distribution of the values of the exponents in the data sets (see Table.~\ref{table:SI_data_sets_fits_heaps} for reference) is about two orders of magnitude higher than the median standard error and one order higher than its $99$\textsuperscript{th} percentile.
Moreover, the $p$-values of the fits are all zero (not shown in the table). \newtext{This proves we can consider the fits valid, and thus the $n$\textsuperscript{th}-order Heaps' laws valid in almost all the sequences of the data sets considered.}
}
\label{table:SI_data_sets_error_fits_heaps}
\end{center}
\end{table}
\endgroup

\clearpage
\begin{figure}[!t]
\begin{center}
\includegraphics[width=\linewidth]{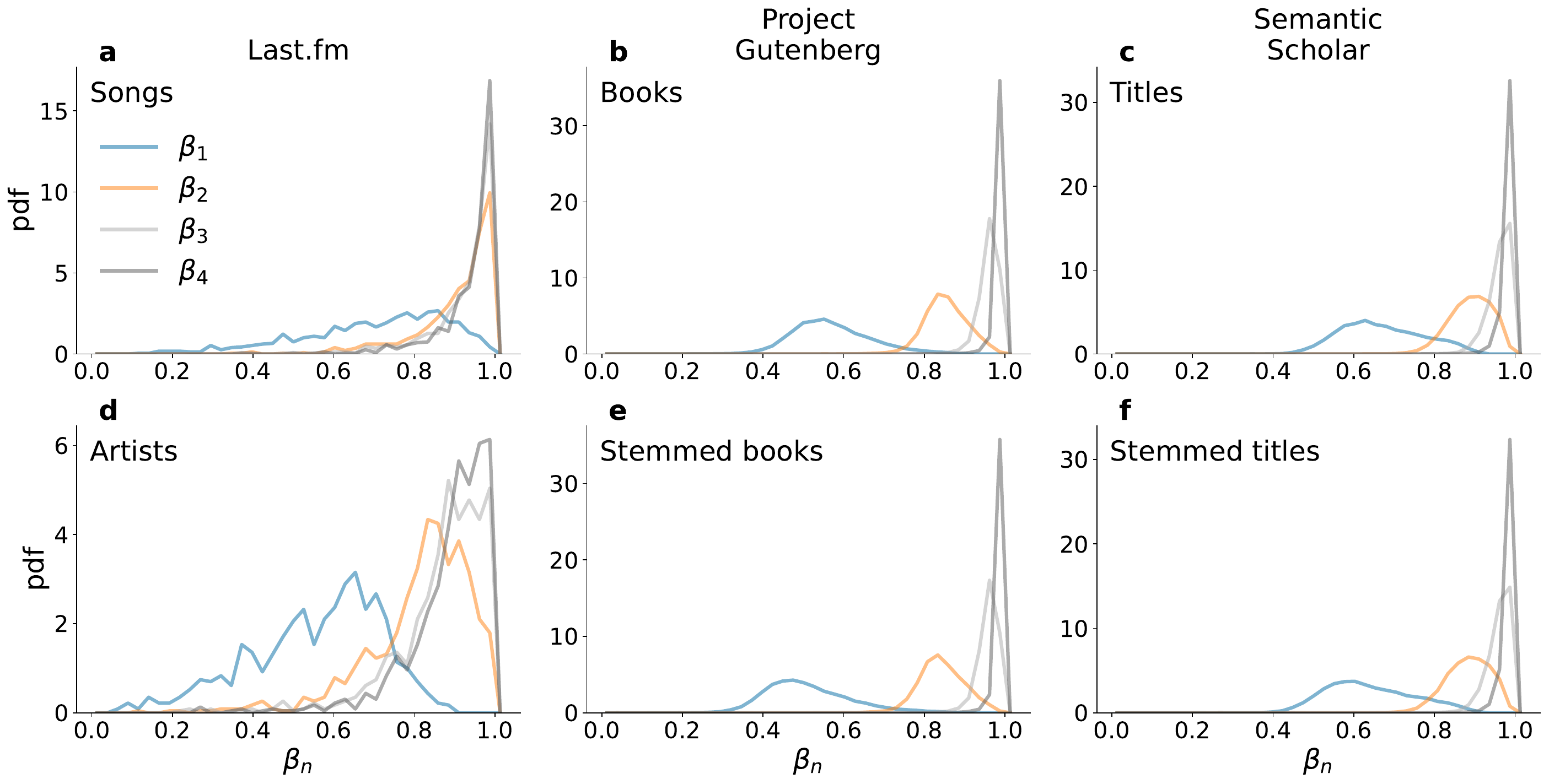}
\caption[Heaps' exponent distribution of the sequences in the data sets.]
{
\textbf{Heaps' exponent distribution of the sequences in the data sets.} 
Probability density functions of the $n$\textsuperscript{th}-order Heaps' exponents $\beta_n$, with $n = 1, 2, 3, 4$, calculated from the empirical sequences (a-c) and respective sequences of labels (d-f). In particular, sequences contain songs (\textbf{a}) and artists (\textbf{d}) in Last.fm, words (\textbf{b}) and stemmed words (\textbf{e}) in Project Gutenberg books, words (\textbf{c}) and stemmed words (\textbf{f}) in Semantic Scholar journal titles. 
}
\label{fig:SI_Heaps_data_betas}
\end{center}
\end{figure}

\begingroup
\definecolor{Gray}{gray}{0.9}
\begin{table}[!hb]
\begin{center}
\rowcolors{0}{white}{Gray}
\begin{tabular}{cccccccccc}
\toprule
         data set ($\beta_n$) &   mean &    std &     min &  $1$\textsuperscript{st} perc. &  $25$\textsuperscript{th} perc. &  median &  $75$\textsuperscript{th} perc. &  $99$\textsuperscript{th} perc. &    max \\
\midrule
          Last.fm ($\beta_1$) & 0.7029 & 0.1797 &  0.1063 &                         0.2010 &                          0.5965 &  0.7395 &                          0.8436 &                          0.9761 & 0.9959 \\
          Last.fm ($\beta_2$) & 0.9048 & 0.1014 &  0.3342 &                         0.5725 &                          0.8699 &  0.9388 &                          0.9754 &                          0.9999 & 0.9999 \\
          Last.fm ($\beta_3$) & 0.9286 & 0.0862 &  0.3664 &                         0.6123 &                          0.9041 &  0.9583 &                          0.9861 &                          0.9999 & 1.0000 \\
          Last.fm ($\beta_4$) & 0.9411 & 0.0759 &  0.3837 &                         0.6643 &                          0.9195 &  0.9679 &                          0.9896 &                          0.9999 & 1.0000 \\
Project Gutenberg ($\beta_1$) & 0.5699 & 0.0973 & -0.0000 &                         0.3678 &                          0.5026 &  0.5594 &                          0.6285 &                          0.8302 & 0.9527 \\
Project Gutenberg ($\beta_2$) & 0.8527 & 0.0547 &  0.0304 &                         0.7118 &                          0.8191 &  0.8509 &                          0.8883 &                          0.9706 & 0.9919 \\
Project Gutenberg ($\beta_3$) & 0.9589 & 0.0300 &  0.0307 &                         0.8648 &                          0.9480 &  0.9627 &                          0.9765 &                          0.9968 & 0.9998 \\
Project Gutenberg ($\beta_4$) & 0.9882 & 0.0203 &  0.0304 &                         0.9242 &                          0.9870 &  0.9923 &                          0.9958 &                          0.9995 & 1.0000 \\
 Semantic Scholar ($\beta_1$) & 0.6695 & 0.1019 &  0.2225 &                         0.4673 &                          0.5923 &  0.6590 &                          0.7436 &                          0.8889 & 0.9293 \\
 Semantic Scholar ($\beta_2$) & 0.8895 & 0.0536 &  0.2587 &                         0.7509 &                          0.8550 &  0.8936 &                          0.9303 &                          0.9803 & 0.9942 \\
 Semantic Scholar ($\beta_3$) & 0.9612 & 0.0305 &  0.2665 &                         0.8686 &                          0.9478 &  0.9680 &                          0.9825 &                          0.9972 & 0.9999 \\
 Semantic Scholar ($\beta_4$) & 0.9847 & 0.0198 &  0.2790 &                         0.9177 &                          0.9807 &  0.9901 &                          0.9954 &                          0.9995 & 1.0000 \\
\bottomrule
\end{tabular}
\caption[Statistics of the fitted higher-order Heaps' exponents in the data.]
{\textbf{Statistics of the fitted higher-order Heaps' exponents in the data.}
Various statistics of the fitted $n$\textsuperscript{th}-order Heaps' exponents $\beta_n$ of the sequences in the three data sets, with $n = 1$, $2$, $3$, and $4$. 
}
\label{table:SI_data_sets_fits_heaps}
\end{center}
\end{table}
\endgroup

\begin{figure}[!hb]
\begin{center}
\includegraphics[width=\linewidth]{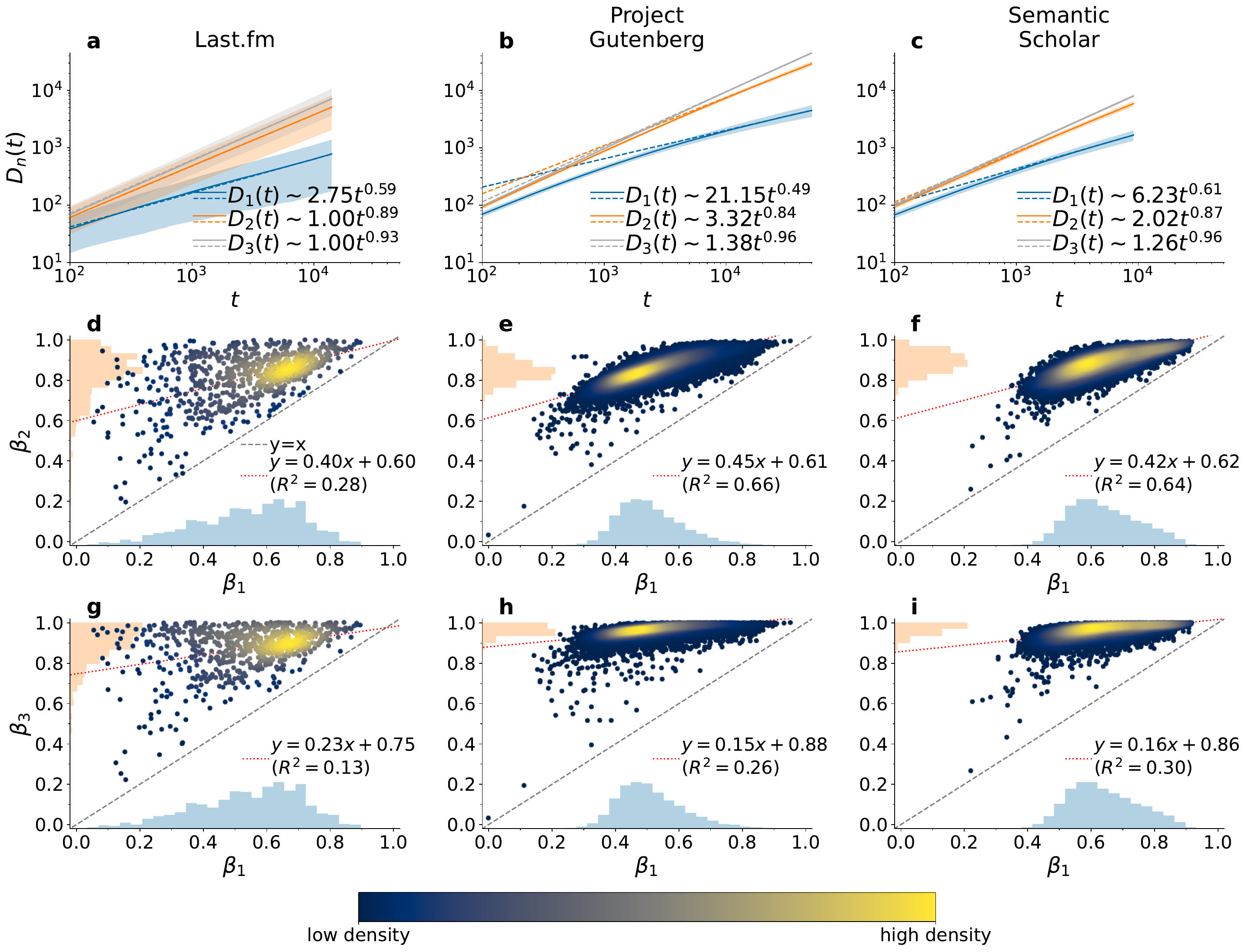}
\caption[Higher-order Heaps' exponents and their correlations in real-world data sets.]
{
\textbf{Higher-order Heaps' exponents\deletetext{ and their correlations} in real-world data sets.}
(\textbf{a-c}) Average number $D_n(t)$ of novelties of order $n$, with $n=1,\,2,\,3$, as a function of the sequence length $t$, and fit of the associated Heaps' laws (dashed lines), with estimated exponents shown in the legend. Shaded area represents one standard deviation above and below the average.
(\textbf{d-i}) Scatter plots between the ($1$\textsuperscript{st}-order) Heaps' exponents $\beta_1$ and the $n$\textsuperscript{th}-order exponents $\beta_n$, with $n = 2$ (\textbf{d-f}) and $3$ (\textbf{g-i}).
Each point refers to a different  sequence, with colors representing the density of points (see color bar). 
Each panel also reports histograms of exponents distributions, the bisector $y=x$ (dashed gray line), as well as the fitted linear model (dotted red line) with the value of its coefficient of determination $R^2$.
Each column refers to a different data set: (\textbf{a},\textbf{d},\textbf{g}) sequences of artists listened on Last.fm, (\textbf{b},\textbf{e},\textbf{h}) sequences of stemmed words of books from Project Gutenberg and (\textbf{c},\textbf{f},\textbf{i}) sequences of stemmed words of titles in journals of Semantic Scholar, respectively.
\newtext{Notice that the sequences analyzed here are different from those shown in Fig.~1 in the main manuscript, since here we consider the sequence of artists or stems instead of the related songs or words.}
}
\label{fig:SI_Heaps_data_labels}
\end{center}
\end{figure}

\begin{figure}[!hb]
\begin{center}
\includegraphics[width=\linewidth]{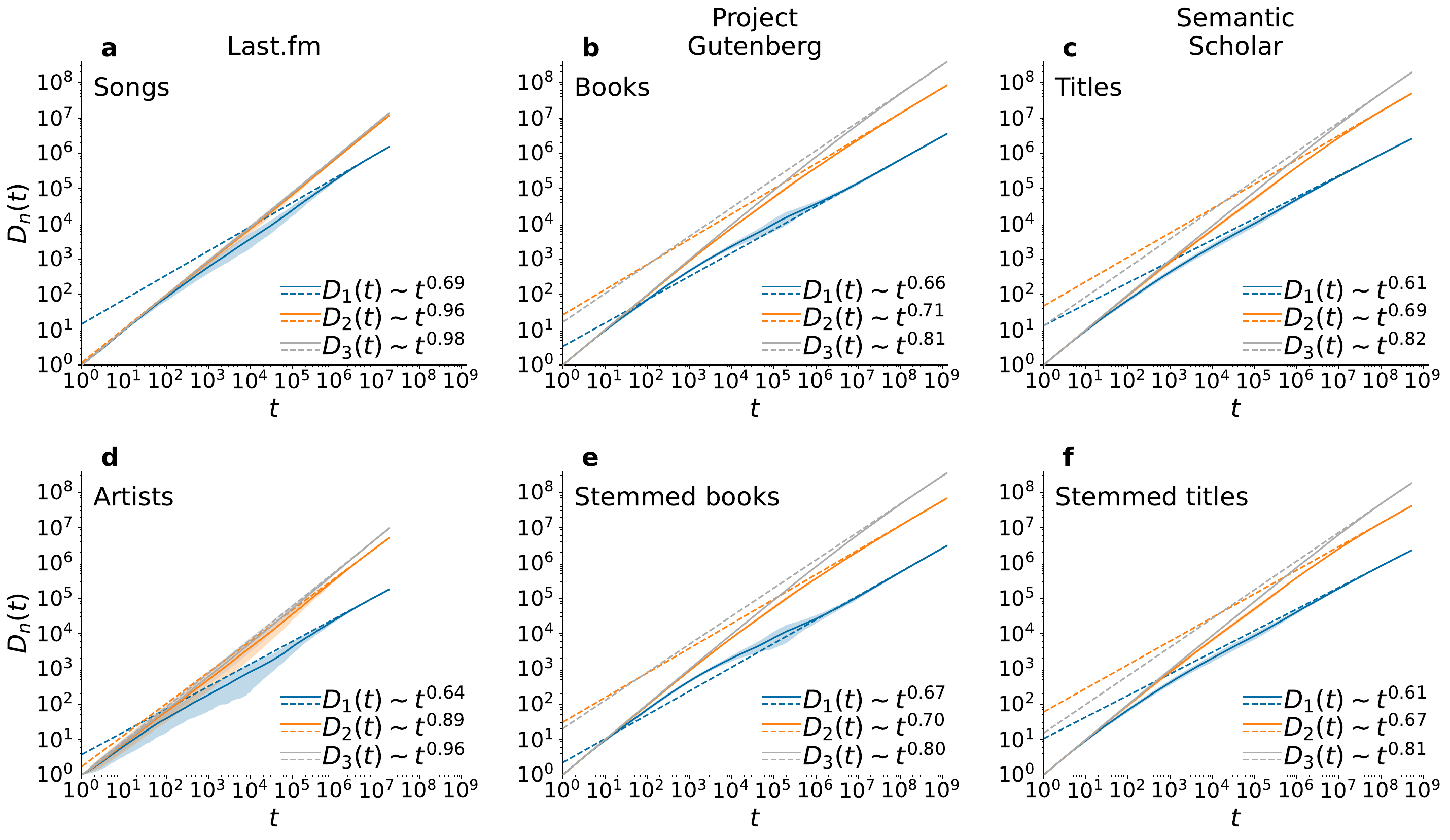}
\caption[Higher-order Heaps' laws for the collective sequence in real-world data sets.]
{
\newtext{\textbf{Higher-order Heaps' laws for the collective sequence in real-world data sets.}
Average number $D_n(t)$ of novelties of order $n$, with $n=1,\,2,\,3$, as a function of the sequence length $t$, and fit of the associated Heaps' laws (dashed lines), with estimated exponents shown in the legend. 
For each data set we compute the Heaps' laws over a collective sequence obtained by consecutively adding all individual sequences in a random order.
This procedure has been repeated 100 times for each data set. 
Shaded area represents one standard deviation above and below the average.
Each panel refers to the collective sequence of a different data set: sequences of songs (\textbf{a}) and artists (\textbf{d}) listened by users on Last.fm, sequences of words (\textbf{b}) and their stems (\textbf{e}) of books from Project Gutenberg, and sequences of words (\textbf{c}) and their stems (\textbf{f}) from titles in journals of Semantic Scholar, respectively.}
}
\label{fig:SI_Heaps_data_collective}
\end{center}
\end{figure}

\begin{figure}[!hb]
\begin{center}
\includegraphics[width=0.95\linewidth]{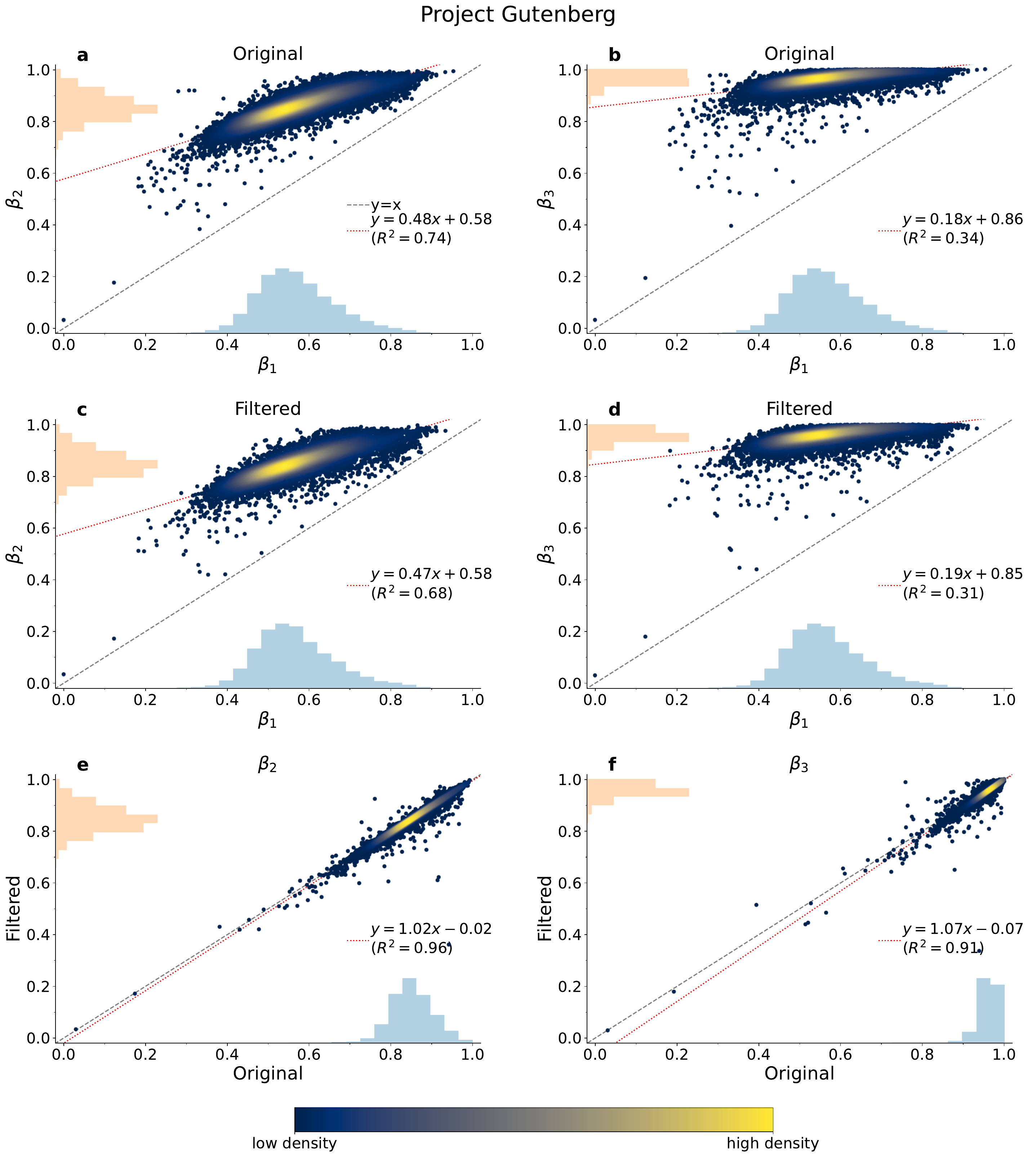}
\caption[Higher-order Heaps' exponents for Project Gutenberg, filtering out pairs and triplets of words between punctuation marks.]
{
\newtext{\textbf{Higher-order Heaps' exponents for Project Gutenberg, filtering out pairs and triplets of words between punctuation marks.}
(\textbf{a-d}) Scatter plots between the ($1$\textsuperscript{st}-order) Heaps' exponents $\beta_1$ and the $n$\textsuperscript{th}-order exponents $\beta_n$, with $n = 2$ (\textbf{a,c}) and $3$ (\textbf{b,d}).
Panels (\textbf{a-b}) refer to the original sequences of words analyzed in Fig.~1(\textbf{e,h}), while the sequences of panels (\textbf{c-d}) are obtained from the original ones filtering out all pairs or triplets of words between punctuation marks. Notice that this filtering only removes elements from the higher-order sequences.
(\textbf{e-f}) Scatter plots between the $n$\textsuperscript{th}-order Heaps' exponents for the original and filtered Project Gutenberg sequences.
Each point refers to a different  sequence, with colors representing the density of points (see color bar). 
Each panel also reports histograms of exponents distributions, the bisector $y=x$ (dashed gray line), as well as the fitted linear model (dotted red line) with the value of its coefficient of determination $R^2$.
Notice how there are no significant differences between higher-order Heaps' exponents in the original and filtered sequences.
}
}
\label{fig:SI_gutenberg_punct}
\end{center}
\end{figure}

\clearpage
\section{Heaps' exponents in simulations of existing models}
\label{sec:SI_Heaps_UMT_simulations}

\begin{figure}[!b]
\begin{center}
\includegraphics[width=\linewidth]{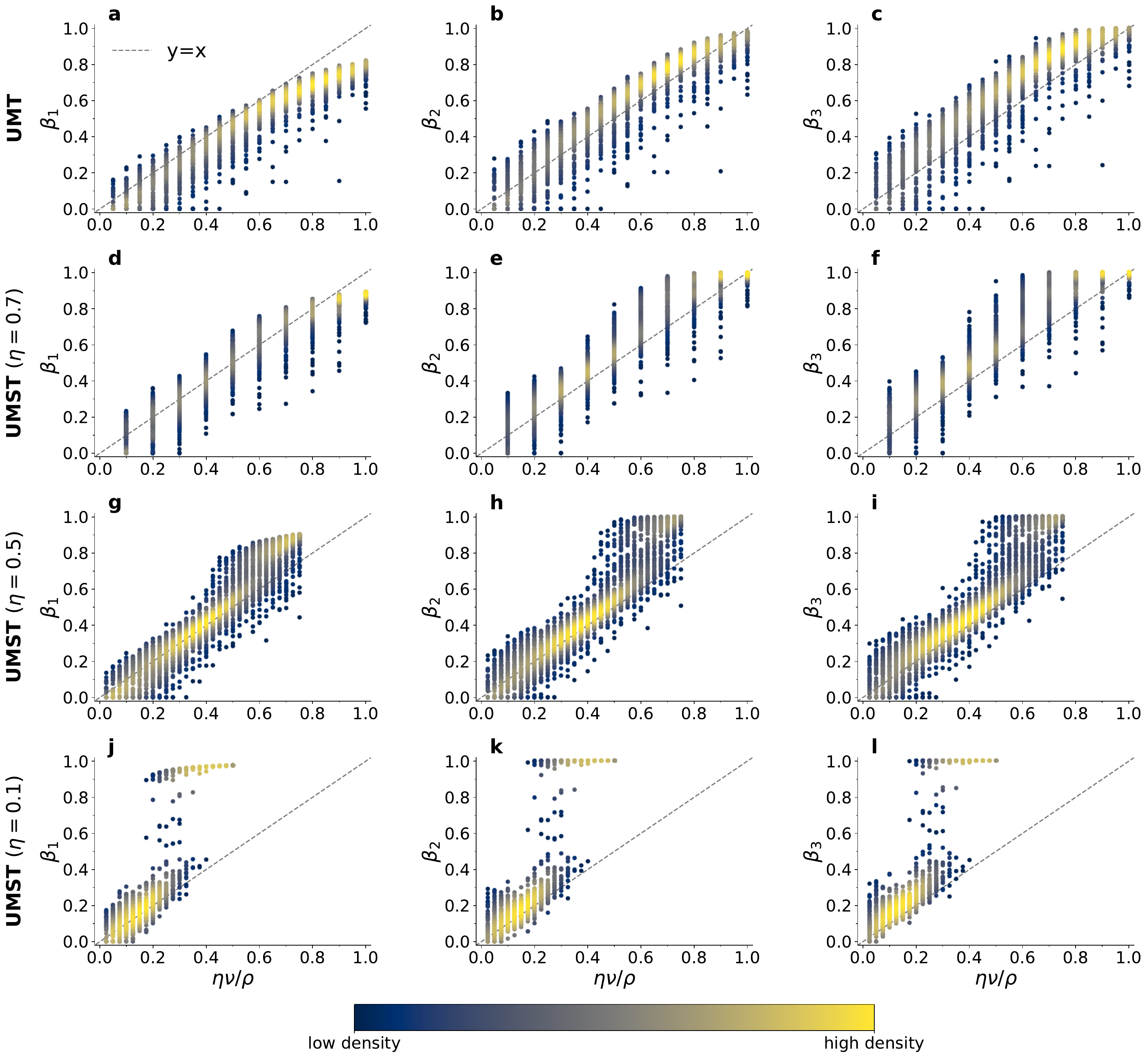}
\caption[Higher-order Heaps' exponents\deletetext{ and their correlations} with the expected asymptotic value in urn model simulations.]
{
\textbf{Higher-order Heaps' exponents\deletetext{ and their correlations} with the expected asymptotic value in urn model simulations.}
Scatter plots between the analytically expected lower bound $\eta \nu / \rho$ for the asymptotic $1$\textsuperscript{st}-order Heaps' exponent---the theoretical upper bound being $\min(1,\nu/\rho$)---and the Heaps' exponents $\beta_1$ (\textbf{a}), $\beta_2$ (\textbf{b}), $\beta_3$ (\textbf{c}), $\beta_4$ (\textbf{d}). Each point refers to a different simulation of $10^5$ time steps, colored according to the density of points (see color bar). Each panel reports the value of the correlation coefficient $r$.
The first row refers to the Urn Model with Triggering (UMT), with no semantic correlations ($\eta=1$) and $\rho=20$. 
The second row refers to the Urn Model with Semantic Triggering (UMST), with $\rho=4$ and $\eta=0.1$.
Here, we show the results of 100 simulations for each set of parameters.
}
\label{fig:SI_Heaps_simulations_UMT_vs_expected_5}
\end{center}
\end{figure}

The Urn Model with Triggering (UMT) features a triggering mechanism for the growth of the adjacent possible~\cite{tria2014dynamics}. In particular, whenever a new color is drawn for the first time, $\nu+1$ new colors are triggered and added into the urn. 
Together with the reinforcement mechanism introduced in Polya's urn~\cite{polya1930quelques}, the UMT manages to reproduce various features of \deletetext{innovation}\newtext{discovery} processes, including the Heaps' law. 
In particular, varying the parameters, the UMT produces different rates of discovery, which can be measured by the power-law exponent $\beta_1$ of the Heaps' law. According to analytical results on the asymptotic Heaps' exponent, we have that $\beta_1 \to \nu/\rho$. We check if this relation holds true at finite times in Fig.~\ref{fig:SI_Heaps_simulations_UMT_vs_expected_5}(\textbf{a}), where we show the scatter plots between $\nu/\rho$ and the fitted value of $\beta_1$ for simulations of the UMT with $\rho = 20$ and $\nu = 1, \dots,20$, run for $T=10^5$ time steps. Each point refers to a different simulation, and we analyze 100 simulations for each set of parameters.
We notice how the relationship holds true in most cases, although the fitted values are less than the theoretical ones, especially for high values of $\nu/\rho$.
We repeat this check also for higher-order Heaps' exponents in Fig.~\ref{fig:SI_Heaps_simulations_UMT_vs_expected_5}(\textbf{b-c}), finding that also in this case there is not so much difference between the theoretical value $\nu/\rho$ and the fitted $\beta_2$ and $\beta_3$, if only that the points in the plot are slightly higher than the bisector.

We repeat the same analysis for the Urn Model with Semantic Triggering (UMST), which introduces semantic triggering between the colors in the urn. In particular, two colors are considered semantically related if they have been triggered by the same color (siblings) or if one has triggered the other (parent and child). 
Then, whenever a new color needs to be extracted, a ball of a certain color in the UMST has a different weight depending on the semantic relationship with the previous color. If the two colors are related, then the ball has weight $1$, otherwise it gets weight $\eta \leq 1$.
\newtext{Notice that the UMT is equivalent to the UMST with $\eta = 1$.}
\newtext{Analytical results on the Heaps' law from the SI in Ref.~\cite{tria2014dynamics} show that the asymptotic Heaps' exponent is found between $\eta \nu / \rho$ and $\min\left(1, \nu/\rho \right)$. We test this in Fig.~\ref{fig:SI_Heaps_simulations_UMT_vs_expected_5}(\textbf{d-l}), where we show the scatter plots  between the expected lower bound $\eta\nu/\rho$ and the fitted value of the Heaps' exponents for simulations of the UMST with various sets of parameters:
(\textbf{d-f}) $\eta=0.7$, $\rho=7$, $\nu = 1,\,2, \dots,\, 10$;
(\textbf{g-i}) $\eta=0.5$, $\rho=20$, $\nu = 1,\,2, \dots,\, 30$;
(\textbf{j-l}) $\eta=0.1$, $\rho=4$, $\nu = 1,\,2, \dots,\, 20$.
Let us start our analysis by looking at panel (\textbf{j}), since the UMST can only reproduce the semantic correlations seen in the data for low values of $\eta$~\cite{dibona2022social}.}
We see that for low values of the $\eta\nu/\rho$ the value of $\beta_1$ corresponds to the \newtext{theoretical lower bound} (points are found on the bisector). 
However, starting from about $\eta\nu/\rho = 0.2$ there are simulations in which the value of $\beta_1$ goes abruptly up to 1. Notice that for these values, we have that $\nu/\rho = 2$.
Interestingly, up to $\eta\nu/\rho = 0.3$ and sometimes up to $\eta\nu/\rho = 0.4$, there are both simulations with Heaps' exponent $\beta_1 \approx \eta\nu/\rho$ and others with $\beta_1 \approx 1$, but almost none in between. After that, there remain only simulations with linear Heaps' law.
We repeat the analysis for higher-order Heaps' exponents, finding the same behavior.
In Table~\ref{table:SI_number_of_simulations_finished} we also report the number of simulations with either of the two behaviors.%
\begingroup
\definecolor{Gray}{gray}{0.9}
\setlength{\tabcolsep}{6pt}
\begin{table}[b]
\begin{center}
\rowcolors{0}{white}{Gray}
\begin{tabular}{ccccc}
\toprule
 $\eta$ &   $\rho$ &   $\nu$ &   $\beta_1 \approx \eta\nu/\rho$ &   $\beta_1 \approx 1$ \\
\midrule
    0.1 &        4 &       1 &                100 &                  0 \\
    0.1 &        4 &       2 &                100 &                  0 \\
    0.1 &        4 &       3 &                100 &                  0 \\
    0.1 &        4 &       4 &                100 &                  0 \\
    0.1 &        4 &       5 &                100 &                  0 \\
    0.1 &        4 &       6 &                100 &                  0 \\
    0.1 &        4 &       7 &                 99 &                  1 \\
    0.1 &        4 &       8 &                 91 &                  9 \\
    0.1 &        4 &       9 &                 80 &                 20 \\
    0.1 &        4 &      10 &                 51 &                 49 \\
    0.1 &        4 &      11 &                 44 &                 56 \\
    0.1 &        4 &      12 &                 21 &                 79 \\
    0.1 &        4 &      13 &                 12 &                 88 \\
    0.1 &        4 &      14 &                  0 &                100 \\
    0.1 &        4 &      15 &                  0 &                100 \\
    0.1 &        4 &      16 &                  1 &                 99 \\
    0.1 &        4 &      17 &                  0 &                100 \\
    0.1 &        4 &      18 &                  0 &                100 \\
    0.1 &        4 &      19 &                  0 &                100 \\
    0.1 &        4 &      20 &                  0 &                100 \\
\bottomrule
\end{tabular}
\caption[Statistics on the number of simulations of urn models.]
{
\textbf{Statistics on the number of simulations of urn models.}
Number of simulations of the UMST ($\eta = 0.1$, $\rho = 4$, $\nu = 1, \dots, 20$) that have an exponent approximately equal to the lower bound $\eta \nu / \rho$ or to the upper bound $1$.
For each set of parameters, 100 simulations have been launched.
}
\label{table:SI_number_of_simulations_finished}
\end{center}
\end{table}
\endgroup
Notice how the number of simulations with $\beta_1 \approx 1$ increases with higher values of $\nu$.

\begin{figure*}
\begin{center}
\includegraphics[width=\linewidth]{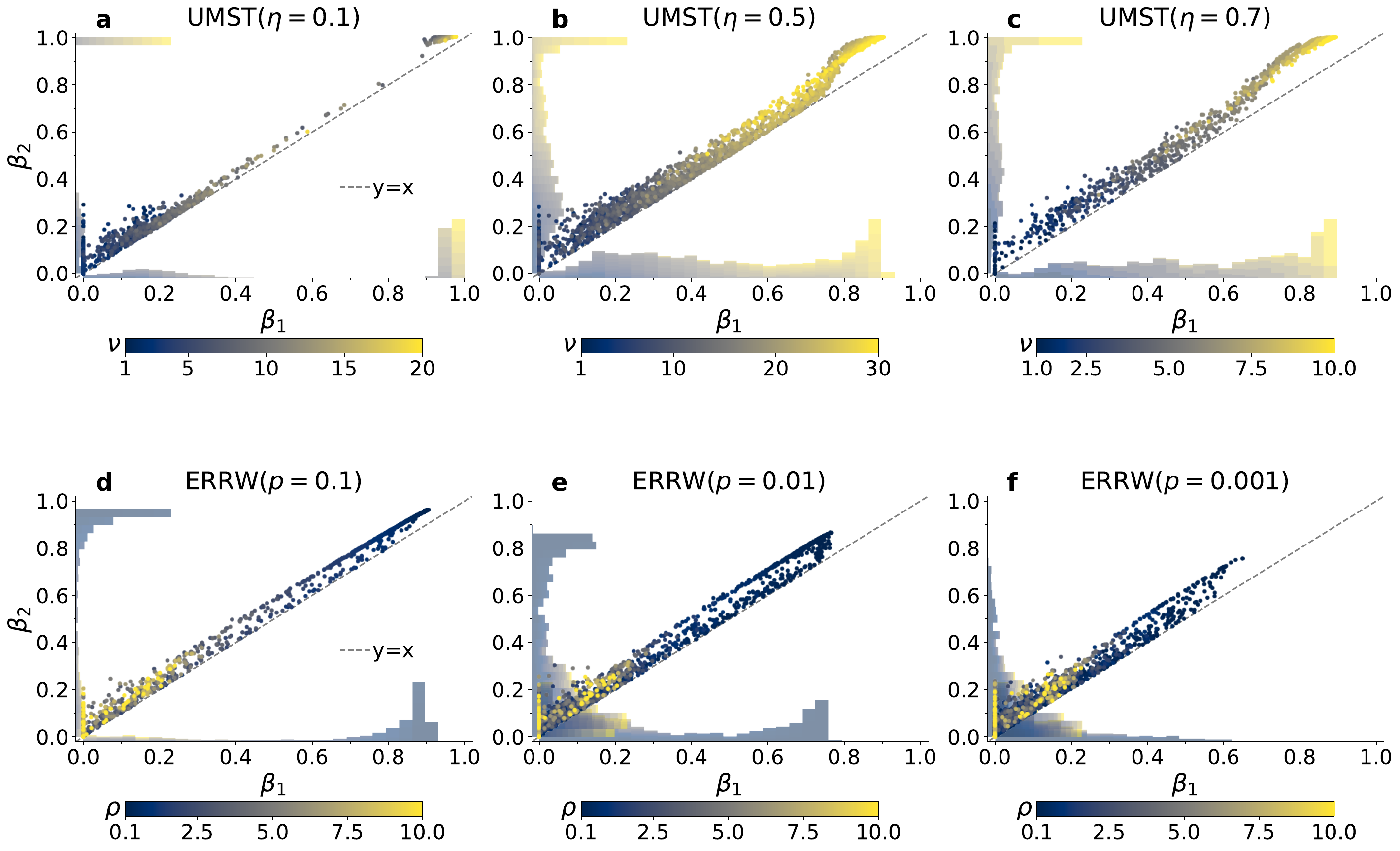}
\caption[Higher-order Heaps' exponents in existing models.]
{
\textbf{Higher-order Heaps' exponents in existing models with other parameters.} 
Scatter plots between the (1\textsuperscript{st}-order) Heaps' exponent $\beta_1$ and the $2$\textsuperscript{nd}-order exponent $\beta_2$ in 
(\textbf{a-c}) the urn model with semantic triggering (UMST) and (\textbf{d-e}) the edge-reinforced random walk (ERRW) on a small-world network with average degree $\langle k \rangle = 4$ and edge reinforcement $\rho$ ranging (geometrically) from $0.1$ to $10$.
Each panel refers to a different set of parameters:
(\textbf{a}) $\eta=0.1$, $\rho=4$, $\nu = 1,\,2, \dots,\, 20$;
(\textbf{b}) $\eta=0.5$, $\rho=20$, $\nu = 1,\,2, \dots,\, 30$;
(\textbf{c}) $\eta=0.7$, $\rho=7$, $\nu = 1,\,2, \dots,\, 10$;
(\textbf{d}) rewiring probability of the small-world network $p = 0.1$;
(\textbf{e}) $p = 0.01$; and
(\textbf{f}) $p = 0.001$.
Each point in these plots refers to a different simulation of the related model, with colors representing the value of the free parameter (see color bar). 
Each panel also reports histograms of exponent distributions on the respective axes, and the bisector $y=x$ (dashed gray line).
All simulations have run for $10^5$ time steps.
}
\label{fig:Heaps_simulations_existing_models_other}
\end{center} 
\end{figure*}

This analysis shows the inadequacy of the UMST to reproduce the whole spectrum of paces of discovery. 
In fact, we are not able to obtain Heaps' exponents in between the two theoretical limits. 
For example, we cannot obtain Heaps' exponents between roughly 0.4 and 0.9 when $\eta = 0.1$.
Instead, simulations actually only produce exponents very close to these two bounds.
Moreover, the higher the theoretical value $\eta\nu/\rho$, the higher the chance of having a Heaps' exponent close to 1.
A possible explanation of why this could happen lies on the way semantic triggering happens.
In the UMST, indeed, when a color is drawn for the first time, $\nu+1$ balls of new colors are added to the urn, and they become semantically connected to the triggering color. Then, the probability to draw a ball of a color semantically close to the previous one is $1/\eta = 10$ times higher with respect to balls of other colors. 
This brings about two possible scenarios.
On the one hand, if a small cluster of colors is highly reinforced in the beginning of the simulation, after one of them is drawn it is very likely that another of these colors is extracted in the next time step. 
On the other hand, if a new color is drawn, since it is highly probable to move to a semantic close color and almost all of them are new, if $\nu$ is high enough the next extracted color is also almost surely new.
Then, once inside one of the two scenarios, it is very unlikely to break the loop, producing the two groups of Heaps' exponent we observe.
This also explains why the likelihood of being in the linear case increases with $\nu$, even though the two behaviors can coexist in the same set of parameters.
Furthermore, this is also confirmed by simulations with higher number steps---we tested with $10^7$ steps---, which show the same results, indicating that the behavior has already reached a stationary state.

\newtext{A similar result is obtained also for higher values of $\eta$, as shown in Fig.~\ref{fig:SI_Heaps_simulations_UMT_vs_expected_5}(\textbf{d-i}). Looking for example at panel (\textbf{g}), we can see the same behavior, where all simulations have their exponents at the theoretical lower bound for low values of $\eta\nu/\rho$ (points in the bisector), while for higher values of $\eta\nu/\rho$ the majority of simulations are closer to the upper bound instead. 
Therefore, there are very few simulations around $\beta_1 = 0.7$.
Considering the higher value of $\eta$ with respect to that shown in panel (\textbf{j}), here this behavior is less evident, but it is still present
}

\newtext{Finally, in Fig~\ref{fig:Heaps_simulations_existing_models_other} we show the scatter plot between the $1$\textsuperscript{st}-order Heaps' exponent $\beta_1$ and the $2$\textsuperscript{nd} one $\beta_2$ in various simulations of UMST and ERRW with different parameters. In particular, the parameters chosen for the UMST are the same as those discussed above in Fig.~\ref{fig:SI_Heaps_simulations_UMT_vs_expected_5}. The parameters for the ERRW are the same as those chosen in the main manuscript, i.e., the edge reinforcement goes from $\rho = 0.1$ to $10$, running on synthetic random small world networks with a rewiring probability of 0.1 (panel~\textbf{d}), 0.01 (\textbf{e}), and 0.001 (\textbf{f}). Notice how reducing the rewiring in the network, so that the network is more similar to a ring, the discovery rates at both first and second orders are lower. Nevertheless, in all cases, both UMST and ERRW cannot reproduce the empirical range of pairs ($\beta_1$, $\beta_2$) seen in the empirical data.
}

\clearpage
\section{Analytic results for higher-order Heaps' exponents in UMT simulations}
\label{sec:SI_Heaps_UMT_analytic}
In this section we provide a complete analytical analysis of the higher-order Heaps' laws for the Urn Model with Triggering (UMT).
Let us consider an urn with parameters $\rho$ and $\nu$ and initially composed by $N_0 \geq 1$ balls of different colors.  

\subsection{First-order Heaps' law}

The evolution of the number $D_1(t)$ of different colors that have appeared in the first $t$ positions of the sequence $\mathcal{S}$ is ruled by the following master equation: 
\begin{align}\label{eq:SI_master_equation_D1t}
\begin{split}
D_1(t+1) = D_1(t) + \mathbb{P}\left(\mathbb{N}^{(t+1)}\right) %\\
= D_1(t) + \frac{N_{0}+vD_1(t)}{N_{0}+\rho t+(v+1)D_1(t)},
\end{split}
\end{align}
where $\mathbb{N}^{(t+1)}$ is the event of drawing at time $(t+1)$
a ball of a color that has not been observed before. 
Its probability $\mathbb{P}\left(\mathbb{N}^{(t+1)}\right)$ can be expressed as the number of colors in the urn yet to be discovered, $N_0 + (\nu + 1) D_1(t) - D_1(t)$, divided by the total number of balls available at time $t$ in the urn.
In the long time limit,  
Eq.~\eqref{eq:SI_master_equation_D1t}
can be approximated by a differential equation, which leads to an analytical expression for $D_1(t)$ (see Refs.~\cite{tria2014dynamics,tria2018zipf} for the analytical calculations):
\begin{equation}\label{eq:SI_sol_D_1_approx}
\begin{cases}
\dfrac{\mathrm{d}D_1(t)}{\mathrm{d}t}=\dfrac{N_{0}+vD_1(t)}{N_{0}+\rho t+(v+1)D_1(t)}\\
D_1(0) = 0
\end{cases}
\implies \quad D_1(t) \underset{t\to\infty}{\approx}
\begin{cases}
b t^{\beta_1}           & \text{if } \nu < \rho, \\
b \dfrac{t}{\log t}     & \text{if } \nu = \rho, \\
b t                     & \text{if } \nu > \rho,
\end{cases}
\end{equation}
where $\beta_1 = \nu / \rho$ and $b$ is a constant depending on $\nu$ and $\rho$.
In other words, in the sublinear case $\nu < \rho$, the Heaps' law is analytically verified, with asymptotic exponent $\beta_1 = \nu/\rho$~\cite{tria2014dynamics}. 

%
% PAIRS ANALYTICAL
%
\subsection{Second-order Heaps' law}
\label{sec:SI_second_order_heaps}

In order to write down an equation similar to Eq.~\eqref{eq:SI_sol_D_1_approx} for the the number $D_2(t)$ of different pairs that have appeared in the sequence  $\mathcal{S}_2$ of length $t$, i.e., 
\begin{equation}\label{eq:SI_dD2t_initial}
\begin{cases}
\dfrac{\mathrm{d}D_2(t)}{\mathrm{d}t} = %
\mathbb{P}(\text{``The }t\text{-th pair is new''}) ,\\
D_2(0) = 0,
\end{cases}
\end{equation}
we need to calculate the probability to observe a new pair.
However, differently from Eq.~\eqref{eq:SI_sol_D_1_approx}, such a probability depends not only on the total number of balls and on the number of extracted colors, but also on the number of balls of each extracted color.
Notice that the $t$-th pair $(x_1,x_2)$ 
of $\mathcal{S}_2$ is composed by the color $x_1$ drawn at time $t$ in $\mathcal{S}$ and the color $x_2$ drawn in the next time step. 
Hence, there are three separate events
in which the $t$-th pair $(x_1,x_2)$ is a novelty in $\mathcal{S}_2$: 
the event $\mathbb{A}$ in which $x_1$ is a novelty, i.e. it appears for the first time in the sequence $\mathcal{S}$ at time $t$; 
the event $\mathbb{B}$ in which $x_1$ is not a novelty but $x_2$ is a novelty; 
the event $\mathbb{C}$ in which both colors $x_1$ and $x_2$ are not novel, but the combination $(x_1,x_2)$ appears for the first time.
Consequently, the probability that the $t$-th pair is new is equal to the sum of the probabilities of such events. Using Eq.~\eqref{eq:SI_sol_D_1_approx}, 
for large values of $t$ 
the probability of event $\mathbb{A}$ can be written as
\begin{equation}\label{eq:SI_prob_A}
\mathbb{P}(\mathbb{A}) = \mathbb{P}\left(\mathbb{N}^{(t)}\right) = \dfrac{\mathrm{d}D_1(t)}{\mathrm{d}t}  
\underset{t\to\infty}{\approx} b\beta_1 t^{\beta_1-1}.
\end{equation}
%
% EVENT B
Similarly, denoting with $\overline{\mathbb{N}^{(t)}}$ the opposite event of $\mathbb{N}^{(t)}$,	
the probability of event $\mathbb{B}$ reads
\begin{align}\label{eq:SI_prob_B}
\begin{split}
\mathbb{P}(\mathbb{B}) \approx
\mathbb{P}\left(\overline{\mathbb{N}^{(t)}}\right)
\mathbb{P}\left(\mathbb{N}^{(t+1)}\right) 
\approx
\left(1-\dfrac{\mathrm{d}D_1(t)}{\mathrm{d}t}\right)
\dfrac{\mathrm{d}D_1(t+1)}{\mathrm{d}t} %\\
\approx b\beta_1 t^{\beta_1-1} = \mathbb{P}(\mathbb{A}),
\end{split}
\end{align}
Thirdly, we can compute the probability of the event $\mathbb{C}$ by calculating the probability that each possible pair of old colors is a novelty in this time step.
Since the number of old colors up to time $t$ is $D_1(t)$, indicating with $\mathbb{C}_{i,j}^{(t)}$ the event in which $i$ and $j$ are two already extracted colors and their pair $(i,j)$ is a novelty at time $t$ in $\mathcal{S}_2$, we can write:
	\begin{align}\label{eq:SI_prob_C}
	\begin{split}
	\mathbb{P}(\mathbb{C}) \approx
    \mathbb{P}\left(\overline{\mathbb{N}^{(t)}}\right)
    \mathbb{P}\left(\overline{\mathbb{N}^{(t+1)}}\right)
	\mathbb{P}\left( \bigcup_{i,j = 1}^{D_1(t)} \mathbb{C}_{i,j}^{(t)}\right) %\\
	\approx
	\left( 1- b\beta_1 t^{\beta_1-1} \right)^2\;
	\mathbb{P}\left(\bigcup_{i,j = 1}^{bt^{\beta_1}} \mathbb{C}_{i,j}^{(t)}\right) 
	%\\
	\approx
	\sum_{i,j = 1}^{bt^{\beta_1}} \mathbb{P}\left(\mathbb{C}_{i,j}^{(t)}\right).
	\end{split}
	\end{align}
	The last equality in Eq.~\eqref{eq:SI_prob_C} holds true because for any $(i_1,j_1)\neq(i_2,j_2)$ we have $\mathbb{C}_{i_1,j_1}^{(t)} \cap \mathbb{C}_{i_2,j_2}^{(t)} = \text{\O}$, since only one pair can be extracted at each time step, and we have disregarded lower infinitesimals.

Let us now concentrate on computing the probability of $\mathbb{C}_{i,j}(t)$.
	Defining the event $\mathbb{E}_{ij}^{(\tau)}$ = ``pair $(i,j)$ appears (not necessarily for the first time) in the sequence at time $\tau$", we can rewrite $\mathbb{C}_{i,j}^{(t)}$ as
	\begin{equation}\label{eq:SI_rewrite_C}
	    \mathbb{C}_{i,j}^{(t)} = \overline{\mathbb{E}_{ij}^{(1)}} \cap \overline{\mathbb{E}_{ij}^{(2)}} \cap \cdots  \cap \overline{\mathbb{E}_{ij}^{(t-1)}}  \cap \mathbb{E}_{ij}^{(t)},
	\end{equation}
	where we denote with $\overline{\mathbb{E}_{ij}^{(\tau)}}$ the opposite event of $\mathbb{E}_{ij}^{(\tau)}$.
	We can hence compute its probability as
	\begin{align}\label{eq:SI_rewrite_P_C}
	    \begin{split}
	    \mathbb{P}\left(\mathbb{C}_{i,j}^{(t)}\right) 
        &= 
    	    \mathbb{P}\left(
    	    \overline{\mathbb{E}_{ij}^{(1)}} \cap \overline{\mathbb{E}_{ij}^{(2)}} \cap \cdots  \cap \overline{\mathbb{E}_{ij}^{(t-1)}}  \cap \mathbb{E}_{ij}^{(t)}
    	    \right) \\
        &= \mathbb{P}\left(
    	    \overline{\mathbb{E}_{ij}^{(1)}}
    	    \right)
    	    \mathbb{P}\left(\overline{\mathbb{E}_{ij}^{(2)}} \,|\, \overline{\mathbb{E}_{ij}^{(1)}} \right)
    	    \cdots
    	   \mathbb{P}\left(\overline{\mathbb{E}_{ij}^{(t-1)}} \,|\, \overline{\mathbb{E}_{ij}^{(1)}} \cap \cdots \cap \overline{\mathbb{E}_{ij}^{(t-2)}} \right)
    	    \mathbb{P}\left(\mathbb{E}_{ij}^{(t)} \,|\, \overline{\mathbb{E}_{ij}^{(1)}} \cap \cdots \cap \overline{\mathbb{E}_{ij}^{(t-1)}} \right).
	    \end{split}
	\end{align}
	First, we notice that we can simplify the expressions in Eq.~\eqref{eq:SI_rewrite_P_C}, since
	\begin{align}\label{eq:SI_rewrite_prob_cond_all}
	    \begin{split}
	    &\mathbb{P}\left(\mathbb{E}_{ij}^{(\tau)} \,|\, \overline{\mathbb{E}_{ij}^{(1)}} \cap \cdots \cap \overline{\mathbb{E}_{ij}^{(\tau-1)}} \right)
	    =
	    \mathbb{P}\left(\mathbb{E}_{ij}^{(\tau)} \,|\, \overline{\mathbb{E}_{ij}^{(\tau-1)}} \right).
	    \end{split}
	\end{align}
	This equality in Eq.~\eqref{eq:SI_rewrite_prob_cond_all} holds true because, the probability of extracting the pair $(i,j)$ at time $\tau$ can only be influenced by what has happened at time $(\tau-1)$, disregarding all previous times.

    Without loss of generality, let us index the colors in the urn in the same order they first appeared in the sequence, i.e., let us suppose that the $i$-th color has appeared at time $t_i$, with $t_{i+1} > t_i$, for $i = 1,\,2,\dots, D_1(t)$. 
    Let us also suppose that the rate at which a new color appears is given exactly by the approximated solution given by Eq.~\eqref{eq:SI_sol_D_1_approx}. Then, it would be 
    \begin{equation}\label{eq:SI_approx_t_i}
    i=D(t_i) \approx b t_i^{\beta_1} \implies t_i \approx \left(\dfrac{i}{b}\right)^\frac{1}{\beta_1}. %\quad,\quad i = 1,2,3,\cdots.
    \end{equation}
    With Eq.~\eqref{eq:SI_approx_t_i} we are assuming that the \deletetext{behaviour}\newtext{behavior} of $D_1(t)$ at finite times can be approximated with the asymptotic one, and that colors appear deterministically at these expected moments. Even though strong, this assumption makes sense if we consider that, as it has been observed before, there is a good correspondence between this analytical solution and simulations at finite times. 
    Moreover, we will confirm {\it a posteriori} the suitability of this assumption since, as we will see, there is correspondence between the analytical solution of $D_2(t)$ we obtain here and the results of model simulations.

    Let us now define $n_i(t)$ as the number of times the color $i$ has appeared before time $t$, supposing it has first appeared at time $t_i \leq t$. 
    If $\mathbb{E}_{i}^{(t)}$ = ``$i$ appears at time $t$" (not necessarily for the first time), then we have that
    $
    \dfrac{dn_i}{\mathrm{d}t} = \mathbb{P}\left(\mathbb{E}_{i}^{(t)}\right)
    $. Thus, we can write:
    \begin{align}\label{eq:SI_dn_i}
    \begin{split}
        \begin{cases}
        \dfrac{dn_i(t)}{\mathrm{d}t} = 
        \dfrac{\rho n_i(t) + 1}{N_0 + a D(t) +\rho t}  \underset{t\to\infty}{\approx} \dfrac{n_i}{t}, \\
        n_i(t_i) = 1,
        \end{cases}
        \!\!\!\!\!\!\!\!\!\implies \!\!
        \begin{cases}
        n_i(t) \underset{t\to\infty}{\approx} \dfrac{t}{t_i}  & \text{if } t\geq t_i , \\
        n_i(t) = 0 & \text{if } t < t_i ,
        \end{cases}
        \!\!\implies \!\!
        \begin{cases}
        \dfrac{dn_i(t)}{\mathrm{d}t}  \underset{t\to\infty}{\approx} \dfrac{1}{t_i} =  \left(\dfrac{b}{i}\right)^\frac{1}{\beta_1}  & \text{if } t\geq t_i , \\
        \dfrac{dn_i(t)}{\mathrm{d}t} = 0 & \text{if } t < t_i .
        \end{cases}
    \end{split}
    \end{align}
    Let us observe that under these assumptions $dn_i/\mathrm{d}t$ is actually constant in time, depending just on $t_i$. 

    Then, supposing that the number of balls $n_i(\tau),\;n_j(\tau)$ of the two colors in the urn follows exactly Eq.~\eqref{eq:SI_dD2t_initial}, 
	we can calculate the probability of $\mathbb{E}_{ij}^{(\tau)}$ as
	\begin{equation}\label{eq:SI_P_ij}
    \mathbb{P}\left(\mathbb{E}_{ij}^{(\tau)}\right) = 
    \mathbb{P}\left(\mathbb{E}_{i}^{(\tau)}\right) \mathbb{P}\left(\mathbb{E}_{j}^{(\tau+1)}\right) = 
    \dfrac{dn_i(\tau)}{d\tau}\dfrac{dn_j(\tau+1)}{d\tau} \underset{t\to\infty}{\approx} 
    \begin{cases}
    \dfrac{1}{t_i t_j} &\text{if } \tau\geq \max(t_i, t_j-1), \\
    0 &\text{if } \tau < \max(t_i, t_j-1).
    \end{cases}
	\end{equation}
	Furthermore, if $\tau\geq \max(t_i, t_j-1)$, we can write
    {\small
	\begin{align}
    \begin{split}
	\mathbb{P}\left(\mathbb{E}_{ij}^{(\tau)} \cap \overline{\mathbb{E}_{ij}^{(\tau-1)}}\right) 
	&= 
	\mathbb{P}\left(\left[\left( \mathbb{E}_{ij}^{(\tau+1)} \cap \overline{\mathbb{E}_{ij}^{(\tau)}} \right) \cap \mathbb{E}_j^{(\tau)} \right] \cup \left[\left(\mathbb{E}_{ij}^{(\tau+1)} \cap \overline{\mathbb{E}_{ij}^{(\tau)}} \right) \cap \overline{\mathbb{E}_j^{(\tau)}} \right]\right) \\
	&= 
	\mathbb{P}\left(\left[\mathbb{E}_{i}^{(\tau)} \cap \mathbb{E}_{j}^{(\tau+1)} \cap \overline{\mathbb{E}_{i}^{(\tau-1)}} \cap \mathbb{E}_j^{(\tau)} \right] \cup \left[\mathbb{E}_{i}^{(\tau)} \cap \mathbb{E}_{j}^{(\tau+1)} \cap \overline{\mathbb{E}_j^{(\tau)}} \right]\right) \\
	&= 
	\mathbb{P}\left(\mathbb{E}_{i}^{(\tau)} \cap \mathbb{E}_j^{(\tau)} \right) \mathbb{P}\left( \mathbb{E}_{j}^{(\tau+1)}  \right) \mathbb{P}\left( \overline{\mathbb{E}_{i}^{(\tau-1)}} \right) + 
	\mathbb{P}\left(\mathbb{E}_{i}^{(\tau)} \cap \overline{\mathbb{E}_j^{(\tau)}} \right) \mathbb{P}\left(\mathbb{E}_{j}^{(\tau+1)} \right) \\
	&= 
	\delta(i,j) \dfrac{1}{t_i} \dfrac{1}{t_j} \left(1-\dfrac{1}{t_i} \right) + \left(1-\delta(i,j)\right)\dfrac{1}{t_i} \dfrac{1}{t_j}.
	\end{split}
	\end{align}
	}
	Therefore, we get
	\begin{align}\label{eq:SI_pair_prob_cond_no_ind}
    \begin{split}
	\mathbb{P}\left(\mathbb{E}_{ij}^{(\tau)} \,|\, \overline{\mathbb{E}_{ij}^{(\tau-1)}}\right)
	&= \frac
	{\mathbb{P}\left(\mathbb{E}_{ij}^{(\tau)} \cap \overline{\mathbb{E}_{ij}^{(\tau-1)}}\right)} {\mathbb{P}\left(\overline{\mathbb{E}_{ij}^{(\tau-1)}}\right)} = \frac
	{
	\delta(i,j) \dfrac{1}{t_i t_j} \left(1-\dfrac{1}{t_i} \right) + \dfrac{1-\delta(i,j)}{t_i t_j} 
	} 
	{1-\dfrac{1}{t_i t_j}} \\
	&= \frac
	{
	\delta(i,j)  \left(1-\dfrac{1}{t_i} \right) + \left(1-\delta(i,j)\right)
	} 
	{t_i t_j - 1} = 
     \delta(i,j)  \frac{\left(1-\dfrac{1}{t_i} \right)}{t_i t_j - 1} + 
     \left(1-\delta(i,j)\right)\frac{1} 	{t_i t_j - 1} \\
	&= \delta(i,j)  \frac{1}{t_i^2 + t_i} + 
    \left(1-\delta(i,j)\right)\frac{1} {t_i t_j - 1} .
	\end{split}
	\end{align}
	In the following of this discussion, we make the following approximation: 
	\begin{equation}\label{eq:SI_assume_independence}
	\delta(i,j)  \frac{1}{t_i^2 + t_i} + 
    \left(1-\delta(i,j)\right)\frac{1} {t_i t_j - 1}  \approx \dfrac{1}{t_i t_j}.
	\end{equation}
    Because of Eq.~\eqref{eq:SI_P_ij}, the approximation in Eq.~\eqref{eq:SI_assume_independence} implies in Eq.~\eqref{eq:SI_pair_prob_cond_no_ind} that  
	\begin{align}\label{eq:SI_assume_independence_consequence}
	\begin{split}
	    &\mathbb{P}\left(\mathbb{E}_{ij}^{(\tau)} \,|\, \overline{\mathbb{E}_{ij}^{(\tau-1)}}\right) % = 
     \approx 
	\begin{cases}
    \dfrac{1}{t_i t_j} & \text{if } \tau\geq \max(t_i, t_j-1)\\
	0 & \text{if } \tau<\max(t_i, t_j-1)
	\end{cases}
	\implies \mathbb{P}\left(\mathbb{E}_{ij}^{(\tau)} \,|\, \overline{\mathbb{E}_{ij}^{(\tau-1)}}\right) \approx \mathbb{P}\left(\mathbb{E}_{ij}^{(\tau)}\right),
	\end{split}
	\end{align}
	which is equivalent to assume that $\mathbb{E}_{ij}^{(\tau)}$ and  $\overline{\mathbb{E}_{ij}^{(\tau-1)}}$ are statistically independent, i.e. that the extraction of a certain pair $(i,j)$ at time $\tau$ is independent of its extraction at the previous time $(\tau-1)$. 
	Therefore, 
    using Eq.~\eqref{eq:SI_rewrite_P_C},  Eq.~\eqref{eq:SI_rewrite_prob_cond_all}, Eq.~\eqref{eq:SI_P_ij} and Eq.~\eqref{eq:SI_assume_independence_consequence}, 
    the probability of the event $\mathbb{C}_{ij}^{(t)}$ that the pair $(i,j)$ is extracted at time $t$ for the first time can be approximated to
	\begin{align}\label{eq:SI_Heaps_pairs_third_case_ij}
	\begin{split}
	\mathbb{P}\left(\mathbb{C}_{ij}^{(t)}\right)  
        &= 
            \mathbb{P}\left(
    	    \overline{\mathbb{E}_{ij}^{(1)}}
    	    \right)
    	    \mathbb{P}\left(\overline{\mathbb{E}_{ij}^{(2)}}\right)
    	    \cdots
    	    \mathbb{P}\left(\overline{\mathbb{E}_{ij}^{(t-1)}}\right)
    	    \mathbb{P}\left(\mathbb{E}_{ij}^{(t)}\right) %\\
        =
            \prod_{\tau = 1}^{t-1} \mathbb{P}\left(\overline{\mathbb{E}_{ij}^{(\tau)}}\right)   \mathbb{P}\left(\mathbb{E}_{ij}^{(t)}\right) \\
        &=
            \prod_{\tau = \max(t_i,t_j-1)}^{t-1} \left(1-\mathbb{P}\left(\mathbb{E}_{ij}^{(\tau)}\right)\right) \mathbb{P}\left(\mathbb{E}_{ij}^{(t)}\right) %\\
	= \left(1-\frac{1}{t_i t_j}\right)^{t-\max(t_i,\,t_j-1)}  \frac{1}{t_i t_j},
	\end{split}
	\end{align}
    which can be used in Eq.~\eqref{eq:SI_prob_C} to obtain an approximated expression for the probability of event $\mathbb{C}$, i.e., 
\begin{equation}\label{eq:SI_Heaps_pairs_third_case}
\mathbb{P}\left(\mathbb{C}\right) \approx
\sum_{i,j = 1}^{bt^{\beta_1}} 
\mathbb{P}\left(\mathbb{C}_{ij}^{(t)}\right)  
\approx
\sum_{i,j = 1}^{bt^{\beta_1}} 
\left(1-\frac{1}{t_i t_j}\right)^{t-\max(t_i,\,t_j-1)}  \frac{1}{t_i t_j}.
\end{equation}
Summing up, by inserting Eq.~\eqref{eq:SI_prob_A}, Eq.~\eqref{eq:SI_prob_B}, and Eq.~\eqref{eq:SI_Heaps_pairs_third_case} into Eq.~\eqref{eq:SI_dD2t_initial}, we get the following differential equation for the number of new pairs in time in the UMT:
\begin{multline}\label{eq:SI_differential_equation_heaps_pairs}
\frac{\mathrm{d}D_2}{\mathrm{d}t}  
 \underset{t\to\infty}{\approx} 2b\beta_1 t^{\beta_1-1} 
 % \\
 + \underbrace{\sum_{i,j = 1}^{bt^{\beta_1}} \left(1-\frac{1}{t_i t_j}\right)^{t-\max(t_i,\,t_j-1)} \frac{1}{t_i t_j}}_{\mathcal{C}(t)}.
\end{multline}

In order to have an estimate of $\mathcal{C}(t)$, let us approximate the sum with the related integral:
\begin{equation}
\label{eq:SI_to_integral}
\mathcal{C}(t)  \underset{t\to\infty}{\approx} \int\displaylimits_1^{bt^{\beta_1}} \int\displaylimits_1^{bt^{\beta_1}} \left(1-\frac{1}{t_x t_y}\right)^{t-\max(t_x,\,t_y-1)} \frac{1}{t_x t_y} \mathrm{d}x\mathrm{d}y.
\end{equation}
This way, using the change of variables $u=t_x = \left(\dfrac{x}{b}\right)^\frac{1}{\beta_1}$, $v = t_y = \left(\frac{y}{b}\right)^\frac{1}{\beta_1}$ ,
we get
\begin{align}
\begin{split}
\mathcal{C}(t)   \underset{t\to\infty}{\approx} 
(b\beta_1)^2 \!\!\int\displaylimits_{\frac{1}{\rho - \nu}}^t \int\displaylimits_{\frac{1}{\rho - \nu}}^t \left(1-\frac{1}{u v}\right)^{t-\max(u,\,v-1)} \frac{\mathrm{d}u\mathrm{d}v}{(u v)^{2-\beta_1}},
\end{split}
\end{align}
where we have substituted the initial value $\left(\dfrac{1}{b}\right)^\frac{1}{\beta_1} = \dfrac{1}{\rho - \nu}$, since $b=(\rho-\nu)^{\beta_1}$ when $\nu < \rho$, that is in this case~\cite{tria2014dynamics}.
Moreover, considering that $u$ and $v$ represent time variables, with $\tau \in (1,t)$, since between $t=0$ and $t=1$ there are no colors extracted yet, we can change the lower integral border to 1, 
i.e.,
\begin{align}\label{eq:SI_C(t)}
\begin{split}
\mathcal{C}(t)   \underset{t\to\infty}{\approx} 
(b\beta_1)^2 \!\!\int\displaylimits_{1}^t \!\!\int\displaylimits_{1}^t \left(1-\frac{1}{u v}\right)^{t-\max(u,\,v)} \frac{\mathrm{d}u\mathrm{d}v}{(u v)^{2-\beta_1}},
\end{split}
\end{align}
where we have also simplified the exponent in the integrand, so that we can more easily calculate it as
\begin{align}\label{eq:SI_C(t)_approximated}
\begin{split}
\mathcal{C}(t)   \underset{t\to\infty}{\approx} 
2\,(b\beta_1)^2 \!\!\int\displaylimits_{1}^t \!\!\int\displaylimits_{1}^u \left(1-\frac{1}{u v}\right)^{t-u} \frac{\mathrm{d}u\mathrm{d}v}{(u v)^{2-\beta_1}}.
\end{split}
\end{align}

We numerically solve the integral in Eq.~\eqref{eq:SI_C(t)_approximated} on specified points $t_i$ using the command \texttt{NIntegrate} of \textit{Mathemathica}~\cite{Mathematica}. The points $t_i$ have been chosen on a fine logarithmically spaced grid of $N=1601$ points $1 = t_0 < t_1 < \cdots < t_N = 10^{16}$. 
By plugging the numerical approximation $\mathcal{C}(t_i)$ into Eq.~\eqref{eq:SI_differential_equation_heaps_pairs}, we also obtain a numerical approximation of $\mathrm{d}D_2/\mathrm{d}t$ in these points. 
We also obtain an analytical approximation of $\mathrm{d}D_2/\mathrm{d}t$ by fitting a function of the type $a t^{b+c/(d+\log_2(t))}$ using \texttt{curve\_fit} (in \textit{Python}'s package \texttt{scipy}), where the minimization of the error has been done in logarithm scale. 
Finally, integrating Eq.~\eqref{eq:SI_differential_equation_heaps_pairs} over $t$, we obtain a solution for $D_2(t)$.
Again, we are not able to solve this integral analytically, so we solve it numerically using the analytical fit of $\mathrm{d}D_2/\mathrm{d}t$. In particular, we integrate using \textit{Python}'s command \texttt{odeint} in the \texttt{scipy} package. 
We find that the numerical integration for $D_2(t)$ can also be fitted by a function of the type $a t^{\beta_1+c/(d+\log_2(t))}$.

\medskip

To sum up, we have derived a solution of Eq.~\eqref{eq:SI_differential_equation_heaps_pairs} of the type
\begin{equation}\label{eq:SI_analytical_integration_D2}
    D_2(t) \approx a \,t^{\beta_2}, \qquad \text{ with }
     \beta_2 = \beta_1 + \frac{c}{d+\log(t)} ,
\end{equation}
where $a$, $c$, and $d$ depend on the parameters $\rho$ and $\nu$.
\begin{figure*}[bt]
\begin{center}
\includegraphics[width=\linewidth]{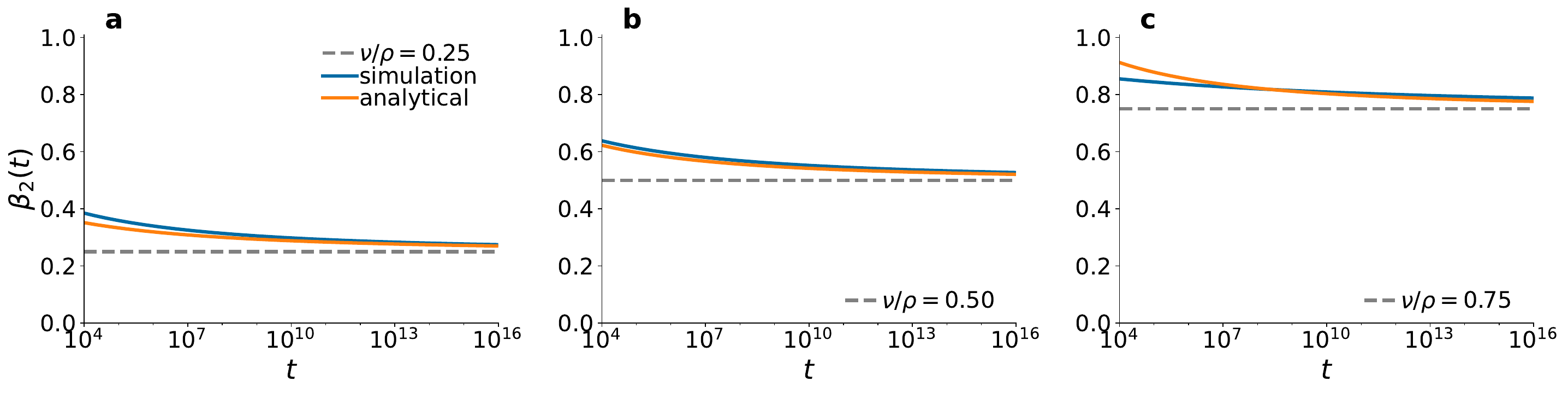}
\caption[$2$\textsuperscript{nd}-order Heaps' exponent in the urn model with triggering.]
{
\textbf{$2$\textsuperscript{nd}-order Heaps' exponent in the urn model with triggering.}
Temporal evolution of the $2$\textsuperscript{nd}-order Heaps' exponents $\beta_2(t)$ of the urn model with triggering according to the simulations (continuous blue line) and the numerical integration of Eq.~\eqref{eq:SI_differential_equation_heaps_pairs} (continuous orange line). 
Parameters are set to $\rho=4$ and $\eta=1$, while $\nu$ varies across panels: $\nu= 1$ (\textbf{a}), $2$ (\textbf{b}), and $3$ (\textbf{c}). 
Continuous lines are obtained by fitting $D_2(t)$ with a function $at^{\beta_1+c/(d+\log(t))}$, as in Eq.~\eqref{eq:SI_analytical_integration_D2}. 
The expected $1$\textsuperscript{st}-order Heaps' exponent in each panel, respectively equal to $\beta_1=\nu/\rho=0.25$, $0.5$, $0.75$, is displayed as a dashed gray horizontal line.
}
\label{fig:numeric_vs_analytic_simulations}
\end{center}
\end{figure*}
Fig.~\ref{fig:numeric_vs_analytic_simulations} shows that the analytical expression of $\beta_2$ we have found is in good agreement with the numerical simulations.
From left to right, we consider parameters $\rho = 4$ and $\nu = 1, 2, 3$, and we run simulations until $T = 10^7$.
In each plot, continuous lines represent the $2$\textsuperscript{nd}-order Heaps' exponents of the power-law fits as a function of time $t$.
The continuous blue line is obtained by fitting the best parameters $a$, $c$ and $d$ that minimize the error between the points $D_2(t)$ of the simulations with a function of the type $a \,t^{\beta_1 + \frac{c}{d+\log(t)}}$. 
The continuous orange line instead represents the result of our analytical approach in Eq.~\eqref{eq:SI_analytical_integration_D2}. 
The expected value of $\beta_1 = \nu / \rho$ is represented as a horizontal dashed gray line.
Our results further confirm that 2\textsuperscript{nd}-order Heaps' exponents differ from the $1$\textsuperscript{st}-order ones at finite times.
However, they also highlight that in the UMT the difference between $\beta_2$ and $\beta_1$ slowly decays in time.

%
% 3rd ORDER ANALYTICAL
%
\subsection{Higher-order Heaps' law}\label{sec:SI_higher_order_heaps}

Finally, we point out that an analytical solution for higher-order Heaps' exponents can also be obtained by induction, with assumptions similar to those used for the 2\textsuperscript{nd}-order one. 
For example, for the $3$\textsuperscript{rd}-order, we can repeat the same process as in Sec.~\ref{sec:SI_second_order_heaps} to compute the probabilities of obtaining a new triplet. 
In particular, supposing that
\begin{align}
\begin{split}
    & \dfrac{\mathrm{d}D_1(t)}{\mathrm{d}t} \approx a_1 t^{\beta_1-1}, \quad
    \dfrac{\mathrm{d}D_2(t)}{\mathrm{d}t} \approx a_2 t^{\beta_1 - 1 +\frac{c_2}{d_2+\log_2(t)}},
\end{split}
\end{align}
we can obtain a new triplet in the three following distinct cases.
\begin{enumerate}
	\item[$(\mathbb{A})$:] when at time $(t-1)$ a new pair is drawn, which happens with probability 
	\begin{equation}\label{eq:SI_prob_A_triplet}
	    \mathbb{P}(\mathbb{A}) = \dfrac{\mathrm{d}D_2(t-1)}{\mathrm{d}t} \approx a_2 t^{\beta_1 - 1 + \frac{c_2}{d_2+\log_2(t)}}.
	\end{equation}
	\item[$(\mathbb{B})$:] when at time $(t-1)$ an old pair is drawn, and at time $t$ a new color is drawn, which happens with probability 
	\begin{equation}\label{eq:SI_prob_B_triplet}
	\mathbb{P}(\mathbb{B}) = \left(1-\dfrac{\mathrm{d}D_2(t-1)}{\mathrm{d}t}\right)\dfrac{\mathrm{d}D_1(t)}{\mathrm{d}t}
	\approx 
	\left( 1- a_2 t^{\beta_1 - 1 +c_2/(d_2+\log_2(t))} \right) a_1 t^{\beta_1-1}
	\approx
	a_1 t^{\beta_1-1}
	.
	\end{equation}
	\item[$(\mathbb{C})$:] when at both times $(t-1)$ and $t$ an old pair and an old color are extracted, but the corresponding triplet has never appeared in the sequence before. 
	Following the same steps of the $2$\textsuperscript{nd}-order case, we get the probability
	\begin{align}\label{eq:SI_prob_C_triplet}
	\begin{split}
	    \mathbb{P}\left(\mathbb{C}\right) &\approx \sum_{i,j,k=1}^{bt^{\beta_1}} 
	    \left(1-\frac{1}{t_i t_j t_k}\right)^{t-\max(t_i,\,t_j,\,t_k)}  \frac{1}{t_i t_j t_k}
	    \approx \\
	    & \approx (a_1)^3 \!\!\int\displaylimits_{1}^t \!\!\int\displaylimits_{1}^t \!\!\int\displaylimits_{1}^t \left(1-\frac{1}{u v w}\right)^{t-\max(u,\,v-1,\,w-2)} \!\!\frac{\mathrm{d}u\mathrm{d}v\mathrm{d}w}{(u v w)^{2-\beta_1}}
	    \approx \\
	    & \approx 3! \, (a_1)^3 \!\!\int\displaylimits_{1}^t \!\!\int\displaylimits_{1}^u \!\!\int\displaylimits_{1}^v \left(1-\frac{1}{u v w}\right)^{t-u} \!\!\frac{\mathrm{d}u\mathrm{d}v\mathrm{d}w}{(u v w)^{2-\beta_1}},
	    \end{split}
	\end{align}
	where $3! = 3\cdot2\cdot1$.
\end{enumerate}
Then, summing up Eq.~\eqref{eq:SI_prob_A_triplet}, Eq.~\eqref{eq:SI_prob_B_triplet} and Eq.~\eqref{eq:SI_prob_C_triplet}, the probability to have a new triplet can be approximated as
\begin{equation}\label{eq:SI_triplet_heaps_analytical}
    \dfrac{\mathrm{d}D_3(t)}{\mathrm{d}t} \approx
        a_2 t^{\beta_1 - 1 +\frac{c_2}{d_2+\log_2(t)}} + 
        a_1 t^{\beta_1-1} + 
        3! \, (a_1)^3 \!\!\int\displaylimits_{1}^t \!\!\int\displaylimits_{1}^u \!\!\int\displaylimits_{1}^v \left(1-\frac{1}{u v w}\right)^{t-u} \!\!\frac{\mathrm{d}u\mathrm{d}v\mathrm{d}w}{(u v w)^{2-\beta_1}}.
\end{equation}

\medskip

In general, for the $n$\textsuperscript{th}-order Heaps' law, 
let us suppose by induction that all lower orders are known, i.e., for all orders $k = 1,\dots,n-1$, with $n\geq2$, we have
\begin{equation}
    \dfrac{\mathrm{d}D_k(t)}{\mathrm{d}t} \approx a_k t^{\beta_1 - 1 +\frac{c_k}{d_k+\log_2(t)}},  
    \quad 
    D_k(t) = \tilde{a}_k t^{\beta_1+\frac{c_k}{d_k+\log_2(t)}},
\end{equation}
with $a,c,d > 0$. Then, following the same procedure used for the $3$\textsuperscript{rd}-order Heaps' law, the probability of extracting a new $n$-tuple is given by:
\begin{multline}
    \dfrac{\mathrm{d}D_n(t)}{\mathrm{d}t} \approx
        \dfrac{\mathrm{d}D_{n-1}(t)}{\mathrm{d}t} + 
        \dfrac{\mathrm{d}D_1(t)}{\mathrm{d}t} 
        + n! \, (a_1)^n \!\!\underbrace{\int\displaylimits_{1}^t \int\displaylimits_{1}^{u_1} \!\cdots \!\int\displaylimits_{1}^{u_{n-1}}}_{n\text{ integrals}} \!\left(1-\frac{1}{u_1 \cdots u_n}\right)^{t-u_1}
        \!\frac{\mathrm{d}u_1\cdots\mathrm{d}u_n}{(u_1 \cdots u_n)^{2-\beta_1}},
\end{multline}
which approximately gives
\begin{equation}
    \dfrac{\mathrm{d}D_n(t)}{\mathrm{d}t} \approx a_n t^{\beta_1 - 1 +\frac{c_n}{d_n+\log_2(t)}},  
    \quad 
    D_n(t) = \tilde{a}_n t^{\beta_1+\frac{c_n}{d_n+\log_2(t)}}.
\end{equation}

\clearpage

\section{Analytical details of ERRWT model}
\label{sec:SI_ERRWT_analytic}
    
    In this section we provide an analytical insight of the model proposed in this manuscript, the Edge-Reinforced Random Walk with Triggering, or ERRWT.
    In particular, we refer to the definition of the ERRWT model given in \textit{Materials and Methods} in the main manuscript, and try to build differential equations for the evolution of $D_1(t)$ and $D_2(t)$. 
    From now on, we omit the explicit time dependence of the variables, e.g., $D_1 \equiv D_1(t)$, so that the mathematical expressions are easier to read.

    In the following analysis we make an important simplification, that is, we do not consider an undirected update as defined the main text.  Undirected update means that at any time a new link $(i,j)$ is reinforced or triggered, the link $(j,i)$ is updated as well; here we consider the directed version of the model (only the visited link $(i,j)$ is updated).

    We start from considering variables referring to single nodes: $D_{1i}^{in}, D_{1i}^{out},D_{2i}^{in},D_{2i}^{out}$ represent respectively the number of times a new node is discovered \textit{arriving} ($in$) in node $i$, and \textit{leaving} ($out$) from node $i$, and the same for the number of times a new link is discovered arriving or leaving from node $i$. Notice that $D_{1i}^{in}$ becomes 1 as soon as node $i$ is visited for the first time.
    
    These micro variables can be aggregated to obtain the macro variables $D_1$ and $D_2$, considering either $in$ or $out$ variables and summing over all the nodes:
    \begin{equation}
        D_1 = \sum_i D_{1i}^{in} = \sum_i D_{1i}^{out} \qquad \qquad D_2 = \sum_i D_{2i}^{in} = \sum_i D_{2i}^{out}
        \label{}
    \end{equation}

    Let us now build differential equations for the evolution of the micro variables, which will be aggregated to obtain self-consistent equations for $D_1$ and $D_2$. 
    Let us consider the probability of exploring a new node starting from node $i$, i.e., the probability that the variable $D_{1i}^{out}$ increases by $1$. 
    On the one hand, the total weight of the links outgoing from node $i$ is equal to 
    \begin{equation}\label{eq:SI_numerator_D1i_ERRWT}
        M_{0i}+\rho n_i + (\nu_1+1) D_{1i}^{in} + (\nu_2 + 1) D_{2i}^{in},
    \end{equation}
    where $M_{0i}$ is the initial number of links connected with node $i$ at time $t=0$, and $n_i \equiv n_i(t)$ is the number of times node $i$ has been visited up to time $t$. The other two terms refer to the new links triggered when arriving in node $i$. Indeed, when $i$ is visited for the first time, $(\nu_1+1)$ links outgoing from $i$ to new nodes are triggered. Moreover, whenever a link ending in $i$ is traversed for the first time, $(\nu_2 + 1)$ new links from $i$ to other explored nodes are triggered.
    On the other hand the total weight of links connecting $i$ and never explored nodes is equal to 
    \begin{equation}\label{eq:SI_denominator_D1i_ERRWT}
        M_{0i} + (\nu_1+1) D_{1i}^{in} - D_{1i}^{out},
    \end{equation}
    i.e., the initial number of nodes connected to $i$ yet to be discovered, plus the number of nodes triggered when discovering node $i$, minus the number of nodes already discovered starting from $i$. 
    These considerations make possible to write that
    \begin{equation}
        \dfrac{d D_{1i}^{out}}{dt} = p(i,t) \dfrac{M_{0i} + (\nu_1+1) D_{1i}^{in} - D_{1i}^{out}}{M_{0i}+\rho n_i + (\nu_1+1) D_{1i}^{in} + (\nu_2 + 1) D_{2i}^{in}},
        \label{eq:SI_ERRWT_D1_out}
    \end{equation}
    where $p(i,t)$ is the probability of being on node $i$ at time $t$, which is a needed condition for $D_{1i}^{out}$ to evolve.
    Using the same argument, we can also write an equation for the evolution of $D_{2i}^{out}$:
    \begin{equation}
        \dfrac{d D_{2i}^{out}}{dt} = p(i,t) \dfrac{M_{0i} + (\nu_1+1) D_{1i}^{in} + (\nu_2 + 1) D_{2i}^{in} - D_{2i}^{out}}{M_{0i}+\rho n_i + (\nu_1+1) D_{1i}^{in} + (\nu_2 + 1) D_{2i}^{in}}
        \label{eq:SI_ERRWT_D2_out}  
    \end{equation}

    Notice that Eq.~\eqref{eq:SI_ERRWT_D1_out} and Eq.~\eqref{eq:SI_ERRWT_D2_out} cannot be obtained so easily in the undirected case. In fact, here we have implicitly assumed that any link in the adjacent possible that has never been visited before has weight $1$. However, if the update is undirected, we may reinforce some link $(j,i)$ never traversed before only because the walker might have visited the edge $(i,j)$, making impossible to know the actual weight of never traversed links. 
    
\begin{figure*}[!tb]
        \begin{center}
        \includegraphics[width=\linewidth]{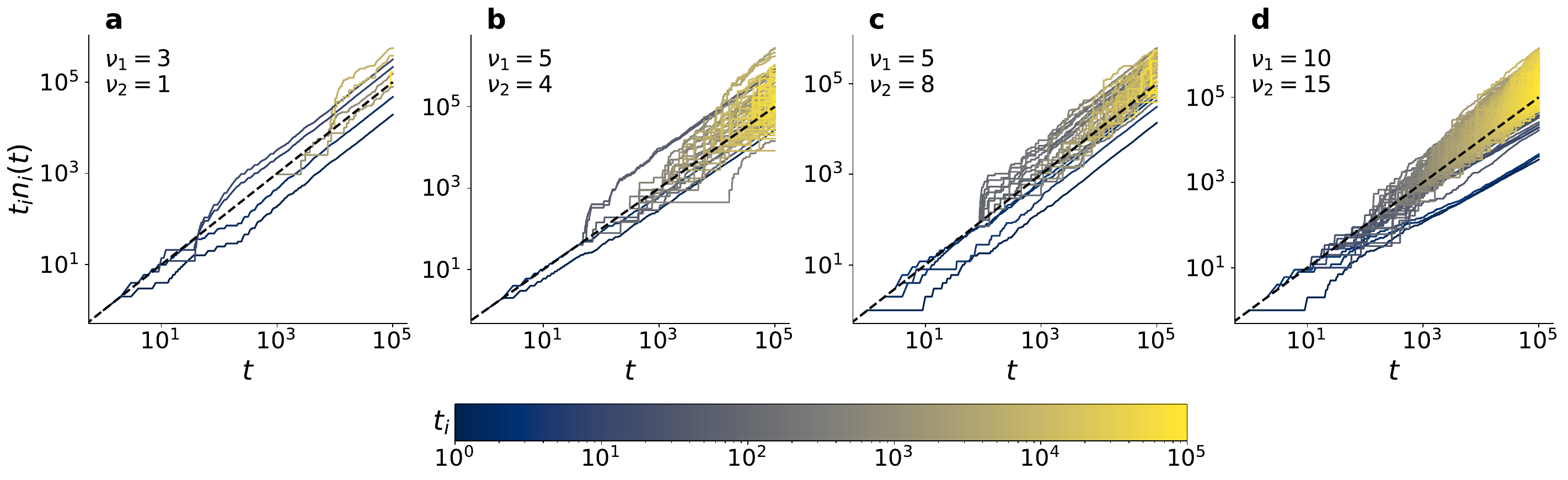}
        \caption[Temporal evolution of the quantities $\boldsymbol{n_i(t)}$.]
        {\textbf{Temporal evolution of the quantities $\boldsymbol{n_i(t)}$.} 
        In these figures we show the temporal evolution of $n_i(t)$, i.e. the number of times node $i$ has been explored at time $t$, for many choices of the node $i$. In order to check the assumption $n_i(t) \sim t/t_i$, where $t_i$ is the time when node $i$ is discovered for the first time, we actually plotted $t_i n_i(t)$ vs $t$. We expect this quantity to go like $t$, which is represented by the dotted black line. As we can see from the four panels, the assumption is valid for a wide range of the parameters $\nu_1$ and $\nu_2$.
        }
        \label{fig:SI_ERRWT_n_i_time_behavior}
        \end{center}
    \end{figure*}

    At this point we make another assumption in order to make the equations solvable. In particular, we assume a precise expression for the variable $n_i$. 
    In fact, as we have seen in Sec.~\ref{sec:SI_Heaps_UMT_analytic}, in the UMT, at least in the sublinear regime, we have $n_i(t) \sim t/t_i$, where $t_i$ is the first time item (node) $i$ has been visited~\cite{tria2014dynamics}. Exploiting the analogy between the UMT and the ERRWT model we assume that $n_i(t)$ has the same \deletetext{behaviour}\newtext{behavior}. 
    We also checked numerically the validity of this assumption.
    We have indeed measured the evolution of $n_i(t)$ in simulations of the ERRWT model, showing that the assumption is reasonable for very different values of the parameters $\nu_1$ and $\nu_2$, as shown in Fig.~\ref{fig:SI_ERRWT_n_i_time_behavior}.
    Further notice that $p(i,t) = dn_i(t-1)/dt \approx 1/t_i$, since the probability of being on $i$ at time $t$ is equal to the probability to move to node $i$ in the previous time step. With all these elements we can rewrite Eq.~\eqref{eq:SI_ERRWT_D1_out} and Eq.~\eqref{eq:SI_ERRWT_D2_out} as
    \begin{equation}
        \dfrac{d D_{1i}^{out}}{dt} \approx \dfrac{1}{t_i} \dfrac{M_{0i} + (\nu_1+1) D_{1i}^{in} - D_{1i}^{out}}{M_{0i}+\rho \dfrac{t}{t_i} + (\nu_1+1) D_{1i}^{in}+ (\nu_2 + 1) D_{2i}^{in}}
        \label{}  
    \end{equation}
    and
    \begin{equation}
        \dfrac{d D_{2i}^{out}}{dt}  \approx \dfrac{1}{t_i}  \dfrac{M_{0i} + (\nu_1+1) D_{1i}^{in} + (\nu_2 + 1) D_{2i}^{in} - D_{2i}^{out}}{M_{0i}+\rho \dfrac{t}{t_i} + (\nu_1+1) D_{1i}^{in} + (\nu_2 + 1) D_{2i}^{in}}.
        \label{}  
    \end{equation}
    The last step before aggregating the equations is to further simplify the denominator. First notice that $D_{1i}^{in}$ is a variable which can only take values $0$ or $1$, since an arriving node can result to be new only one time (this is not true for $D_{1i}^{out}$, which can be larger than $1$). 
    So we can neglect it with respect to the term with $D_{2i}^{in}$, because also this can be larger than $1$ and can go to infinity with time with a pace dependent on the parameters as we will see later. Finally, we assume $D_{2i}^{in} \approx D_2 / t_i$; this is a reasonable assumption given the fact that $n_i(t) \approx t/t_i$. Indeed, if a node $i$ is visited with a frequency depending on the inverse of $t_i$, it is reasonable to assume that also the number of new links traversed arriving in node $i$ occurs with the same frequency as well.

    We can finally aggregate the equations summing over all nodes $i$ obtaining 
 a self consistent equation for the evolution of $D_1$:
    \begin{equation}
    \begin{split}
        \dfrac{d D_1}{dt} &= \sum_{i=1}^{D_1} \dfrac{d D_{1i}^{out}}{dt} 
        \approx \sum_i \dfrac{1}{t_i} \dfrac{M_{0i} + (\nu_1+1) D_{1i}^{in} - D_{1i}^{out}}{\rho \dfrac{t}{t_i} + (\nu_2 + 1) \dfrac{D_2}{t_i}}
        \approx \sum_i  \dfrac{M_{0i} + (\nu_1+1) D_{1i}^{in} - D_{1i}^{out}}{\rho t + (\nu_2 + 1) D_2}
        \\
        &= \dfrac{M_0 + (\nu_1+1) D_1 - D_1}{\rho t + (\nu_2 + 1) D_2} \approx 
        \dfrac{\nu_1 D_1}{\rho t + (\nu_2 + 1) D_2},
    \end{split}
    \label{eq:SI_ERRWT_D1}
    \end{equation}
    where in the last approximation we have disregarded $M_0$ in the numerator since $D_1(t)\to \infty$ is the leading term in the numerator. Similarly for the $2$\textsuperscript{nd}-order Heaps' law we can write
    \begin{equation}
    \begin{split}
        \dfrac{d D_2}{dt}&= \sum_{i=1}^{D_1} \dfrac{d D_{2i}^{out}}{dt} 
        \approx \sum_{i=1}^{D_1} \dfrac{1}{t_i} \dfrac{M_{0i} + (\nu_1+1) D_{1i}^{in} + (\nu_2 + 1) D_{2i}^{in} - D_{2i}^{out}}{\rho \dfrac{t}{t_i} + (\nu_2 + 1) \dfrac{D_2}{t_i}} \\
        &= \dfrac{M_0 + (\nu_1 + 1) D_1  + (\nu_2 + 1) D_2-D_2}{\rho t + (\nu_2 + 1) D_2} \approx \dfrac{(\nu_1 + 1) D_1 + \nu_2 D_2}{\rho t + (\nu_2 + 1) D_2}.
    \end{split}
    \label{eq:SI_ERRWT_D2}
    \end{equation}
    
\begin{figure*}[!tb]
        \begin{center}
        \includegraphics[width=\linewidth]{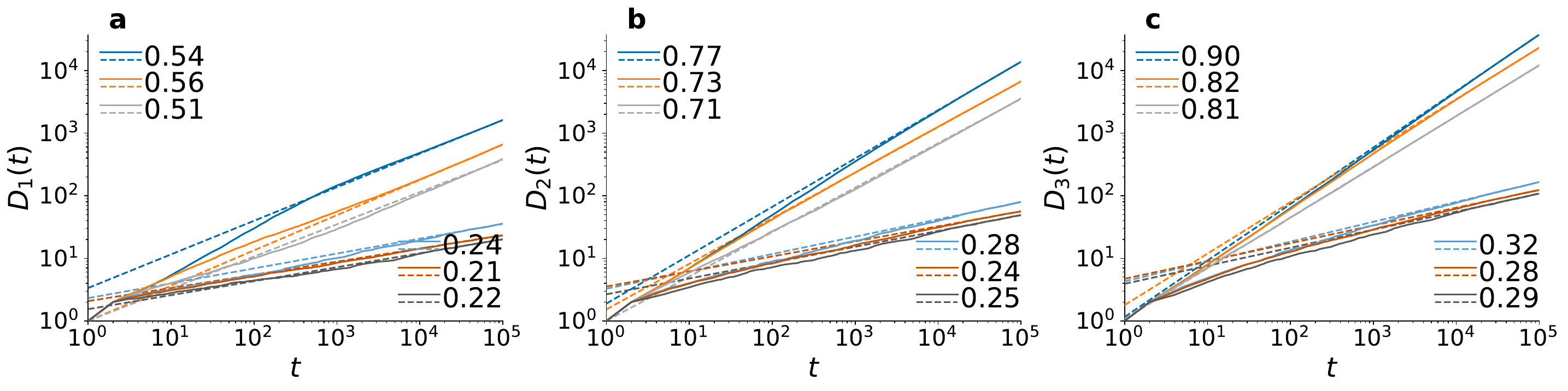}
        \caption[Heaps' exponents for different choices of the initial conditions.]
        {
        \textbf{Heaps' exponents for different choices of the initial conditions.} 
        The three panels show the \deletetext{behaviour}\newtext{behavior} of $D_1(t)$, $D_2(t)$ and $D_3(t)$ versus $t$ for three different initial conditions (i.e., initial structure of the network). 
        In particular, we consider a regular tree with branching parameter $\nu_1+1$ and number of levels equal to 1 (gray lines), 2 (orange lines), 3 (blue lines). 
        All nodes apart from the leaves are considered known (or discovered, or triggered) by the ERRWT at the beginning of the simulation.
        In each panel, the lines with higher Heaps' law (see top left legend), refer to simulations with $\rho = 10$, $\nu_1 = 8$, and $\nu_2 = 8$, while the other lines (see bottom right legend) with $\rho = 10$, $\nu_1 = 3$, and $\nu_2 = 1$.
        In the legend, the related extracted power-law exponent is reported.
        As we can see, the exponents measured in the three cases are similar across order, thus showing that the initial structure of the network is not relevant for the asymptotic \deletetext{behaviour}\newtext{behavior} of the ERRWT.
        }
        \label{fig:SI_ERRWT_different_initial_conditions}
        \end{center}
    \end{figure*}

\begin{figure}[!tb]
    \begin{center}
    \includegraphics[width=0.35\linewidth]{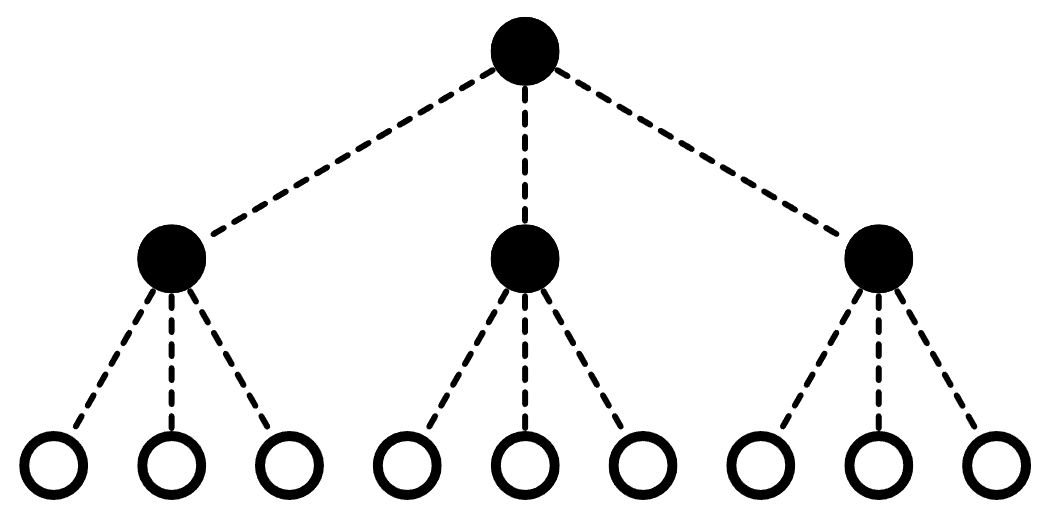}
    \caption[Representation of the initial network structure used in simulations of the ERRWT model in the main.]
    {
    \textbf{Representation of the initial network structure used in simulations of the ERRWT model in the main.}
    Although any initial network structure can be used for the ERRWT model, in the simulations shown in the main manuscript we consider a regular tree with branching parameter $\nu_1+1$ (equal to $3$ in the figure) and 2 levels. This structure resembles the way new nodes are triggered during the exploration, so that the root and first layer (full nodes) are considered triggered and known, while the leaves (empty nodes) are considered new.
    All links are regarded as new (represented as dashed).
    The choice of this tree has been done to ensure that the triggering of new edges finds nodes that are already known by the random walker.
    }
    \label{fig:SI_initial_structure}
    \end{center}
    \end{figure}

    Notice that the initial structure of the network only enters in the equations through the constant $M_0 \equiv \sum_i M_{0i}$, and as we already said this term can be safely neglected with respect to the other variables. This means that the asymptotic \deletetext{behaviour}\newtext{behavior} of $D_1$ and $D_2$, and so the exponents $\beta_1$ and $\beta_2$, should not depend on the initial structure of the network. We checked this fact running simulations with different initial conditions and measuring the exponents $\beta_1$ and $\beta_2$, checking that we obtain similar result in all cases. The results of this analysis is shown in Fig.~\ref{fig:SI_ERRWT_different_initial_conditions}.
    In particular, we consider regular trees with different number of levels, but same branching size.
    The idea is that we start from a node and trigger nodes, adding new levels of the tree.
    Therefore, the first initial network we consider is made by a root, considered triggered, connected to $\nu_1+1$ new nodes.
    The second one adds another level to the first one. Therefore, it is a regular tree with branching size $(\nu_1+1)$ and 2 levels. Here, the root and the first level are considered triggered, while the leaves are still new.
    This structure is the same used in the simulations of the main text (see also Fig.~\ref{fig:SI_initial_structure}).
    Finally, the third one adds one more level, thus being much bigger than the previous ones.
    In the panels we show only two sets of parameters, but we find comparable results for other choices of the parameters. 
    For both sets of parameters, we find that by increasing the number of levels (and hence the number of nodes and links) in the initial network, the higher-order Heaps' exponents slightly increase. 
    Moreover, the bigger the network, the longer we see a transient time in which there is a much higher Heaps' exponent. For example, see the blue line in Fig.~\ref{fig:SI_ERRWT_different_initial_conditions}(\textbf{a}), where we can clearly find the initial higher slope. Nevertheless, notice that after this period, the pace of discovery, i.e., the exponent, seems to be similar across different initial conditions, thus showing that the initial structure of the network is not relevant for the asymptotic \deletetext{behaviour}\newtext{behavior} of the ERRWT model.
    % INITIALIZATION
    \medskip
    
    Now, using Eq.~\eqref{eq:SI_ERRWT_D1} and Eq.~\eqref{eq:SI_ERRWT_D2}, we are able to work out an analytical expression for the two exponents $\beta_1$ and $\beta_2$.
    Let us consider various cases.
    First, assuming a sublinear regime for $D_2$ (so that it can be neglected with respect to $t$), in the large time limit we can further simplify the equations and get
    \begin{equation}
    % \begin{cases}
        \dfrac{d D_1}{dt} \approx  \dfrac{ \nu_1 D_1}{ \rho t},
        % \\
    \end{equation}
    \begin{equation}
        \dfrac{d D_2}{dt} \approx \dfrac{(\nu_1+1)D_1+\nu_2 D_2}{\rho t }.
    % \end{cases}
        \label{}  
    \end{equation}
    Solving both of these equations, we obtain an explicit expression for the two exponents $\beta_1$ and $\beta_2$; in the sublinear case we hence have asymptotically 
    \begin{equation}
        \beta_1 = \dfrac{\nu_1}{\rho} \quad,\quad \beta_2 = \max \left(\dfrac{\nu_1}{\rho} , \dfrac{\nu_2}{\rho} \right)
        \label{eq:SI_ERRWT_exponents_sublinear_regime}
    \end{equation}
    From the expression of $\beta_1$ and $\beta_2$ in Eq.~\eqref{eq:SI_ERRWT_exponents_sublinear_regime}, we get that the sublinear regime holds only if $\nu_1 < \rho$ and $\nu_2<\rho$.

    \medskip
    
    Before moving on to the other regimes, let us notice that $\beta_2$ is constrained to be at most equal to $2 \beta_1$. This is because if the number of nodes available to explore in the network is $O(N)$, then the number of available edges is $O(N^2)$. This means that $D_2$ can at most grow as the square of $D_1$ in the large time limit, imposing a constraint on the related exponents.  
    Let us now consider the case in which $D_2$ grows linearly in time, but not $D_1$.
    Notice that this can happen only provided that $\nu_1 > \rho/2$; in fact, since $\beta_2$ is constrained to be smaller or equal than $2 \beta_1$, then it would not be possible for $\beta_2$ to be equal to $1$ if $\beta_1 = \nu_1 /\rho < 1/2$. 
    This regime can be obtained substituting a linear expression for $D_2 \sim a t$ into Eq.~\eqref{eq:SI_ERRWT_D2}. 
    In this case, if we assume a sublinear \deletetext{behaviour}\newtext{behavior} for $D_1$, we can neglect the second term in the numerator, obtaining
    \begin{equation}
        \dfrac{dD_2}{dt} \approx \dfrac{\nu_2 a t}{\rho t + (\nu_2 + 1) a t} = \dfrac{\nu_2 a }{\rho  + (\nu_2 + 1) a } = a \qquad \implies \qquad a = \dfrac{\nu_2-\rho}{(\nu_2 + 1)},
        \label{}  
    \end{equation}
    thus showing that the condition for this regime to exist is $\nu_2 > \rho$, otherwise the coefficient $a$ would be negative.
    Then, we can substitute $D_2 = \dfrac{\nu_2-\rho}{(\nu_2 + 1)} t$ into Eq.~\eqref{eq:SI_ERRWT_D1}, to get the actual value of $\beta_1$: 
    \begin{equation}
        \dfrac{dD_1}{dt} \approx \dfrac{\nu_1 D_1}{\rho t + (\nu_2-\rho) t} = \dfrac{\nu_1 D_1}{\nu_2 t} \qquad \implies \qquad \beta_1 = \dfrac{\nu_1}{\nu_2}.
        \label{}  
    \end{equation}
    Therefore, $D_1$ keeps growing sublinearly provided that $\nu_1 < \nu_2$.
    Notice that in this case there are no conditions on the value of $\nu_1$, which can also be larger than $\rho$. 
    Reminding also the network constraint $\beta_2 \leq 2 \beta_1$, we have that this regime holds provided that $\beta_1 > 1/2$, which means $2 \nu_1 > \nu_2$. 

    \medskip 
    
    Finally, there is one last regime, in which both $D_1$ and $D_2$ are linear, i.e., with exponents $\beta_1=\beta_2=1$. 
    Substituting the two linear expressions $D_2 \sim at$ and $D_1 \sim bt$ in Eq.~\eqref{eq:SI_ERRWT_D1} and Eq.~\eqref{eq:SI_ERRWT_D2}, we obtain the following system of equations
    \begin{equation}
        \begin{cases}
            \dfrac{dD_1}{dt} \approx \dfrac{\nu_1 b}{\rho + (\nu_2 + 1) a} = b  \vspace{0.3cm}
            \\
            \dfrac{dD_2}{dt} \approx \dfrac{(\nu_1 + 1)b + \nu_2 a}{\rho + (\nu_2 + 1) a} = a      
        \end{cases}
        \label{}  
    \end{equation}
    from which we can work out the values of the two coefficients:
    \begin{equation}
        a = \dfrac{\nu_1-\rho}{(\nu_2 + 1)} \qquad \qquad b= \dfrac{(\nu_1-\rho)}{(\nu_2 + 1)} \dfrac{\left( \nu_1-\nu_2 \right)}{(\nu_1+1)},
        \label{}  
    \end{equation}
    which give the conditions $\nu_1 > \rho$ and $\nu_1 > \nu_2$ for this regime to hold. 
    This comes out from the fact that, as we have seen before, whenever $\nu_2>\nu_1$ we have a sublinear regime for $D_1$.

    \bigskip 
    
    Summarizing the predicted exponents for the directed version of the model for any choice of the parameter $\nu_1, \nu_2$ and $\rho$, we have:
    \begin{equation}
        \begin{cases}
            \nu_2 < \rho, \ \nu_1 < \rho  & \beta_1 = \dfrac{\nu_1}{\rho}, \ \beta_2 = \min\left(\max \left( \dfrac{\nu_1}{\rho}, \dfrac{\nu_2}{\rho}\right),\dfrac{2\nu_1}{\rho} \right) \\
            \nu_2 \geq \rho, \  \nu_1 \leq \dfrac{\rho}{2} 
            & \beta_1 = \dfrac{\nu_1}{\rho}, \ \beta_2 = \dfrac{2\nu_1}{\rho} \\ 
            \nu_2 \geq \rho, \ \dfrac{\rho}{2}  < \nu_1 < \nu_2
            & \beta_1 = \dfrac{\nu_1}{\nu_2}, \ \beta_2 = 1 \\
            \nu_1 \geq \rho, \  \nu_1 \geq \nu_2
            & \beta_1 = \beta_2 = 1 \\ 
        \end{cases}
        \label{}  
    \end{equation}
    The results above give us an analytical overview of a simplified version of the model, which can still provide the phenomenology we are interested in. In fact, with this analysis we still obtain a different \deletetext{behaviour}\newtext{behavior} for $D_1$ and $D_2$ with two different Heaps' exponents $\beta_1$ and $\beta_2$, which are controlled by the parameters $\nu_1$ and $\nu_2$, given $\rho$.

\end{document}